\documentclass[fleqn,usenatbib]{mnras}

\usepackage{newtxtext,newtxmath}
\usepackage{graphicx}	% Including figure files
\usepackage{amsmath}	% Advanced maths commands
\usepackage{color} 
\usepackage{rotating}
\usepackage{lscape}
\usepackage[normalem]{ulem} 
\usepackage[T1]{fontenc} 

\DeclareRobustCommand{\VAN}[3]{#2}
\let\VANthebibliography\thebibliography
\def\thebibliography{\DeclareRobustCommand{\VAN}[3]{##3}\VANthebibliography}

\usepackage{multirow}
\usepackage{adjustbox}
\usepackage{xcolor}
\usepackage{array}
\usepackage[para]{threeparttable}
\defcitealias{Nair2010}{NA10}
\defcitealias{VazquezMata2022}{VM22}

\title[Morphology via Galaxy Parameters using ML]{Morphological Classification of Galaxies Through Structural and Star Formation Parameters Using Machine Learning}

\author[G. Aguilar-Argüello et al.]{
G. Aguilar-Argüello,$^{1}$\thanks{E-mail: gy\_alagar@ciencias.unam.mx}
G. Fuentes-Pineda,$^{1}$
H. M. Hernández-Toledo,$^{2}$
L. A. Martínez-Vázquez$^{2}$
\newauthor{
J. A. Vázquez-Mata,$^{2}$\thanks{E-mail: jvazquez@astro.unam.mx}
S. Brough,$^{3}$
R. Demarco,$^{4}$
A. Ghosh,$^{5}$
Y. Jiménez-Teja,$^{6}$
G. Martin,$^{7}$
}
\newauthor{
W. J. Pearson,$^{8}$
and
C. Sifón$^{9}$
}
\\
$^{1}$Instituto de Investigaciones en Matem\'aticas Aplicadas y en Sistemas, Universidad Nacional Aut\'onoma de M\'exico, A.P. 20-126, 04510, M\'exico, CDMX, M\'exico
\\
$^{2}$Universidad Nacional Aut\'onoma de M\'exico. Instituto de Astronom\'ia. A.P. 70–264, 04510. Ciudad de M\'exico, M\'exico
\\
$^{3}$School of Physics, University of New South Wales, NSW 2052, Australia
\\
$^{4}$Institute of Astrophysics, Facultad de Ciencias Exactas, Universidad Andr\'es Bello, Sede Concepci\'on, Talcahuano, Chile
\\
$^{5}$DiRAC Institute and the Department of Astronomy, University of Washington, Seattle, WA, U.S.A
\\
$^{6}$Instituto de Astrof\'isica de Andaluc\'ia–CSIC, Glorieta de la Astronom\'ia s/n, E–18008 Granada, Spain
\\
$^{7}$School of Physics \& Astronomy, University of Nottingham, University Park, Nottingham NG7 2RD, UK
\\
$^{8}$National Centre for Nuclear Research, Pasteura 7, 02-093 Warszawa, Poland
\\
$^{9}$Instituto de F\'isica, Pontificia Universidad Cat\'olica de Valpara\'iso, Casilla 4059, Valpara\'iso, Chile
}

\date{Accepted XXX. Received YYY; in original form ZZZ}

\pubyear{2024}

\begin{document}
\label{firstpage}
\pagerange{\pageref{firstpage}--\pageref{lastpage}}
\maketitle

% Abstract of the paper
\begin{abstract}
We employ the XGBoost machine learning (ML) method for the morphological classification of galaxies into two (early-type, late-type) and five (E, S0--S0a, Sa--Sb, Sbc--Scd, Sd--Irr) classes, using a combination of non-parametric ($C,\,A,\,S,\,A_S,\,\mathrm{Gini},\,M_{20},\,c_{5090}$), parametric (S\'ersic index, $n$), geometric (axial ratio, $BA$), global colour ($g-i,\,u-r,\,u-i$), colour gradient ($\Delta (g - i)$), and asymmetry gradient ($\Delta A_{9050}$) information, all estimated for a local galaxy sample ($z<0.15$) compiled from the Sloan Digital Sky Survey (SDSS) imaging data. We train the XGBoost model and evaluate its performance through multiple standard metrics. Our findings reveal better performance when utilizing all fourteen parameters, achieving accuracies of 88\% and 65\% for the two-class and five-class classification tasks, respectively. In addition, we investigate a hierarchical classification approach for the five-class scenario, combining three XGBoost classifiers. We observe comparable performance to the ``direct'' five-class classification, with discrepancies of only up to 3\%. Using SHAP (an advanced interpretation tool), we analyse how galaxy parameters impact the model's classifications, providing valuable insights into the influence of these features on classification outcomes. Finally, we compare our results with previous studies and find them consistently aligned.
\end{abstract}

% Select between one and six entries from the list of approved keywords.
% Don't make up new ones.
\begin{keywords}
methods: data analysis -- methods: miscellaneous -- galaxies: general -- galaxies: structure
\end{keywords}

%%%%%%%%%%%%%%%%%%%%%%%%%%%%%%%%%%%%%%%%%%%%%%%%%%

%%%%%%%%%%%%%%%%% BODY OF PAPER %%%%%%%%%%%%%%%%%%

\section{Introduction}

According to the standard Lambda Cold Dark Matter ($\Lambda$CDM) paradigm, galaxies originate in dark matter haloes undergoing a process of continuous mergers in a first stage. However, as the universe expands and redshift decreases, these mergers become less frequent and galaxy evolution becomes internally driven by the so-called secular processes. All these processes collectively shape what we observe today as galaxy morphology.

Since Hubble’s morphological classification scheme \citep[][]{Hubble1926, Hubble1936}, galaxies have been systematically categorized, revealing that morphology is strongly correlated with star formation activity \citep[][]{ Strateva2001, Blanton2005, Blanton2009, Kormendy2009, Conselice2014}. For instance, the bimodal distribution of galaxies in optical colours, characterized by a blue cloud, a red sequence, and an intermediate green valley, is highly linked to morphological types \citep[e.g.][]{Strateva2001, Blanton2003, Baldry2006,Pearson2021}. Furthermore, morphology correlates with various physical properties, such as stellar distribution, colour, environment, gas content, size, and kinematics \citep[e.g.][]{Blanton2005, Zehavi2005, Cappellari2011, PerezMillan2023, Ghosh2024}. Therefore, the classification of galaxies in the nearby universe, based on their physical attributes, is fundamental to understanding their formation and evolution.

For years, many authors have tried to classify galaxies through a traditional visual classification, where they manually assign morphological types directly from images \citep[e.g.][]{Fukugita2007, Nair2010, VazquezMata2022}. However, this method becomes inefficient as the number of galaxies to classify increases. This challenge is amplified by the current and next generation of large-scale surveys, like DESI \citep[][]{DESI2016}, Euclid \citep[][]{Laureijs2011}, Rubin-LSST \citep[][]{Ivezic2008,Ivezic2019, Robertson2019}, and SKA \citep[][]{Braun2019}, which will generate unprecedented amounts of high-resolution data (images and spectra for billions of objects in the Universe), making it essential to develop new sophisticated and efficient tools to process and analyse such a vast information.

In recent years, several approaches have emerged for the large-scale morphological analysis and automatic classification of galaxies. Visually-based methods, such as the Galaxy Zoo project \citep[GZ;][]{Lintott2008,Willett2013,Walmsley2022}, have exploited human pattern recognition capabilities through crowdsourcing to successfully classify thousands of galaxies over a few years. However, even at that rate, classifying the amount of galaxies expected from upcoming surveys would be impossible. Other authors have instead utilized the structural information provided by high resolution imaging in the form of parametric and non-parametric properties of the observed light distribution of galaxies \citep[e.g.,][and references therein]{Lotz2004,Cheng2011,Conselice2014}. More recently, machine learning (ML) techniques have shown significant success in predicting galaxy morphology purely from images \citep[e.g.,][and references therein]{Dieleman2015, DominguezSanchez2022, Ghosh2020}. Among the ML methods employed are Random Forests \citep[RF;][]{delaCalleja2004a}, Locally Weighted Regression \citep[LWR;][]{delaCalleja2004b}, and Convolutional Neural Networks \citep[CNNs; e.g.][]{Huertas-Company2015, Dieleman2015, PerezCarrasco2019, Walmsley2020, Cavanagh2021, DominguezSanchez2022}, all based on supervised ML. More recently, other authors like \citet[]{Martin2019, Martin2020, Cheng2021b, Hayat2021, Zhou2022, Wei2022} and \citet{Dai2023} have explored state-of-the-art techniques based on unsupervised or self-supervised ML methods. \citet{Parker2024} presented AstroCLIP, a sophisticated approach that utilizes a pretrained Vision Transformer (ViT) model capable of embedding both galaxy images and spectra into a shared, physically meaningful latent space. AstroCLIP has different applications, including the morphological classification of galaxies, where it was tested on the GZ questions, achieving accuracies ranging from 0.44 to 0.97, depending on the specific question.

While Deep Learning models have been very successful, particularly for image-based predictions, they require a large amount of data for training and powerful computational resources, including \textit{GPUs}, for effective implementation. Moreover, they are generally harder to interpret compared to simpler ML techniques. In recent years, various authors have investigated an alternative approach that incorporates physical information, utilizing structural properties to predict galaxy morphology through various ML techniques. For instance, \cite{Sreejith2018} employed multiple ML models to classify galaxies into five distinct categories within the Galaxy and Mass Assembly (GAMA) survey \citep{Driver2011, Liske2015}, using photometric structural parameters (e.g., S\'ersic index, ellipticity) and achieving an average accuracy of 75\%. Similarly, \citet{Barchi2020} applied ML techniques using a modified version of the $CAS$ (Concentration, Asymmetry, Smoothness) parameters, along with entropy \citep[][]{Bishop2006,Ferrari2015} and the new Gradient Pattern Analysis (GPA; \citealt{Rosa2018}) parameter, to separate late- from early-type galaxies with a 98\% accuracy. However, this accuracy decreases to $\sim$65\% when attempts are made to distinguish morphological types into finer-grained sub-classes. \citet{Tarsitano2022} proposed constructing a 1D sequence from the elliptical isophotes of the galaxies' light distribution to automatically classify them as early- or late-type. Using the XGBoost ML model, they achieved class accuracies of 86\% and 93\%, respectively. \citet{Mukundan2024} considered a set of only structural parameters to classify the 11 morphological types reported by \citet{Nair2010}, utilizing the k-nearest neighbors algorithm. Their classification achieved an average accuracy of approximately 80--90\% for each morphological type. However, it should be noted that this level of accuracy was only achieved when a prediction is deemed successful if it falls within $\pm$ 2 T-Types of the original classification.

In this paper, we take advantage of previous results and employ not only a set of structural parameters, as used in previous works, but also integrate a set of colour parameters that capture the star formation properties of galaxies. We have compiled a sample of $\sim$18,000 local galaxies, each with detailed visual morphological classifications and a comprehensive set of fourteen galaxy parameters. These include well-known structural parameters such as $CAS$, Gini \citep[][]{Lotz2004}, $M_{20}$ \citep[][]{Lotz2004}, shape asymmetry ($A_S$), Sérsic index ($n$; \citealt{Sersic1963}), and axis ratio ($BA$), as well as newer parameters like asymmetry gradient ($\Delta A_{9050}$; Hernández-Toledo et al. in prep.), along with star-formation indicators like three colour indices ($g-i$, $u-r$, and $u-i$) and a new colour gradient ($\Delta (g-i)$). All parameters were homogeneously estimated from the SDSS images and used to train eXtreme Gradient Boosting (XGBoost; \citealt{Chen2016}) models for automatic galaxy classification. We explore various classification tasks, including binary and five-class classifications, experimenting with different parameter configurations to enhance the performance of the models. Additionally, we assess the effectiveness of both direct and hierarchical classification approaches. To further understand the results of the automated classification, we employ interpretation tools to analyse the influence of different galaxy parameters on the model's predictions. Finally, we discuss possible error sources that could affect model performance, providing a comprehensive evaluation of our methodology and results.

This paper is organized as follows. In Section~\ref{sec:dataset}, we describe the data compilation and provide a brief description of the structural and star formation parameters selected for the classification. Section~\ref{sec:methods} details XGBoost, including the hyperparameter settings used in our experiments, as well as the classification tasks and parameter configurations explored. Section~\ref{sec:results} presents the performance results of the trained models, covering both direct and hierarchical approaches. In section~\ref{sec:discussion}, we discuss the obtained results, including model interpretation (Sec.~\ref{sec:model_interp}), possible error sources (Sec.~\ref{sec:errorSource}), and comparison with existing works (Sec.~\ref{sec:comparison}). Finally, the concluding remarks and a brief summary are presented in Section~\ref{sec:conclusion}.

\section{Dataset} \label{sec:dataset}

\subsection{Galaxy sample} \label{sec:galaxy_sample}
To carry out an automatic classification using supervised ML methods, it is essential to have a robust training sample with a detailed and trustworthy morphological classification. In this sense, direct visual classification is the most reliable method, as expert classifiers visually examine each object and assign a morphological type following a standard classification scheme. In this work, we considered two existing catalogues with detailed visual classifications: 1) the \citet{Nair2010} catalogue (hereafter NA10), which contains $\sim$14,000 classified galaxies based on SDSS $gri$ colour composited images, and 2) the Visual Morphology Catalogue for the Mapping Nearby Galaxies \citep[MANGA;][hereafter VM22]{VazquezMata2022}, which includes $\sim$10,500 galaxies classified using digitally post-processed images from the DESI Legacy Imaging \textit{r}-band Survey \citep[]{Dey2019}. Both catalogues followed a similar classification scheme using the Hubble T-Type number codes as described in Table~\ref{tab:ttypes}. We have merged the morphological results from both catalogues, obtaining a total sample of $\sim$24,500 galaxies, with $\sim$3,000 galaxies in common. Notice that, although the morphological classifications in these catalogues are based on different image surveys, \citetalias{VazquezMata2022} have shown that the morphological classification for the galaxies in common differs by $\Delta T_{Type} = |T_{VM22} - T_{NA10}|$ = 1.3, in agreement with the expected differences between classifiers found in other works \citep[e.g. in][]{Naim1995}. Therefore, for those galaxies in common we adopt the classification reported by \citetalias{VazquezMata2022}.

%------------------------Table----------------------------------%
\begin{table*}
    \centering
    \caption{Hubble morphologies and T-Type convention adopted in \citetalias{Nair2010} and \citetalias{VazquezMata2022}.}
    \label{tab:ttypes}
    \begin{tabular}{ c c c c c c c c c c c c c c c }
        \hline
        Class & E & S0$^-$ & S0 & S0a & Sa & Sab & Sb & Sbc & Sc & Scd & Sd & Sdm & Sm & Irr \\
        T-Type & -5 & -3 & -2 & 0 & 1 & 2 & 3 & 4 & 5 & 6 & 7 & 8 & 9 & 10 \\
        \hline
    \end{tabular}
\end{table*}

\begin{figure}
     \centering
        \includegraphics[width=\columnwidth]{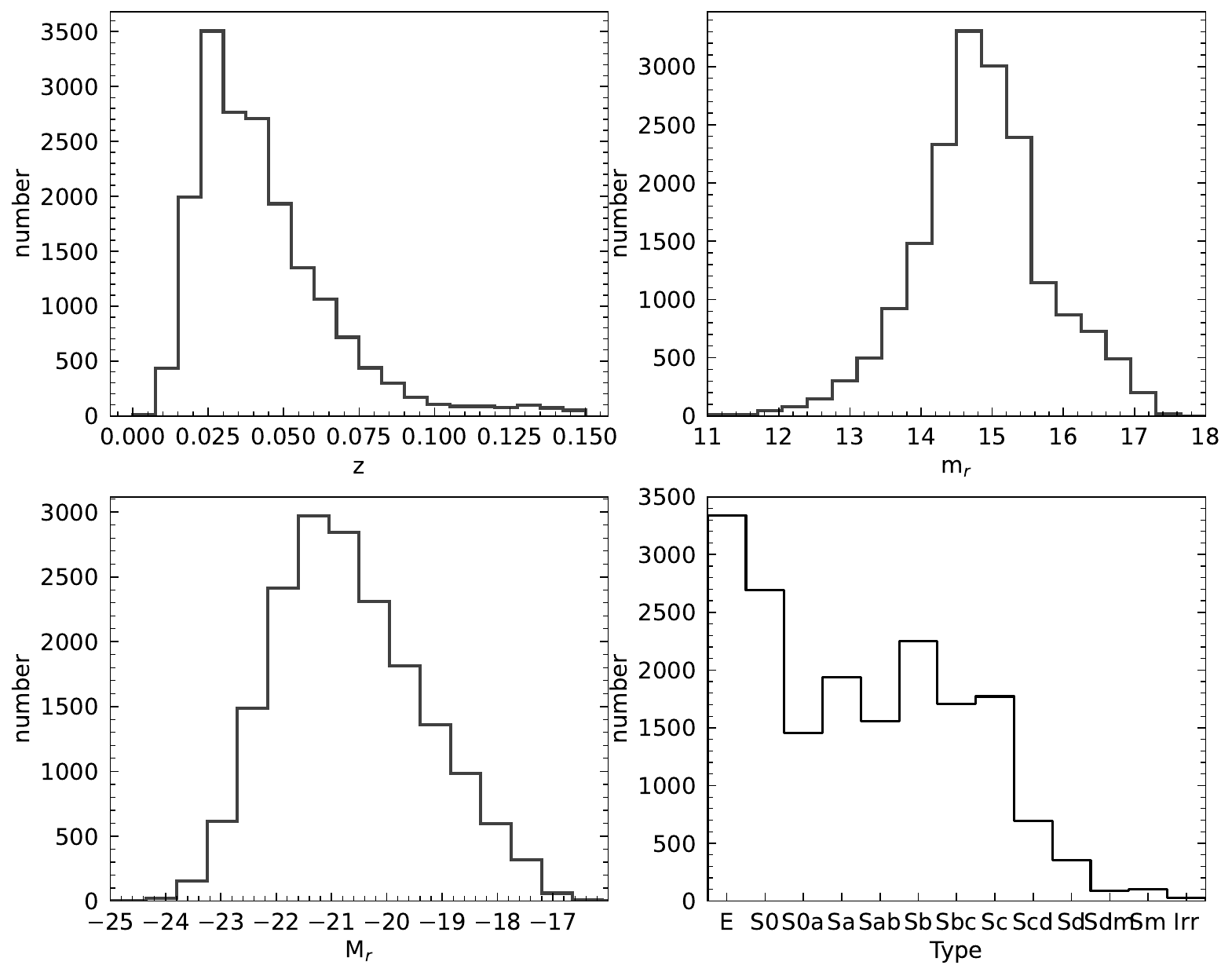} \\
\caption{Histograms of some general properties for the final sample in this work. This includes redshift (upper left), apparent (upper right) and absolute $r$--band magnitude (lower left) adopted from the NSA catalog \citealt{Blanton2011}, and the corresponding morphology (lower right) adopted from the \citetalias{Nair2010} and \citetalias{VazquezMata2022} catalogues.}
\label{fig:histo}
\end{figure}

Aside from 3,000 galaxies in common, the merged sample was also refined by eliminating galaxies coded in both \citetalias{Nair2010} and \citetalias{VazquezMata2022} as showing evidence of advanced mergers and strongly perturbed galaxies, both lacking a clear and well-identifiable Hubble morphology, which is a main requisite for the present study. We note that the correct identification of mergers and their morphological analysis is highly relevant due to their role in galaxy evolution; however, this is beyond the scope of the present work. Our final sample consists of 19,012 galaxies with detailed morphological classification. The redshift ($z$) limit is 0.15 with an apparent $r$--band magnitude limit of 17.2, and an absolute $M_{r}$ magnitude in the range (-24,-16). The overall distributions of the combined sample are presented in Figure~\ref{fig:histo}, where the $z$ and magnitudes information is coming from the SDSS database (NASA-Sloan Atlas, NSA catalogue, \citealt{Blanton2011}). 

In the following section, we briefly summarize the definitions and corrections adopted for the structural, gradient, and colour parameters used in the present paper.

\subsection{Structural and star formation parameters} \label{sec:m_params}
Our approach to predict (parametric and non-parametric) structural parameters requires the analysis of the surface brightness distribution of galaxies on the images. The $r$--band structural parameters were adopted from an homogeneous estimate by \citet{Nevin2023} for a massive set of galaxies from the SDSS-DR16 photometric catalogue. Since the reliability of these parameters as classification tools is closely tied to the properties of the intervening images (namely their resolution and average signal-to-noise ratio per pixel, $\left<S/N\right>$; e. g., \citealt{Lotz2004}), we implemented a minimum cutoff of $\left<S/N\right> > 2.5$. Most of the SDSS galaxies in our final sample have $\left<S/N\right>$ values between 5 and 10, corresponding roughly to $r$--band magnitudes brighter than 16 mag, which is well below the $r$--band flux limit of the SDSS images (17.7 mag), guaranteeing reliable estimates for the present study. Note that \citet{Tarsitano2018} have generated one of the largest samples of structural and morphological parameters for galaxies based on deeper images from the DES survey; unfortunately the overlap with our sample is minimal.

We also estimated a complementary set of parameters related to the star formation activity, specifically the colour gradient and asymmetry gradient parameters. For that purpose, we retrieved 800 $\times$ 800 pixel \textit{g, r}, and \textit{i}--band cutouts from the SDSS-DR13 database, centred on the right ascension and declination of each galaxy in our sample. We segmented each cutout using SExtractor \citep[][]{Bertin1996} to characterize the background and identify the sources at a given threshold. Once the  background is subtracted and the nearby sources were identified and masked, we proceeded to estimate the colour gradient $\Delta (g-i)$ parameter at two different radii (50\% and 90\% of a Petrosian radius, $R_\mathrm{Pet}$, for each galaxy), following \citet{Park2005}. In a similar way, we estimated a new asymmetry gradient parameter ($\Delta A_{9050}$), which measures the asymmetry of the surface brightness distribution of a galaxy at two different radii (50\% and 90\% of $R_\mathrm{Pet}$), following Hern\'andez-Toledo et al. (in prep., see description below).

Once all structural and star formation parameters were estimated, we further proceeded to a final refinement of the sample by excluding the galaxies with incomplete parameter information. We also excluded galaxies considered as outliers in the colour--colour gradient and colour--asymmetry gradient diagrams, comprising a final sample of 17,966 galaxies.

In the following, we summarize the definitions of the gradient and colour index parameters. 

\begin{itemize}
    \item Colour gradient ($\Delta (g - i)$): Radial colour gradients arise from underlying age and metallicity gradients. Late-type galaxies exhibit a more pronounced stellar colour and age gradients, giving rise to, for example, negative colour gradients (redder cores and bluer outskirts), expected in galaxy mass assembly scenarios. \citet{Park2005} found that the colour--colour gradient diagram is a good morphology classification tool. They calculated the colour gradient, $\Delta (g - i)$ by comparing the $g-i$ colour in two regions of a galaxy. Specifically, they proposed to adopt the difference in $(g - i)$ colour between the region with $R < 0.5 R_\mathrm{Pet}$ and that of the annulus with $0.5 R_\mathrm{Pet} < R < R_\mathrm{Pet}$, where $(g - i)$ is rest-frame \textit{K}-corrected. Hence, a negative colour difference means the galaxy becomes bluer towards the outside.
    
    \item Asymmetry gradient ($\Delta A_{9050}$):     Following the ideas in \citet{Park2005}, we estimate a new gradient parameter involving the difference of asymmetries in two regions of a galaxy, $A_{50}$: $0 < R < 0.5 R_\mathrm{Pet}$ and $A_{90}$: $0.5R_\mathrm{Pet} < R < R_\mathrm{Pet}$, such that $\Delta A_{9050} = A_{90}-A_{50}$. After background subtraction and masking of the galaxy up to an external $r_\mathrm{max} \sim 2 R_\mathrm{Pet}$, we follow \citet{Conselice2003} to estimate what we name as the asymmetry gradient within that region. This metric takes advantage of the fact that as spiral galaxies go from early- to late-types, they gradually open their spiral arms while increasing the presence and resolution of clumpy regions. This results in a more pronounced asymmetry in the outer regions compared to the inner/bulge regions, leading to more negative asymmetry gradients (more symmetric inner regions and more asymmetric outskirts) compared to early-type spirals. In the present paper, we test the ability of $\Delta A_{9050}$ as a morphological segregator. Similar to the concentration index (see comments below), in a forthcoming paper (Hern\'andez-Toledo et al. in prep.) we will be exploring the robustness and stability of $\Delta A_{9050}$. This feature study will examine whether factors like radial extent and other observational errors introduce biases related to image properties such as exposure depth, signal-to-noise ratio, image pre-processing methods, etc.

    \item Colour index parameters: The colours of galaxies reflect their dominant stellar populations. \citet{Strateva2001} proposed the $u-r$ colour to distinguish between early-type and late-type galaxies. In this work, we considered three colour index parameters after correcting for galactic extinction and \textit{K}-correction: $g-i$, $u-r,$ and $u-i$, all obtained from the NSA catalogue.
\end{itemize} 

We also present a brief summary of the definitions behind the non-parametric structural parameters used in this work:

\begin{itemize}
    \item Concentration ($C$): This parameter measures the distribution of light within a galaxy and how centrally concentrated this is. Specifically, the definition adopted in this work is:
    \begin{equation}
        C=5 \log \left( R_{80}/R_{20}\right),
    \end{equation}
    \noindent where $R_{80}$ and $R_{20}$ are the circular radius that contains 80\% and 20\%, respectively, of the total flux of the galaxy \citep[][]{Lotz2004}. The total flux is defined as the flux contained within 1.5 $R_\mathrm{Pet}$ of the galaxy centre \citep[][]{Conselice2003}.
    
    \item Inverse concentration index ($c_{5090}$): It offers an alternative view of light concentration, and it is defined as:
    \begin{equation}
        c_{5090} = R_{50} / R_{90},
    \end{equation}
    \noindent where $R_{50}$ and $R_{90}$ are the radii from the centre of the galaxy containing 50\% and 90\% of the Petrosian flux, respectively. With this definition, higher values of $c_{5090}$ indicate more light contained in the central regions of the galaxy. $R_{50}$ and $R_{90}$ were obtained from the NSA catalogue. In the present paper we adopt these standard definitions of concentration, however, at this point the reader may notice that other definitions showing a robust behavior against changes of external radius (related to exposure depth and other observational errors; c.f. \citealt{Graham2001})  could be used and will be explored in future works.  
    
    \item Asymmetry ($A$): This parameter quantifies the degree of rotational symmetry in the galaxy light distribution. In particular, $A$ is measured by subtracting the galaxy image rotated by 180 deg about the centre, $I_{180}$, from the original image, $I$ \citep[e.g.][]{Conselice2000,Lotz2004}: 
    \begin{equation}
        A = \frac{\Sigma_{i,j} \left| I(i,j) - I_{180}(i,j) \right|}{\Sigma_{i,j} \left| I(i,j) \right|} - B_{180},
    \end{equation}
    \noindent where $B_{180}$ is the average asymmetry of the background, which is a correction for background noise. $A$ is summed over all pixels $(i,j)$ within 1.5 $R_\mathrm{Pet}$. The galaxy centre is determined by minimizing $A$ as in \citet{Lotz2004}. Galaxies with higher $A$ values tend to have more irregular or disturbed structures.

    \item Clumpiness ($S$): By quantifying the fraction of light in a galaxy contained in clumpy distributions, $S$ indicates the degree of small-scale structure in a galaxy. From \citet{Conselice2003} and \citet{Lotz2004}, $S$ is calculated as:
    \begin{equation}
        S = \frac{\Sigma_{i,j} \left| I(i,j) - I_S(i,j) \right|}{\Sigma_{i,j} \left| I(i,j) \right|} - B_S,
    \end{equation}
    \noindent where $I(i,j)$ is the original image and $I_S(i,j)$ is its smoothed counterpart, which is smoothed by a boxcar of width 0.25 $R_\mathrm{Pet}$. $B_S$ is the average smoothness of the background. $S$ is summed over all pixels $(i,j)$ within 1.5 $R_\mathrm{Pet}$ of the galaxy centre. Since the centres of galaxies often are highly concentrated, the central pixels within the smoothing length of 0.25 $R_\mathrm{Pet}$ are excluded from the calculation.  

    \item Shape Asymmetry ($A_S$): It is similar to $A$, but it is calculated using a binary detection mask instead of the image itself \citep[for more details see][]{Pawlik2016,RodriguezGomez2019}. $A_S$ is more sensitive to low surface brightness tidal features than $A$.
    
    \item Gini: This parameter measures the inequality in the distribution of pixel intensities within a galaxy image. Gini correlates with $C$, however, it does not assume that the brightest pixels are in the central region of the galaxy. Gini is defined by \citet{Lotz2004} as:
    \begin{equation}
        Gini = \frac{1}{\left| \overline{f} \right| N(N-1)} \sum_i^N (2i-N-1) \left| f_i \right|,
    \end{equation}
    \noindent where $N$ is the number of pixels assigned to the galaxy, $\overline{f}$ is the average flux value, and $f_i$ is the flux value for each pixel. The pixels are previously ordered by increasing flux value. If Gini is 0, the light is evenly distributed over all galaxy pixels, while if Gini is 1, all the flux is concentrated in just one pixel.

    \item $M_{20}$: It describes the second-order moment of the brightest 20\% of the spatial distribution of the galaxy flux \citep[][]{Lotz2004} and does not assume a central concentration. Mathematically, the total second-order moment, $M_\mathrm{tot}$, is:
    \begin{equation}
        M_\mathrm{tot} = \sum_i^N M_i = \sum_i^N f_i \left[ \left( x_i - x_c \right)^2 + \left( y_i - y_c \right)^2 \right],
    \end{equation}
    \noindent where $f_i$ is the flux in pixel ($x_i,y_i$), and ($x_c,y_c$) is the galaxy centre which is determined by minimizing $M_\mathrm{tot}$. Then, to compute $M_{20}$, the galaxy pixels are ranked by flux in descending order and $M_i$ is summed over the brightest pixels until that sum equals 20\% of the total galaxy flux, $f_\mathrm{tot}$, and then normalized by $M_\mathrm{tot}$:
    \begin{equation}
        M_{20} = \log_{10} \left( \frac{\Sigma_i M_i}{M_\mathrm{tot}} \right),\,\mathrm{while}\,\sum_i f_i < 0.2 f_\mathrm{tot},
    \end{equation}
    $M_{20}$ is similar to $C$, however, a $M_{20}$ value denoting a high concentration (a very negative value) does not imply a central concentration, as the centre of the galaxy is a free parameter. Hence, it provides information about the presence of multiple components, such as bright knots or companion galaxies.

We also describe the definition of the parametric structural S\'ersic index:
    
    \item S\'ersic index ($n$): This parameter describes the shape of the light profile of a galaxy. It is derived from fitting a S\'ersic profile \citep[][]{Sersic1963} to the galaxy brightness distribution:
    \begin{equation}
        I(R) = I_e \exp \left( -b_n \left[ \left( \frac{R}{R_e} \right)^{1/n} - 1 \right] \right),
    \end{equation}
    \noindent where $I(R)$ is the intensity within a circular radius, $R$, and $I_e$ is the intensity at the effective radius ($R_e$), which is the radius that contains half of the total light. $b_n$ is a constant that depends on the S\'ersic index, $n$. The S\'ersic index, $n$, can indicate whether a galaxy has a steep (high $n$) or shallow (low $n$) central brightness profile, providing insights into its bulge or disk dominance.

Finally, we adopt a geometric parameter as a measure of the shape of galaxies:

    \item $BA$: We adopt the axis ratio $b/a$, from a two-dimensional, single component S\'ersic fit in the \textit{r}--band, as reported in the NSA catalogue up to $R_\mathrm{Pet}$. In this case, the definition of the $R_\mathrm{Pet}$ is based on the SDSS \textit{r}--band using elliptical instead of circular apertures.
\end{itemize}

\section{Methodology} \label{sec:methods}

In this section, we introduce the experimental setup to explore the effectiveness of various classification strategies, including different parameter configurations. Additionally, we briefly describe XGBoost and provide the hyperparameter values adopted in this work.

\subsection{Experimental setup} \label{sec:experiments}
To assess the model's ability to distinguish between various morphological types, we explore different re-categorizations of the sample galaxies. Table~\ref{tab:recat} outlines our proposed re-categorizations and details the galaxy distribution for each class. The first re-categorization, referred to as 2cats (first row), involves a broad classification between early-type (T-Types: -5 to 0) and late-type (T-Types: 1 to 10) galaxies, providing a baseline for model performance. The 5cats re-categorization (second row) introduces a finer classification, grouping the galaxies into five morphological classes: (-5), (-3, -2, 0), (1, 2, 3), (4, 5, 6), and (7, 8, 9, 10), to evaluate the model's ability to handle a more complex classification (T-Types are presented in Table~\ref{tab:ttypes}). In the Early re-categorization (third row), we focus only on early-type galaxies, sub-classifying them into ellipticals (T-Type -5) and lenticulars (T-Types: -3, -2, and 0) to test the model's performance in distinguishing between these closely related morphological classes. Finally, the Late re-categorization (fourth row) concentrates exclusively on late-type galaxies, classifying them into three groups: (1, 2, 3), (4, 5, 6) and (7, 8, 9, 10), which helps us to assess the model's capability to differentiate among various types of spirals galaxies.

%------------------------Table----------------------------------%
\begin{table}
    \centering
    \caption{Re-categorization of the sample galaxies adopted in this work. The numbers (-5 to 10) refer to the Hubble T-Type classification from \citetalias{Nair2010} and \citetalias{VazquezMata2022}. Also indicated are the Hubble morphologies and the number of galaxies for each class within the proposed re-categorizations.}
    \label{tab:recat}
    \begin{adjustbox}{width=0.5\textwidth,center}
    \begin{tabular}{ c c c c c c }
        \hline
        \textbf{Category} & \textbf{0} & \textbf{1} & \textbf{2} & \textbf{3} & \textbf{4} \\
        \hline
        \multirow{3}{*}{2cats} & -5,-3,-2,0 & 1 to 10 &  &  & \\
         & (E--S0a) & (Sa--Irr) &  &  & \\
         & 7,485 & 10,481 &  &  & \\
        \hline
        \multirow{3}{*}{5cats} & -5 & -3,-2,0 & 1,2,3 & 4,5,6 & 7,8,9,10 \\
         & (E) & (S0$^-$--S0a) & (Sa--Sb) & (Sbc--Scd) & (Sd--Irr) \\
         & 3,340 & 4,145 & 5,749 & 4,167 & 565 \\
        \hline 
        \multirow{3}{*}{Early} & -5 & -3,-2,0 &  &  &  \\
         & (E) & (S0$^-$--S0a) &  &  &  \\
         & 3,340 & 4,145 &  &  & \\
        \hline 
        \multirow{3}{*}{Late} & 1,2,3 & 4,5,6 & 7,8,9,10 &  & \\
         & (Sa--Sb) & (Sbc--Scd) & (Sd--Irr) &  & \\
         & 5,749 & 4,167 & 565 &  & \\
        \hline
    \end{tabular}
    \end{adjustbox}
\end{table}

In addition to these re-categorizations, we investigate the influence of different galaxy parameters on classification performance by combining them into four distinct groups (detailed in Table~\ref{tab:configs}), each representing a different aspect of the galaxy's physical characteristics. The first configuration, termed ``Colour'', encompasses the parameters related to colour and colour gradients, namely $g-i,\,u-r,\,u-i$ and $\Delta \left( g-i \right)$. The ``Structural1'' configuration comprises eight \textit{classical} structural parameters: $C,\, A,\, S,\, A_S,\, \mathrm{Gini},\, M_{20},\, n,$ and $c_{5090}$. Expanding upon Structural1, the ``Structural2'' configuration adds further structural parameters, including the semi-minor to semi-major axis ratio ($BA$) and the asymmetry gradient ($\Delta A_{9050}$). Finally, the ``S2+C'' configuration combines the Structural2 and Colour parameter sets, encompassing all fourteen parameters. These four configurations allow us to assess the contribution of each parameter set to the model's overall performance.

%------------------------Table----------------------------------%
\begin{table}
    \centering
    \caption{Different galaxy parameter configurations adopted in this work.}
    \label{tab:configs}
    \begin{tabular}{ c l }
        \hline
        \textbf{Configuration} & \textbf{Set of parameters} \\
        \hline
        Colour & $g-i,\,u-r,\,u-i,\,\Delta (g - i)$ \\
        Structural1 & $C,\,A,\,S,\,A_S,\,\mathrm{Gini},\,M_{20},\,n,\,c_{5090}$ \\
        Structural2 & Structural1 + $BA,\,\Delta A_{9050}$ \\
        S2+C & Structural2 + Colour \\
        \hline
    \end{tabular}
\end{table}

Furthermore, seeking to optimize the model’s performance, we also explore two classification approaches for distinguishing between galaxy types using ML models: direct classification and hierarchical classification. In the direct classification approach, a single XGBoost model is trained to perform the entire classification task in one step, distinguishing between all galaxy classes directly according to the chosen re-categorization. This straightforward method serves as a baseline, providing initial insights into the model's capacity to handle different levels of classification tasks.

In contrast, the hierarchical classification approach breaks down the classification process into a series of steps, each focusing on a specific classification task. As a first step, we train an XGBoost model for a binary classification to distinguish between early-type and late-type galaxies (i.e., the 2cats classification). Once the galaxies are separated into these two broad groups, two additional XGBoost models are trained: one to only sub-classify the early-type galaxies into ellipticals and lenticulars (based on the Early re-categorization, see Table~\ref{tab:recat}) and another to just differentiate the late-type galaxies into three spiral classes (according to the Late re-categorization). Note that this hierarchical approach is only implemented to differentiate between the five galaxy groups outlined for the 5cats re-categorization by training three separate models. The performance of the hierarchical approach is evaluated by combining the results of the individually trained models and comparing them to the direct 5cats classification results.

Throughout this work, we evaluate and compare each classification task (galaxy re-categorization), parameter configuration, and classification approach, aiming to explore the model's performance across different levels of classification complexity and input data information. The results of these experiments will provide insights into the relative importance of different types of galaxy parameters in morphological classification, the model's ability to handle diverse classification tasks, and the efficacy of direct versus hierarchical classification approaches.

For all experiments, we adopt a stratified split of the dataset (17,966 galaxies), with 70\% (12,576 galaxies) allocated for training and 30\% (5,390 galaxies) for testing. This stratified split ensures that the distribution of galaxies across different classes remains consistent between the training and testing subsets, preserving the relative proportions for each re-categorization.

Additionally, we apply a stratified 3-fold cross-validation, where the training subset is randomly split into three stratified folds. Two of the three folds are used to train the model, and the remaining fold is used as validation subset to evaluate the model's performance. The above process is repeated for all three permutations, ensuring each fold is used once as the validation subset. To obtain a more robust estimation of the model's performance, we repeated the 3-fold cross-validation process 10 times, where each repetition involves a different stratified random split of the training subset into folds. The resulting performance metrics (e.g., accuracy, precision, etc., see Sec.~\ref{sec:direct_class}) are averaged across the complete process to estimate the model performance. Finally, to assess the model's performance on unseen data, we further evaluate it on the test subset (30\% of the dataset).

\subsection{XGBoost model} \label{sec:xgboost}
XGBoost, short for eXtreme Gradient Boosting \citep[][]{Chen2016}, is a powerful and widely-used ML method known for its high performance and effectiveness in a variety of tasks (e.g. regression, classification, ranking, and recommendation systems). It belongs to the family of gradient boosting methods, which are ensemble learning methods that combine multiple weak predictive models, typically decision trees, to create a strong predictive model.

The idea behind XGBoost is to iteratively train a series of decision trees and combine their predictions to produce a final ensemble model. Each tree is built sequentially, with each subsequent tree attempting to correct the mistakes made by the previous tree. Decision trees are simple models that make predictions based on a series of hierarchical decisions. Additionally, XGBoost incorporates regularization techniques such as L1 and L2 weight penalty terms\footnote{For more information see, e.g., \citet{Murphy2012}} to mitigate overfitting, as well as tree pruning, which prevents the model from becoming overly complex.

In this work, we employ XGBoost to perform the galaxy morphlogical classification due to its effectiveness for handling structured data, its robustness against overfitting, and its capability to model complex non-linear relationships. Specifically, we use the \texttt{XBGClassifier} class from the XGBoost library\footnote{\url{https://xgboost.readthedocs.io/en/stable/}}. The input data of the XGBoost model comprises any of the different parameter combinations in Table~\ref{tab:configs}. Given an input, the model generates a probability vector, with each element representing the probability of belonging to a specific class. Subsequently, the model's prediction is determined by selecting the class with the highest probability value.

\subsection{Selection of hyperparameters} \label{sec:hyperparams}
The parameters that determine the design of a ML method and those that specify its learning process are known as hyperparameters. For the XBGClassifier, we perform an empirical hyperparameter search considering the 5cats classification with the S2+C input data. In particular, we tuned the following hyperparameters: \texttt{learning\_rate} (step size shrinkage to prevent overfitting), \texttt{alpha} (L1 regularization term on weights), \texttt{reg\_lambda} (L2 regularization term on weights), \texttt{max\_depth} (maximum depth of a tree), \texttt{colsample\_bytree} (subsample ratio of columns when constructing each tree), \texttt{max\_delta\_step} (maximum delta step allowed for the weight estimation of each tree), \texttt{min\_child\_weight} (minimum sum of instance weight needed in a child), \texttt{gamma} (minimum loss reduction required to make a further partition on a leaf node of the tree), and \texttt{subsample} (subsample ratio of the training instances).

For each of the 20 hyperparameter configurations investigated, we conduct a stratified 3-fold cross-validation, repeating the process 10 times to obtain a more robust evaluation of performance. This cross-validation process follow the procedure detailed in Section~\ref{sec:experiments}, and the performance metrics are averaged across all repetitions to obtain reliable performance estimates. After evaluating all hyperparameter configurations, we select the hyperparameters that obtained the top performance, namely: \texttt{learning\_rate} = 0.1, \texttt{alpha} = 2, \texttt{reg\_lambda} = 0.5, \texttt{max\_depth} = 5, \texttt{colsample\_bytree} = 0.7, \texttt{max\_delta\_step} = 2, \texttt{min\_child\_weight} = 3, \texttt{gamma} = 0.3, and \texttt{subsample} = 0.9. These hyperparameter values will be used for all the experiments in this work.

\section{Results} \label{sec:results}

In this section, we present the performance results of the classification experiments outlined in Sec.~\ref{sec:experiments} using the XGBoost model. To evaluate and compare the outcomes of the experiments, we compute various performance metrics, including accuracy, precision, recall, F1-score, and the Area Under the ROC Curve \citep[AUC-ROC;][]{Bradley1997, Fawcett2005}, which are described in Appendix~\ref{app:performance_metrics}. In addition, we provide the confusion matrices (CMs) of the highest-performing experiments.

\subsection{Direct classification approach} \label{sec:direct_class}
As introduced in Sec.~\ref{sec:experiments}, the direct classification approach involves training a single XGBoost model to carry out the given classification task. Here, we adopt such an approach and evaluate the model performance for each classification task (see Table~\ref{tab:recat}) using different parameter configurations (Table~\ref{tab:configs}). For each of these experiments, we perform 10 repetitions of a 3-fold cross-validation (on the training subset), as described in Sec.~\ref{sec:experiments}. Table~\ref{tab:metrics} presents the mean accuracy, precision, recall, and AUC-ROC metrics, along with their standard deviations across the cross-validation procedure for each parameter configuration in both the 2cats and 5cats direct classification tasks. For metrics other than accuracy, the reported values represent the averages across all classes. Specifically, they represent the \textit{macro average}, where the metric is calculated for each class independently, and then the unweighted average of these class-wise scores is taken. As a reference, a random classifier with a uniform class distribution would achieve an accuracy, precision, recall, and AUC-ROC of 50\% for the 2cats task, and 20\% accuracy, precision, and recall with a 50\% AUC-ROC for the 5cats task.

%------------------------Table----------------------------------%
\begin{table*}
    \centering
    \caption{Mean and standard deviation of different performance metrics across 10 repetitions of a 3-fold cross-validation for different parameter configurations in the 2cats and 5cats direct classifications. Precision, recall, and AUC-ROC values correspond to the macro average. Note that for a random classification with a uniform class distribution the accuracy, precision, and recall are all at 50\% for 2cats and 20\% for 5cats, while AUC-ROC is at 50\% in both tasks.}
    \label{tab:metrics}
        \begin{tabular}{ *{2}{c} | *{4}{c} }
            \hline
            \multicolumn{2}{c |}{\textbf{Experiment}} & \multicolumn{4}{c}{\textbf{Performance Metrics}} \\
            Classification & Input data & Accuracy & Precision & Recall & AUC-ROC \\
            \hline
            \multirow{4}{*}{2cats} & Colour & $0.821\pm0.006$ & $0.852\pm0.009$ & $0.853\pm0.009$ & $0.893\pm0.005$ \\
             & Structural1 & $0.825\pm0.006$ & $0.855\pm0.011$ & $0.858\pm0.007$ & $0.904\pm0.005$ \\
             & Structural2 & $0.850\pm0.006$ & $0.875\pm0.009$ & $0.880\pm0.009$ & $0.923\pm0.004$ \\
             & S2+C & $0.869\pm0.007$ & $0.892\pm0.008$ & $0.893\pm0.009$ & $0.943\pm0.004$ \\
            \hline
            \multirow{4}{*}{5cats} & Colour & $0.520\pm0.008$ & $0.490\pm0.014$ & $0.455\pm0.009$ & $0.829\pm0.004$ \\
             & Structural1 & $0.541\pm0.009$ & $0.489\pm0.035$ & $0.454\pm0.008$ & $0.831\pm0.004$ \\
             & Structural2 & $0.589\pm0.008$ & $0.534\pm0.026$ & $0.502\pm0.007$ & $0.860\pm0.004$ \\
             & S2+C & $0.634\pm0.007$ & $0.608\pm0.012$ & $0.580\pm0.011$ & $0.897\pm0.003$ \\
            \hline
        \end{tabular}
\end{table*}

Notably, the precision and recall scores in each experiment of Table~\ref{tab:metrics} are similar to each other, indicating a balanced performance of the models. For example, in the 2cats classification, the difference between these two metrics is within the standard deviation, while in the 5cats classification, the difference is up to 1.2\%. This balance implies that the models are equally effective at identifying true positive instances (recall) and ensuring that the identified positive instances are indeed correct (precision), highlighting the reliability of the classification models. Additionally, the consistency of this balance across the different experiments indicates a robust model performance.

Regarding the different parameter configurations, in the 2cats classification, the model performance is quite similar across the configurations, with differences up to 4.1\%. The S2+C configuration yields the best performance, while the Colour configuration yields the worst. However, the differences in performance between Colour and Structural1 are within their standard deviation. For the 5cats classification, the performance differences among the parameter configurations are more pronounced, with variations up to 10.7\%. Similar to the 2cats classification, the S2+C configuration achieves the best performance, and the Colour configuration performs the worst. Again, the differences in performance metrics for Colour and Structural1 are within their standard deviation, except for accuracy where the difference is only 0.4\%.

Hence, across both 2cats and 5cats classification tasks, the Colour parameters provide a baseline performance that is adequate for the 2cats classification but less effective for the more complex 5cats classification. The Structural1 configuration offers a baseline performance which is almost identical to the Colour configuration, regardless of the classification task. This suggests that neither set of parameters is significantly more informative than the other when used in isolation, and both have similar limitations. For instance, Colour parameters lack information about structural properties of galaxies, whereas Structural1 parameters lack information about star-formation history.

The Structural2 configuration improves model performance, especially in the 5cats classification, indicating the importance of shape axis ratio ($BA$) and gradient asymmetry ($\Delta A_{9050}$) in galaxy classification. Finally, the S2+C configuration consistently delivers the best performance, highlighting the advantage of integrating both photometric and structural data. This performance improvement (up to 4.1\% for 2cats and up to 10.7\% for 5cats) indicates that both parameter types (photometric and structural) capture complementary information of the galaxies, leading to a more comprehensive and effective model.

To assess the model's performance on unseen data, we first evaluate the overall metrics (accuracy, precision, recall, F1-socre, and AUC-ROC) on the test subset (unseen by the model during the training process) using the S2+C parameter configuration for both the 2cats and 5cats classification tasks (see Table~\ref{tab:class_report} in Appendix~\ref{app:perclass_metrics}). For the 2cats classification, the model achieves 88\% across accuracy, precision, recall, and F1-score, with an AUC-ROC of 95\%, reflecting the strong performance in binary classification. For the more complex 5cats classification, the model achieves 65\% in both accuracy and recall, 64\% in precision and F1-score, and 90\% in AUC-ROC. The difference in performance between the two tasks reflects their inherent complexity, making them not directly comparable. The 2cats classification is a simpler task, involving only two broad galaxy types, while the 5cats task requires finer differentiation among multiple galaxy morphologies. Therefore, a poorer performance for the 5cats classification is expected since it is more challenging, even for human visual inspection, than the 2cats task.

In addition, comparing the achieved performance with a random classifier, we observe a clear improvement of the XGBoost model over random guessing highlights its ability to capture meaningful patterns in the data and reliably distinguish between galaxy classes.

Next, we analysed the per-class performance (precision, recall, and F1-score) on the test subset for both tasks (see Table~\ref{tab:class_report}). In the 2cats classification, the model shows balanced performance across both classes, without favoring one class over the other. In the 5cats classification, performance varies among the classes, with Class 0 (elliptical) and Class 3 (Sbc--Scd) obtaining the highest scores (e.g., $\sim$71\% in F1-score), while Class 4 (Sd--Irr) records the lowest (e.g., 39\% in F1-score). However, it should be noted that Class 4 has $\sim$85\% fewer galaxies than the other classes, making it more challenging for the model to learn sufficient distinguishing features for this class. In addition, there is a balance between precision and recall metrics within each class for both classification tasks, except for Class 4, which shows a 17\% gap. This underscores the model's difficulty in handling this under-represented class. Despite challenges with specific classes in the 5cats task, the high AUC-ROC value (90\%) demonstrates the model's strong overall ability to differentiate between galaxy types, providing a solid foundation for further improvements. 

\begin{figure}
     \centering
     \begin{tabular}{cc}
        \includegraphics[width=0.72\columnwidth,angle=0, trim={0 -0.1cm 0 0}]{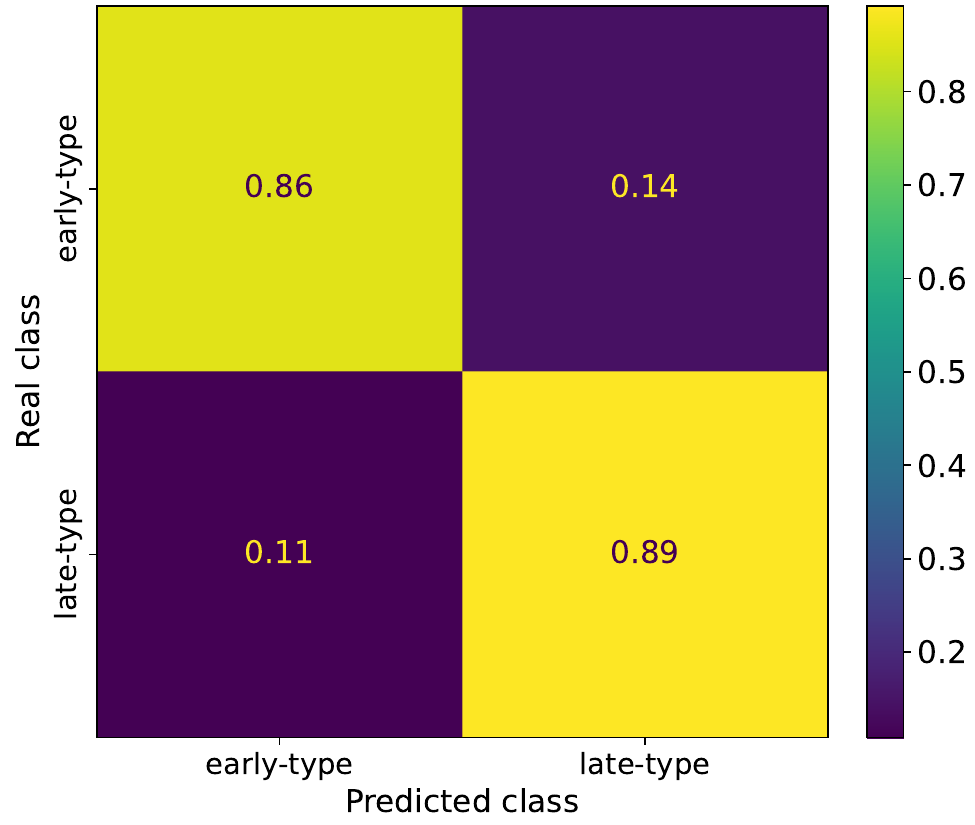} \\
        \includegraphics[width=0.72\columnwidth,angle=0]{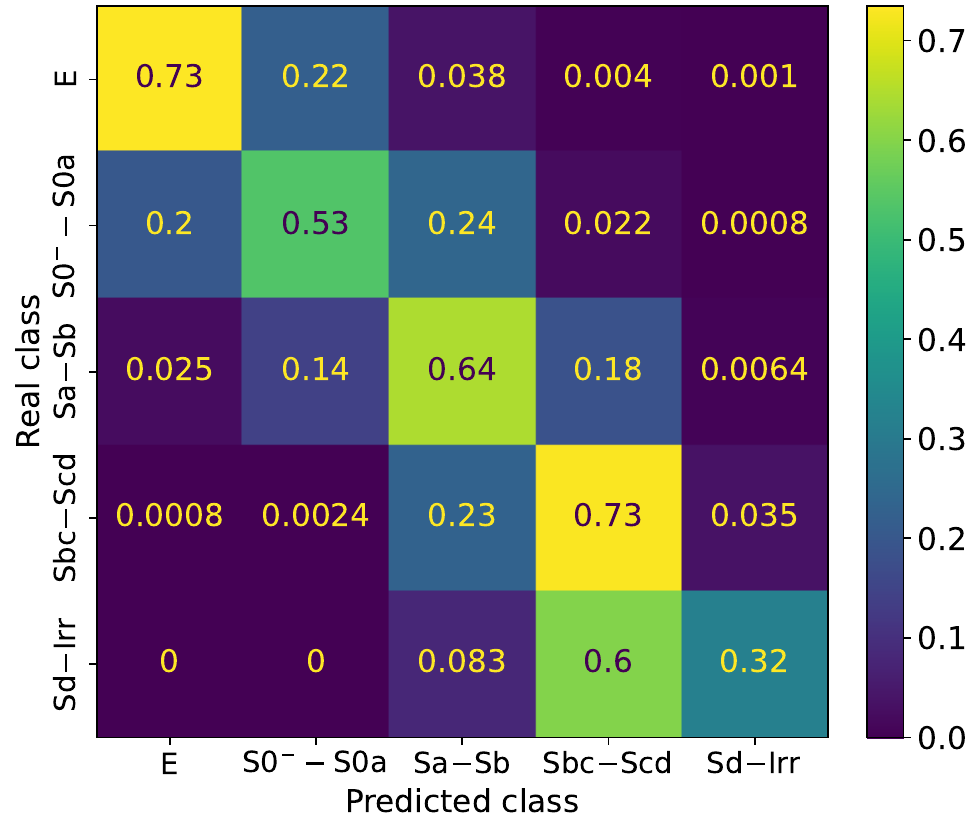}
     \end{tabular}
\caption{Confusion matrices for 2cats (upper) and 5cats (bottom) direct classifications (see Table~\ref{tab:recat}) using the S2+C parameter configuration, calculated with the test subset. Colours are according to the performance of the classification. 2cats shows a balanced performance, whereas in 5cats, performance varies across the classes, with most misclassifications occurring between adjacent classes.}
\label{fig:CM_2p_14p}
\end{figure}

Figure~\ref{fig:CM_2p_14p} shows the confusion matrices (CMs) of the 2cats (upper panel) and 5cats (bottom panel) classifications using the S2+C parameter configuration, obtained using the test subset. The x-axis displays the predicted classes by the model and the y-axis the true classes from the catalogue. Therefore, the diagonal of the CM indicates the success rates of the model's predictions, while off-diagonal values indicate misclassifications. In the 2cats classification, the model performs well in distinguishing between Class 0 (early-type) and Class 1 (late-type) galaxies, with success prediction rates of 86\% and 89\%, respectively. This indicates that the model makes relatively lower misclassification rates (14\% for Class 0 and 11\% for Class 1) for galaxies between these two broad groups.

The CM for the 5cats classification (bottom panel of Fig.~\ref{fig:CM_2p_14p}) shows a varying performance across the classes. For Class 0 (elliptical), the model achieves 73\% success in predictions but has a noticeable misclassification rate of 22\% into Class 1 (lenticular: S0$^-$--S0a). This is explained, in part, since a fraction of lenticulars share structural and colour similarities with elliptical galaxies. In the case of Class 1, the model has moderate performance with a 53\% success rate, but has misclassifications of 24\% and 20\% in the adjacent Class 2 (Sa--Sb) and Class 0, respectively. This is also related to the nature of lenticular galaxies, showing a wide variety of structural properties, some resembling those of ellipticals, but another fraction resembling those of spiral discs (e.g., \citealt{Laurikainen2007,Cappellari2011,Graham2019}).

Furthermore, the model correctly classifies 64\% of Class 2 instances, with misclassifcation rates of 18\% into Class 3 (Sbc--Scd) and 14\% into Class 1, expected this time due to structural and colour similarities among the adjacent (1 and 3) classes. Similar to Class 0, the model performs well for Class 3 with 73\% correct classifications, however it misclassifies 23\% as Class 2, also suggesting some overlapping features between these classes. Finally, for Class 4 (Sd--Irr), the model has the lowest performance with only 32\% of successful predictions and a high misclassification rate of 60\% into Class 3. It is important to highlight that although a fraction of missclassifications, typically 10-20\%, are expected due to both structural and colour similarities with the adjacent Class 3, a higher missclasification rate is probably due to a significant under-representation of Class 4 in our dataset, providing to the model a poor training set for this class. 

The CMs, clearly illustrate that an important part of the model misclassifications occur between adjacent galaxy classes, highlighting the difficulty in distinguishing galaxies with similar features with a combination of parametric and non-parametric approaches but also even after a visual inspection. Therefore, although the model performs well overall, particularly in binary classification, there is room for improvement in accurately distinguishing between galaxy classes, especially in the more complex multi-class classification task. In Section~\ref{sec:discussion}, we will discuss these aspects further.

%------------------------Figure----------------------------------%

\subsection{Hierarchical classification approach} \label{sec:hier_class}
Aiming to improve the performance of the 5cats direct classification, we also explored a hierarchical classification approach, as described in Sec.~\ref{sec:experiments}. In this approach, the classification process is divided into a sequence of steps, each handled by a separate XGBoost model. The first step classifies galaxies according to the 2cats re-categorization (early-type vs. late-type galaxies). Subsequently, two additional models are trained: one to further classify early-type galaxies following the Early re-categorization (elliptical vs. lenticular), and the other to only sub-classify late-type galaxies according to the Late re-categorization (see Table~\ref{tab:recat}). The predictions from these three models are then combined to achieve the final five-class classification as follows: if a galaxy is classified as early-type by the top classifier, it is further subclassified as elliptical (E) or lenticular (S0$^-$--S0a) by the bottom left classifier; else if it is classified as late-type, it is further subclassified as one either Sa--Sb, Sbc--Scd, or Sd--Irr by the bottom right classifier.

As with the direct classification, we evaluated this hierarchical approach by employing different parameter configurations (Table~\ref{tab:configs}), and for each configuration we carry out 10 repetitions of a 3-fold cross-validation (on the training subset). For these hierarchical experiments, we explored two scenarios: one where the same parameter configuration is applied across all three models, and another one where different parameter configurations are allowed for each model. This second scenario is motivated by the structural diversity and varying characteristics of galaxies at different steps of the classification process. For instance, previous studies have shown that combining distinct galaxy parameters, such as concentration ($C_{9050}$), bulge-to-total light ratio ($B/T$), and axial ratio ($b/a$), can help in segregating early-type galaxies into ellipticals and lenticulars more effectively (\citealt{Cheng2011}; \citetalias{VazquezMata2022}). Meanwhile, both colour--colour gradient \citep[][]{Park2005} and colour--asymmetry gradient (Hernández-Toledo et al. in prep.) diagrams offer a more refined segregation of spiral galaxies into subclasses. Given these findings, it is plausible that different parameter configurations may be better suited for specific classification tasks within the hierarchical process. Therefore, this flexible scenario may improve the classification performance by adapting the feature configuration to the specific morphological distinctions being made.

We investigated all possible combinations of the four parameter configurations for each step in the hierarchical process. Table~\ref{tab:hier_metrics} presents the mean values of the performance metrics and their standard deviations, from the cross-validation procedure, for the best-performing experiments in the hierarchical approach. Specifically, it shows two experiments: the first (Hier1) uses the same parameter configuration (S2+C) across all three classifiers, while the second (Hier2) uses different configurations, with S2+C being applied to the 2cats (early-type vs. late-type) and Late (Sa--Sb, Sbc--Scd, Sd--Irr) steps, and Structural2 to the Early (ellipticals vs. lenticulars) step. Additionally, we include the individual performance results for the Late classification step using the S2+C configuration and for the Early step using both the Structural2 and S2+C configurations. Note that the results for the 2cats classification step are provided in Table~\ref{tab:metrics}.

\begin{table*}
    \centering
    \caption{Mean and standard deviation of different performance metrics across 10 repetitions of a 3-fold cross-validation, for the best hierarchical classification experiments (Hier1 and Hier2). Precision, Recall, and AUC-ROC correspond to the Macro average. Note that Late classification correspond to the hierarchical step that subclassifies late-type galaxies into Sa--Sb, Sbc--Scd, and Sd--Irr, whereas Early to the step that only subclassifies early-type galaxies into E and S0$^-$--S0a.}
    \label{tab:hier_metrics}
    \begin{tabular}{ *{2}{c} | *{4}{c} }
        \hline
        \multicolumn{2}{c |}{\textbf{Experiment}} & \multicolumn{4}{c}{\textbf{Performance Metrics}} \\
        Classification & Input data & Accuracy & Precision & Recall & AUC-ROC \\
        \hline
        Late & S2+C & $0.759\pm0.009$ & $0.678\pm0.024$ & $0.623\pm0.015$ & $0.890\pm0.005$ \\
        \hline
        \multirow{2}{*}{Early} & Structural2 & $0.757\pm0.011$ & $0.786\pm0.018$ & $0.772\pm0.016$ & $0.833\pm0.009$ \\
         & S2+C & $0.762\pm0.010$ & $0.787\pm0.016$ & $0.781\pm0.018$ & $0.846\pm0.010$ \\
        \hline
        Hier1 & \textit{all three models}:S2+C & $0.637\pm0.007$ & $0.613\pm0.014$ & $0.583\pm0.011$ &  $0.864\pm0.005$ \\
        Hier2 & \textit{2cats}:S2+C, \textit{Early}:Structural2, \textit{Late}:S2+C & $0.636\pm0.007$ & $0.611\pm0.014$ & $0.583\pm0.011$ &  $0.863\pm0.005$ \\
        \hline
    \end{tabular}
\end{table*}

From Table~\ref{tab:hier_metrics}, we observe that the performance metrics for Hier1 and Hier2 are quite similar, with Hier1 showing only up to a 0.2\% improvement over Hier2. However, accounting for the standard deviation values, both Hier1 and Hier2 metrics fall within the same range. This similar performance can be explained since, for the Early classification, both the Structural2 and S2+C configurations have similar performance. This indicates that for a sub-classification between elliptical and lenticular galaxies, the Structural2 configuration (which includes $BA$ and $c_{5090}$) is highly informative and effective. Hence, the addition of photometric parameters in the S2+C configuration does not significantly improve performance, suggesting that the Colour configuration may not provide substantial complementary information that is already captured by the structural parameters.

Furthermore, the precision and recall metrics are similar within each experiment (considering the standard deviation). In particular, for Hier1, the gap between these two metrics is 0.5\%, and for Hier2, it is 0.3\%, indicating a balanced performance.

Evaluating the Hier1 and Hier2 on unseen data (test subset), they achieved an accuracy of 65\% and 64\%, respectively. In a per-class performance context, on the test subset, we also observe a similar performance between the Hier1 and Hier2 classifications (see Fig.~\ref{tab:hier_class_report} in Appendix~\ref{app:perclass_metrics}), with differences not higher than 2\% in the metrics (precision, recall, and F1-score). 

\begin{figure}
     \centering
     \begin{tabular}{cc}
        \includegraphics[width=0.72\columnwidth,angle=0, trim={0 -0.1cm 0 0}]{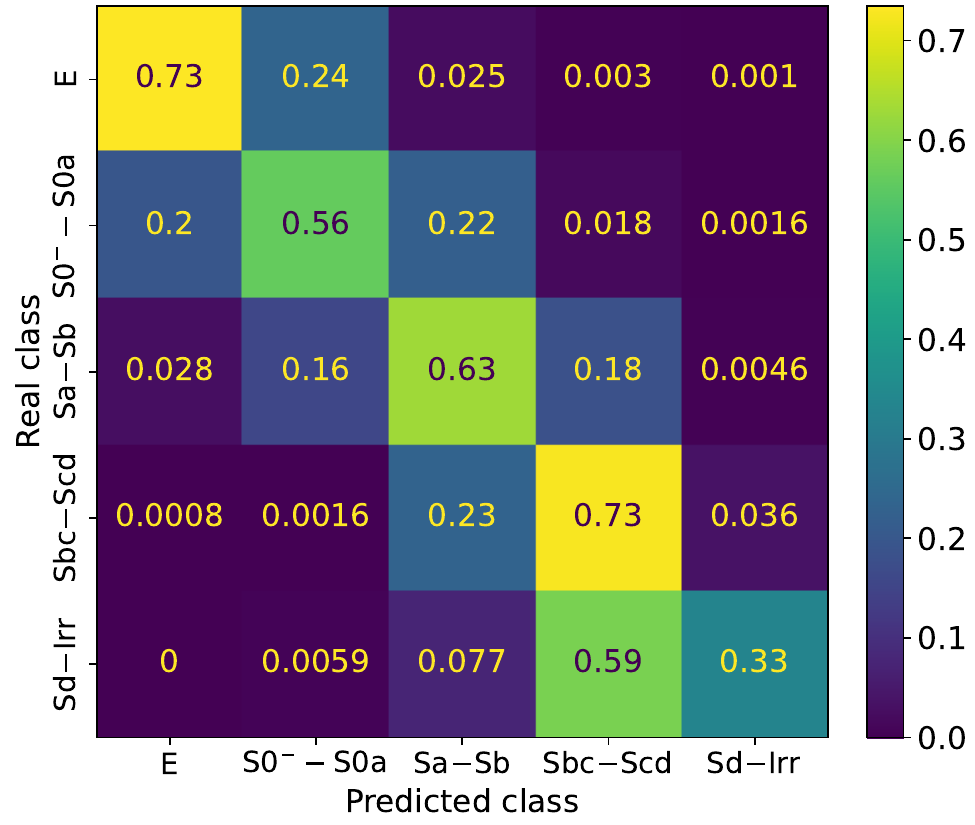} \\
        \includegraphics[width=0.72\columnwidth,angle=0]{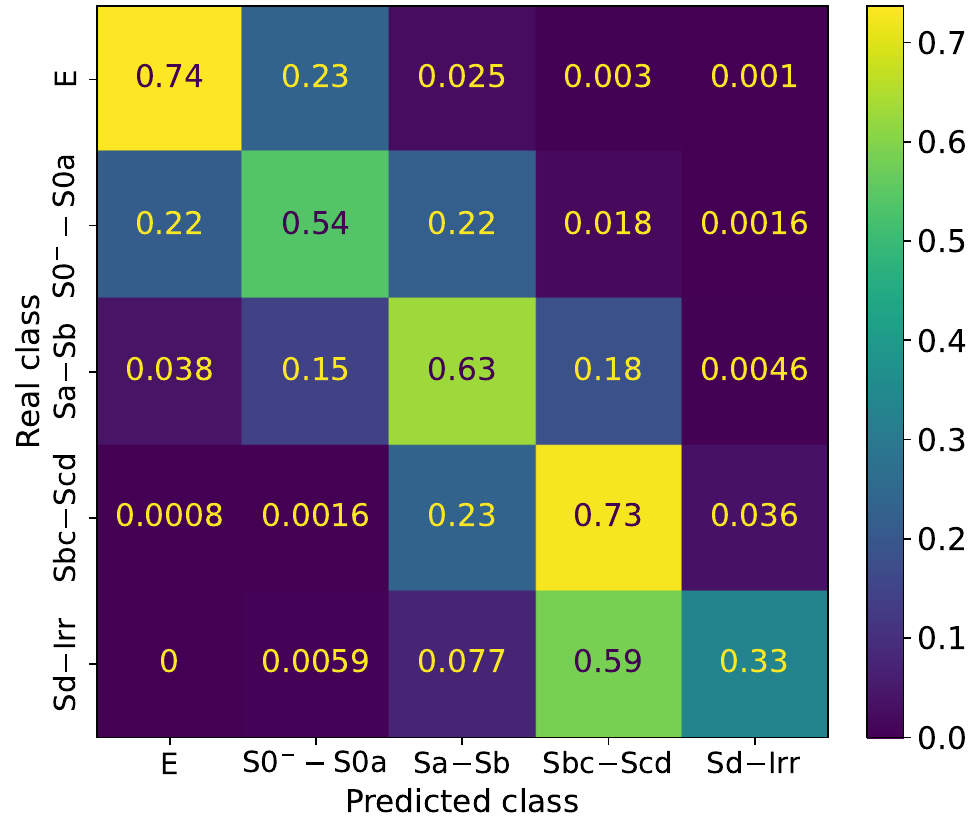}
     \end{tabular}
\caption{Confusion matrices for the Hier1 (upper) and Hier2 (bottom) classifications, calculated with the test subset. Colours are according to the accuracy of the classification. Hier1, Hier2, and 5cats (Fig.~\ref{fig:CM_2p_14p}) classifications yield similar performance.}
\label{fig:CM_hier}
\end{figure}

In Fig.~\ref{fig:CM_hier}, we provide the CM's of the Hier1 (upper panel) and Hier2 (bottom panel) hierarchical classifications, calculated using the test subset. Again, both classifications yield similar results, with only a 1\% difference in the successful prediction of Class 0 (elliptical) and a 2\% difference for Class 1 (lenticular). We also observe that the majority of the model's misclassifications are between adjacent classes. For instance, a 22\% of Class 1 instances are misclassified as Class 2 (Sa--Sb), and $\sim$20\% as Class 0. This underscores the morphological similarities between adjacent galaxy types.

%------------------------Figure----------------------------------%

Comparing the Hier1 and Hier2 performance results with those obtained from the 5cats direct classification using the S2+C parameter configuration (see Table~\ref{tab:metrics} and Fig.~\ref{fig:CM_2p_14p}), we observe that they are closely aligned, with differences of 0.2\%-0.5\% in mean performance metrics and up to 3\% in CM and AUC-ROC. This indicates that hierarchical and direct classification approaches are equally effective for galaxy classification. Given these findings, along with the increased complexity (in both implementation and evaluation) and computational cost of the hierarchical approach, we will continue our discussion adopting the simpler and more efficient 5cats direct classification.

\section{Discussion} \label{sec:discussion}

\subsection{Model interpretation} \label{sec:model_interp}
In the following, we present the interpretative analysis of the trained XGBoost model. For illustrative purposes, we focused on the 5cats direct classification for this analysis, as its greater complexity provides a more comprehensive exploration of feature contributions across a broader range of galaxy types than the 2cats classification. Understanding the relationships between input features and model predictions is important for the model interpretation. For this purpose, we use the SHapley Additive exPlanations \citep[SHAP;][]{Lundberg2017} library\footnote{\url{https://shap.readthedocs.io/en/latest/}}, a powerful visualization tool designed to elucidate the decision-making processes of complex models. SHAP is based on the concept of Shapley values, a game-theoretic approach that offers a unified measure to explain each feature's contribution to a prediction. SHAP values specify both the direction of a feature's impact (whether it increases or decreases the prediction probability) and the magnitude of its contribution. In particular, we use the SHAP functions \texttt{shap.summary\_plot} and \texttt{shap.plots.waterfall} to visualize these contributions.

\subsubsection{SHAP global analysis}
The \texttt{shap.summary\_plot} visualization function provides a global view of feature importance by aggregating the SHAP values across the entire dataset. It displays how much each feature contributes to the prediction classes, identifying the most influential features for the process of distinguishing between different types of galaxies. 

\begin{figure*}
     \centering
     \includegraphics[width=0.33\textwidth,angle=0]{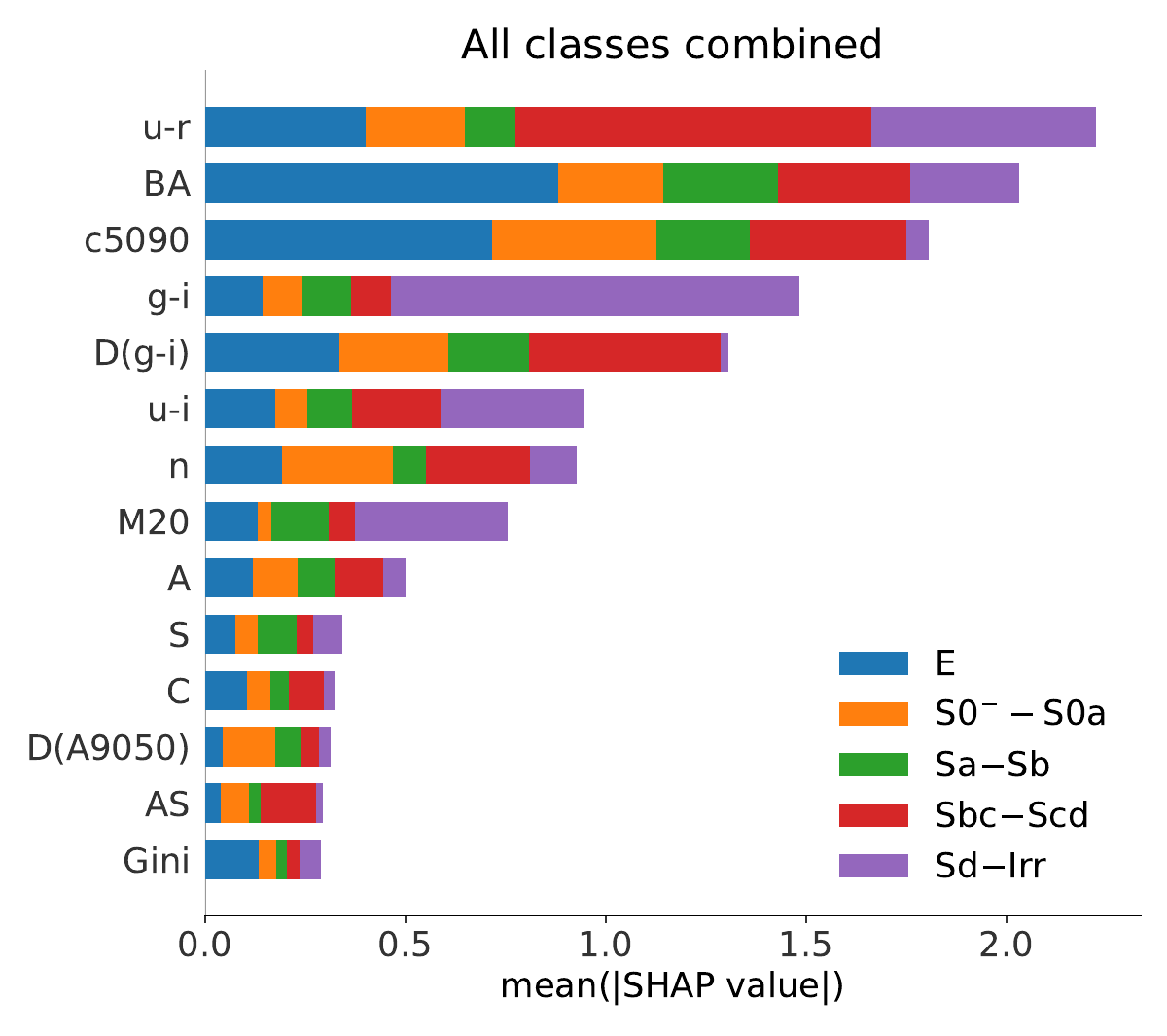}
     \includegraphics[width=0.33\textwidth,angle=0]{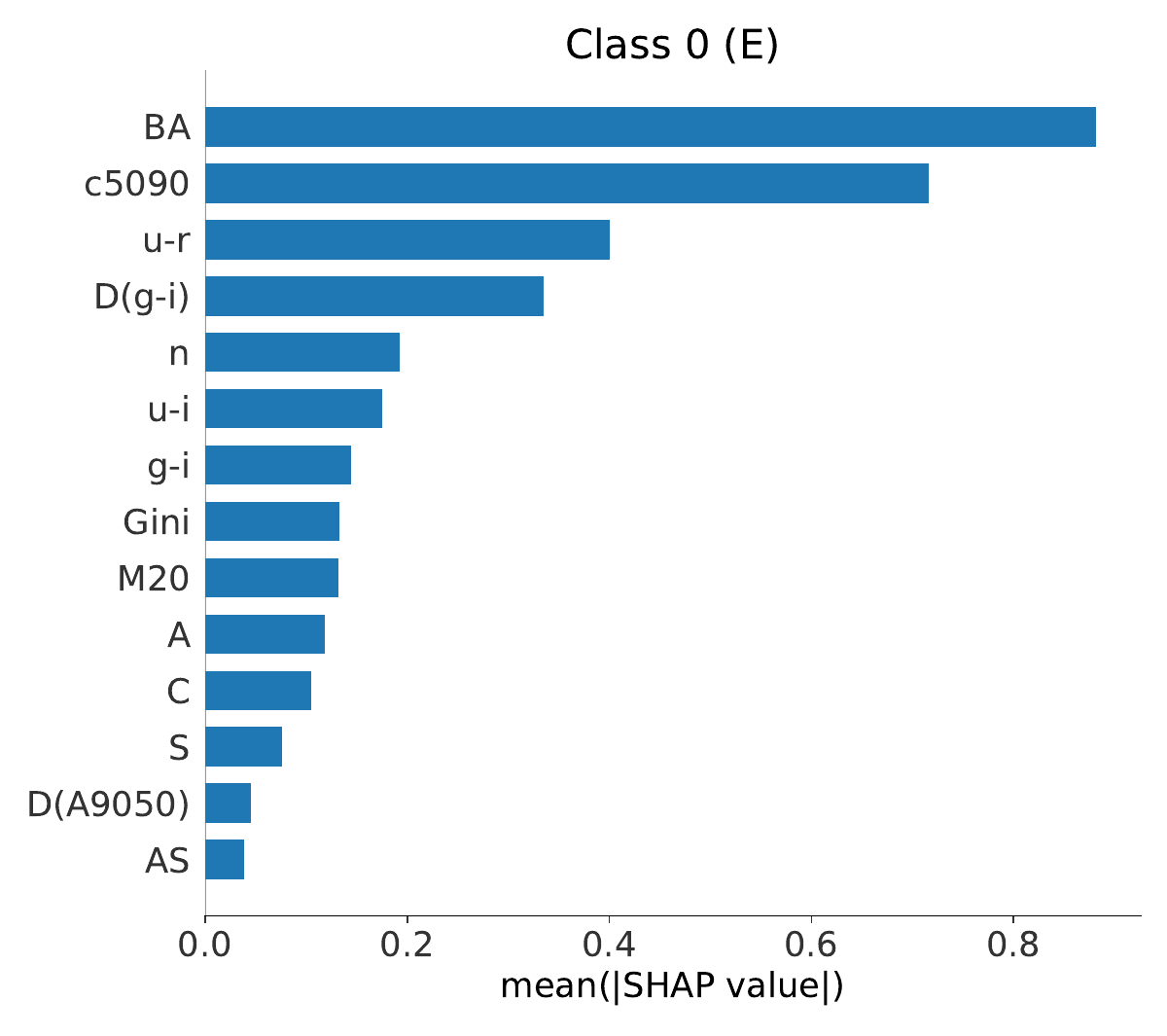}
     \includegraphics[width=0.33\textwidth,angle=0]{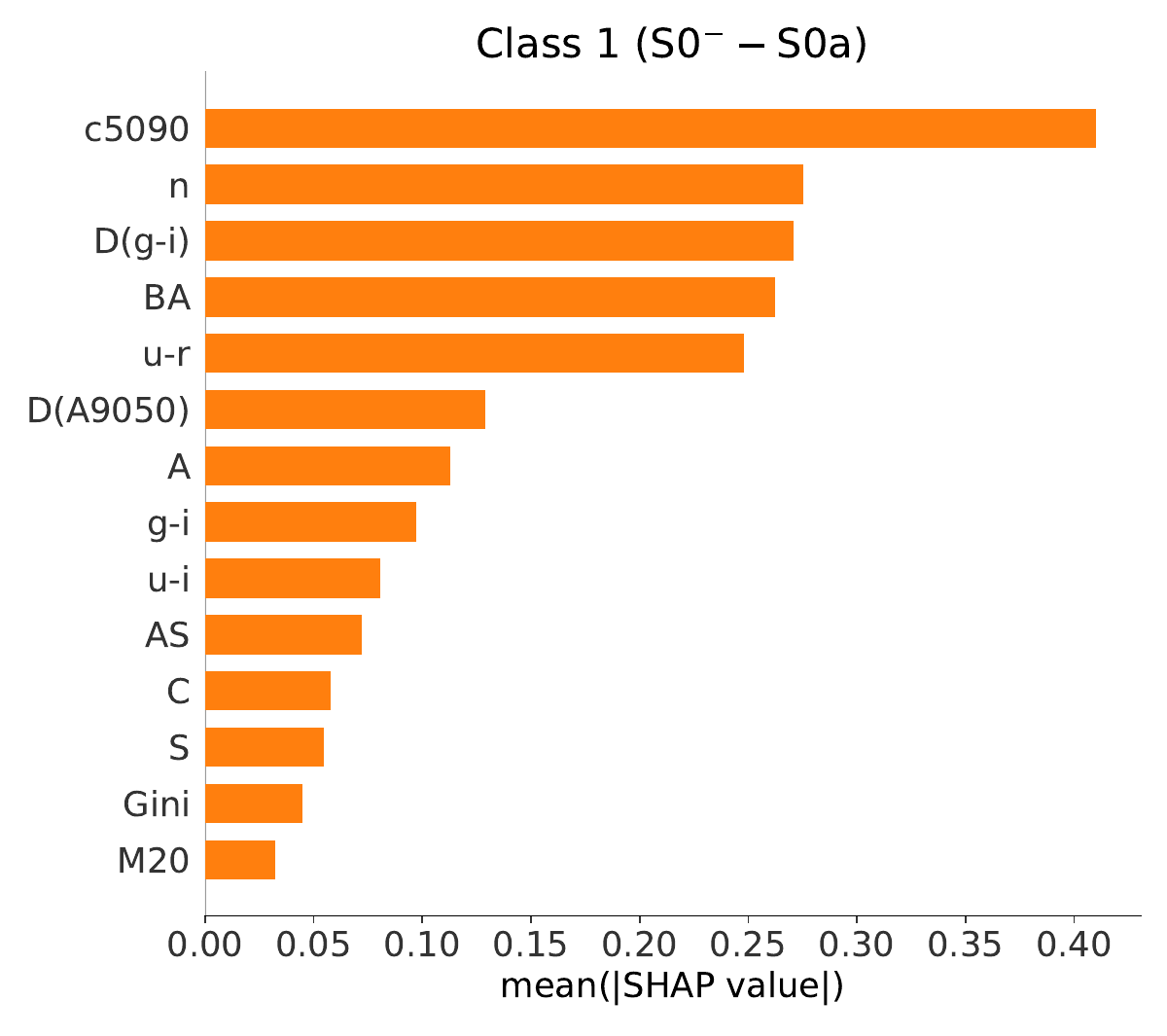} \\
     \includegraphics[width=0.33\textwidth,angle=0]{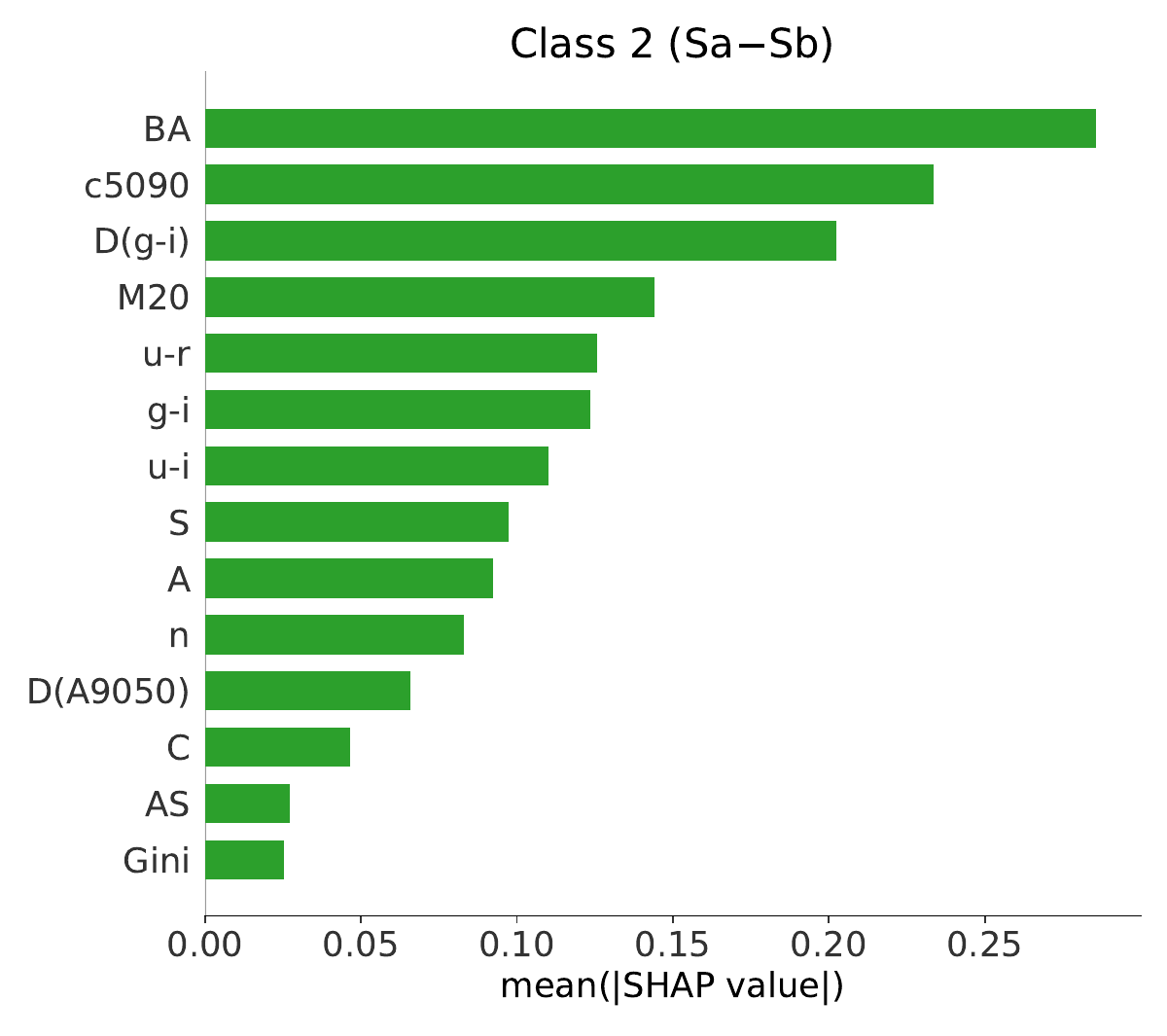}
     \includegraphics[width=0.33\textwidth,angle=0]{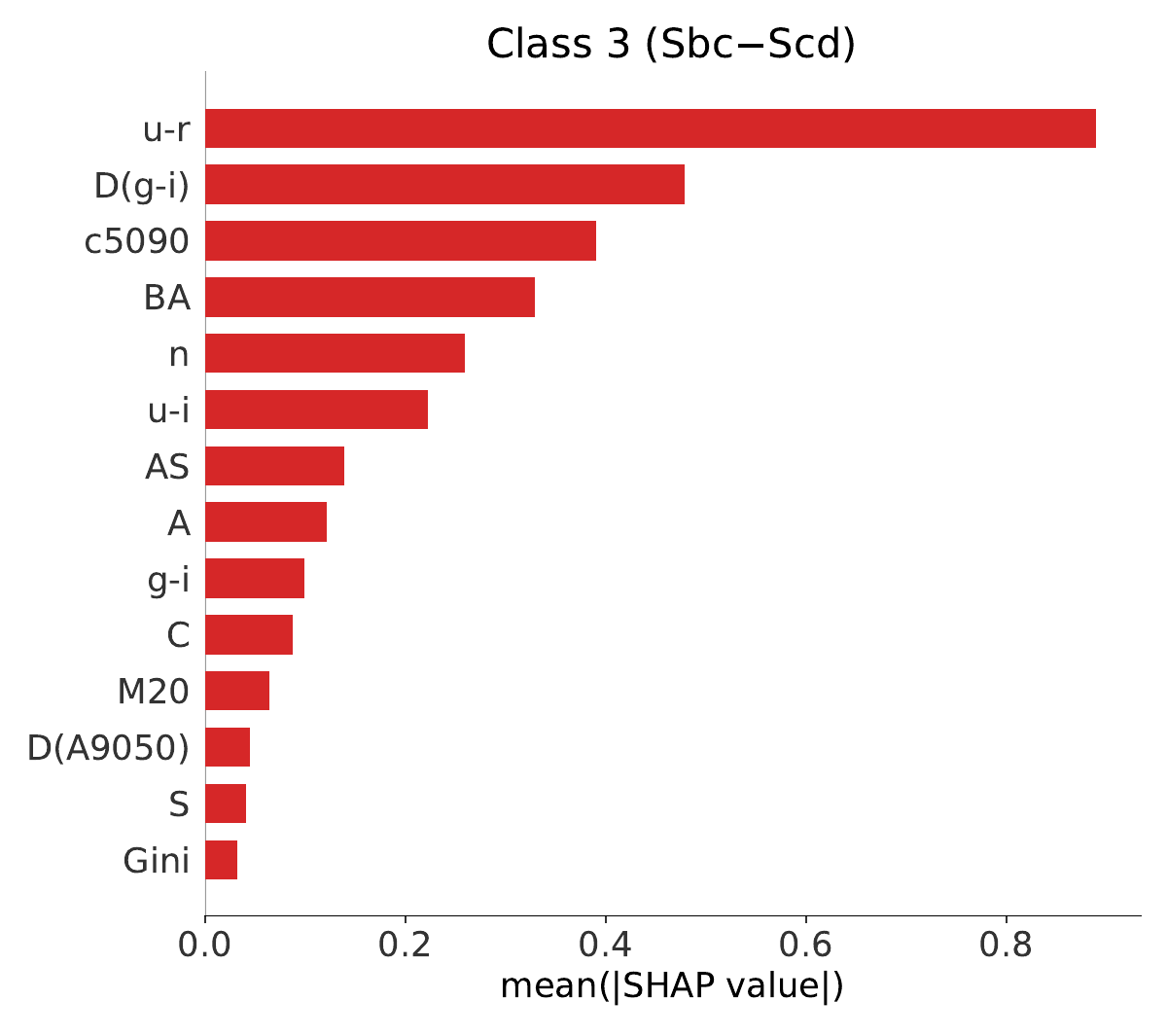}
     \includegraphics[width=0.33\textwidth,angle=0]{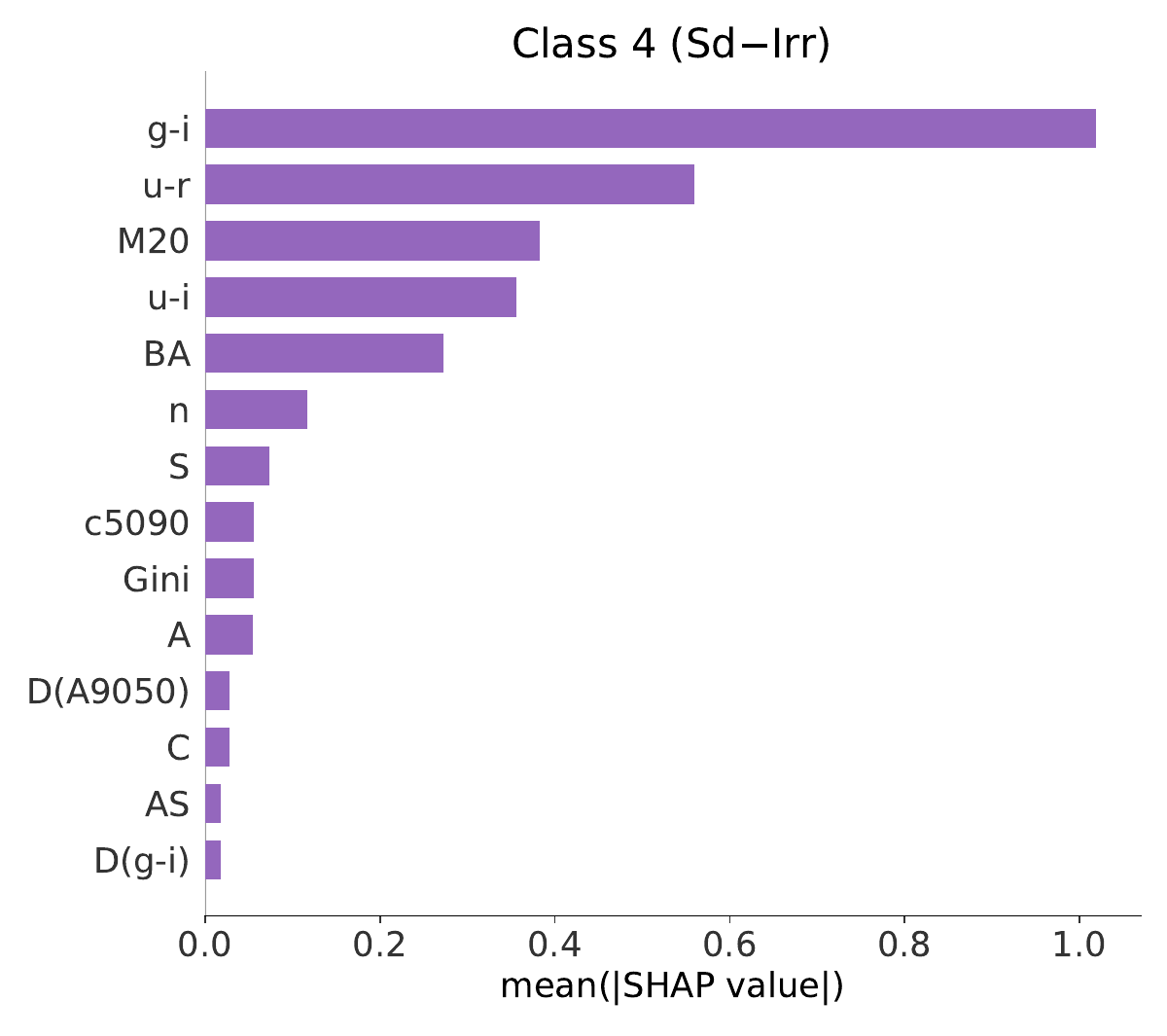}
\caption{Feature importance in the XGBoost model for the 5cats direct classification using the S2+C parameter configuration. The upper-left panel shows the SHAP summary plot for all classes combined, and the other panels the SHAP summary plot for each individual class, where the upper-middle panel corresponds to Class 0 and the bottom-right panel to Class 4. In each plot, the features are ranked vertically by importance, with the mean absolute SHAP value displayed horizontally. $u-r$, $BA$, $c5090$, $g-i$, and $\Delta (g-i)$, are, overall, the parameters with more impact on the model. Additionally, the structural parameters are more important for early-type galaxies whereas the photometric parameters are more important for late-type galaxies. }
\label{fig:shap_14p}
\end{figure*}

Figure~\ref{fig:shap_14p} presents the SHAP summary plots for the 5cats direct classification task using the S2+C parameter configuration. This figure consists of six panels: the upper-left panel shows the combined feature importance for all classes, while the remaining five panels represent the feature importance for each class separately, going from Class 0 up to Class 4. Each panel shows a horizontal bar chart with the morphological parameters ranked by importance on the $y$-axis and the mean absolute SHAP values on the $x$-axis. Parameters with larger bars have more impact on the model's predictions. Moreover, the different colours in the horizontal bars of the upper-left panel correspond to each class. Hence, the extent of the colour within that bar corresponds to the importance of that parameter for the corresponding class. 

%------------------------Figure----------------------------------%

Overall, we can observe that a combination of colour and structural parameters, namely the $u-r$ colour, a shape parameter (the axis ratio $BA$), the surface brightness distribution (the $c_{5090}$ inverse concentration index), and a gradient parameter ($\Delta (g-i)$) are playing a significant role in the model's galaxy classification across all classes (upper-left panel of Fig.~\ref{fig:shap_14p}). This is consistent with the results reported in other works \citep[e.g.,][]{Strateva2001, Graham2019}.

The subsequent panels of Fig.~\ref{fig:shap_14p} also illustrate the relative importance of the features, now split into different galaxy classes. Specifically, for Class 0 (elliptical, upper-middle panel) the most influential feature is $BA$, followed by $c_{5090}$ and $u-r$. Elliptical galaxies tend to have a smoother spheroidal shape, with a centrally concentrated light distribution due to a dominant bulge and older stellar populations. Consequently, $BA$ and $c_{5090}$ capture essential morphological and structural characteristics, while $u-r$, third in importance, correlates with the stellar population ages, helping differentiate Class 0 from other classes.

For lenticular galaxies (S0$^-$--S0a) in Class 1 (upper right-most panel of Fig.~\ref{fig:shap_14p}), the $c_{5090}$, $n$ (Sérsic index), $\Delta \left( g-i \right)$, $BA$, and $u-r$ features are the most relevant. These last four parameters show consistently similar mean SHAP values, suggesting a wide diversity of structural properties in this class. While S0 galaxies show a diversity of bulge components within a definite disk structure, S0a galaxies show, in addition, hints of a very tight spiral structure in the outer disk along with mixed stellar populations with older stars in the bulge and some intermediate-age stars in the disk, supporting the relevance of the colour and colour gradient parameters. 

For Class 2 (bottom-left panel of Fig.~\ref{fig:shap_14p}), $BA$, $c_{5090}$, and a colour gradient $\Delta \left( g-i \right)$ parameter appear as the more relevant, followed by $M_{20}$. This class includes Sa, Sab, and Sb galaxies, which show a diversity of bulges in a disk, hence the relevance of $BA$ and $c_{5090}$ to capture those structural characteristics. This class also exhibit a variety of spiral arms, mixed stellar populations, increasing gas and dust content from Sa to Sb, and higher star formation rates compared to Classes 0 and 1, underscoring the relevance of $M_{20}$ and $\Delta \left( g-i \right)$ in capturing the concentration of the brightest regions in the disk and arms.

In contrast to early-types, in Class 3 (bottom-middle panel of Fig.~\ref{fig:shap_14p}) the $u-r$ colour and $\Delta \left( g-i \right)$ colour gradient parameters play the most relevant role in the model's prediction, followed this time by the structural features ($c_{5090}$, $BA$, and $n$). Sbc, Sc, and Scd galaxies compose this class, which are characterized by a central bulge of decreasing prominence but prominent, loosely wound spiral arms in the disk. These arms are traced by abundant star-forming regions and contain higher amounts of gas and dust, supporting the relevance of the colour and colour gradient parameters as first-order predictors. 

Finally, for Class 4 (bottom-right panel of Fig.~\ref{fig:shap_14p}), the different combinations of colour parameters, represented by the $g-i$, $u-r$, and $u-i$ colour indices, and the $M_{20}$ parameter are the most important for the model prediction. This class is composed by very late Sd, Sdm, Sm, and Irr galaxies, where the bulge component goes from being almost absent to being completely absent, and the spiral structure goes from being very loosely wound and almost absent to completely absent, with very prominent and widespread star-forming regions. Hence, the importance of the colour parameters, capturing the wide range of star formation activity and the youngest stellar populations in these galaxies. $M_{20}$ and $BA$, although with less impact than colour parameters, help to capture localized structures along the disk.

In summary, we note that for Classes 0, 1, and 2, the structural shape ($BA$) and surface brightness distribution ($c_{5090}$) parameters appear among the most relevant for the model predictions, followed by the colour ($u-r$) and/or colour-gradient ($\Delta \left( g-i \right)$) parameters. These results are consistent with the morphological results by \citet{Cheng2011} and \citetalias{VazquezMata2022} who argue that $BA$ and $c_{5090}$ are among the most influential parameters for a light-based morphological classification of early-type galaxies (Classes 0 and 1). In contrast, for Classes 3 and 4, colour and colour gradient parameters are among the most influential for the model's predictions (consistent with, e.g., \citealt{Park2005}), reflecting the star formation and younger stellar populations properties along the disk in these galaxies. It is also noticeable the presence of the $M_{20}$ parameter reflecting the degree of locality of the light distribution in star-forming regions when going from Sbc up to irregular types. 

Overall, the XGBoost model was trained to classify galaxies using a combination of structural and colour parameters associated to their light distribution and star formation properties, rendering results consistent with the known properties of the different galaxy classes. These results, analysed with the SHAP tool, capture the importance of the structural and colour features, aligning with the morphological properties of galaxies, and reinforcing the reliability and interpretability of the XGBoost model's predictions.  

It is important to mention that these feature importance scores are specific to the trained XGBoost model and the dataset it was trained on.

\subsubsection{SHAP individual analysis} 
\label{sec:shap_waterfall}
To understand in more detail the model's predictions, we perform, this time on an individual basis, a SHAP analysis looking at the contribution of each feature to individual predictions and trying to recognize cases where the model either correctly or incorrectly predicts a galaxy class. To that purpose, we use the \texttt{shap.plots.waterfall} visualization function of the SHAP tool. The SHAP waterfall plot decomposes the prediction of a particular instance into the contributions of each feature to the final classification outcome. In the context of a multi-class classification, the waterfall plot would display the contributions of each feature towards assigning the instance to one of the multiple classes. Hence, in a five-class problem, for each instance there will be five waterfall plots, one per class.

\begin{figure*}
     \centering
     $\qquad$
     \includegraphics[width=0.2\textwidth]{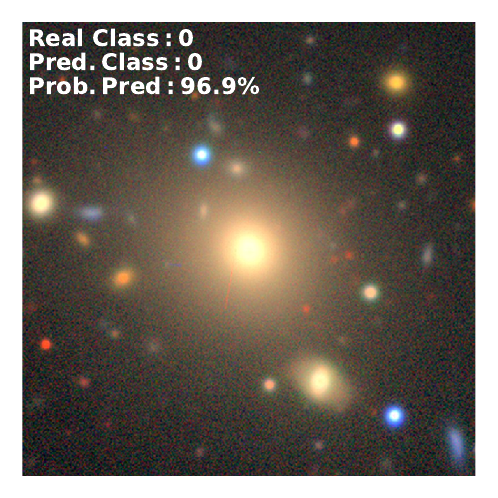} 
     $\qquad\qquad\qquad\qquad$
     \includegraphics[width=0.2\textwidth]{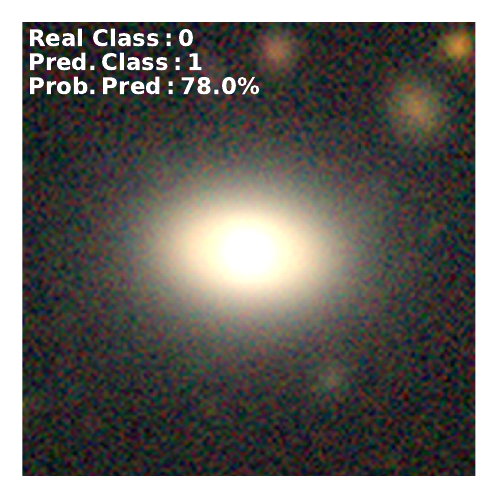} 
     $\qquad\qquad\qquad\qquad$
     \includegraphics[width=0.2\textwidth]{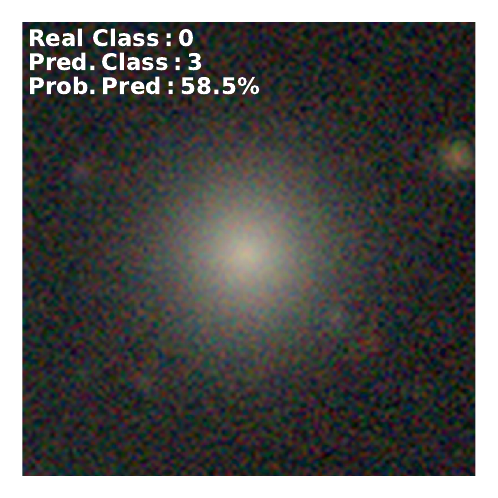} \\
     \rotatebox[origin=l]{90}{$\qquad\qquad\,\,$\textbf{Class 0 (E)}}$\,$
     \adjustbox{cframe=red}{\includegraphics[width=0.31\textwidth, height=0.215\textwidth]{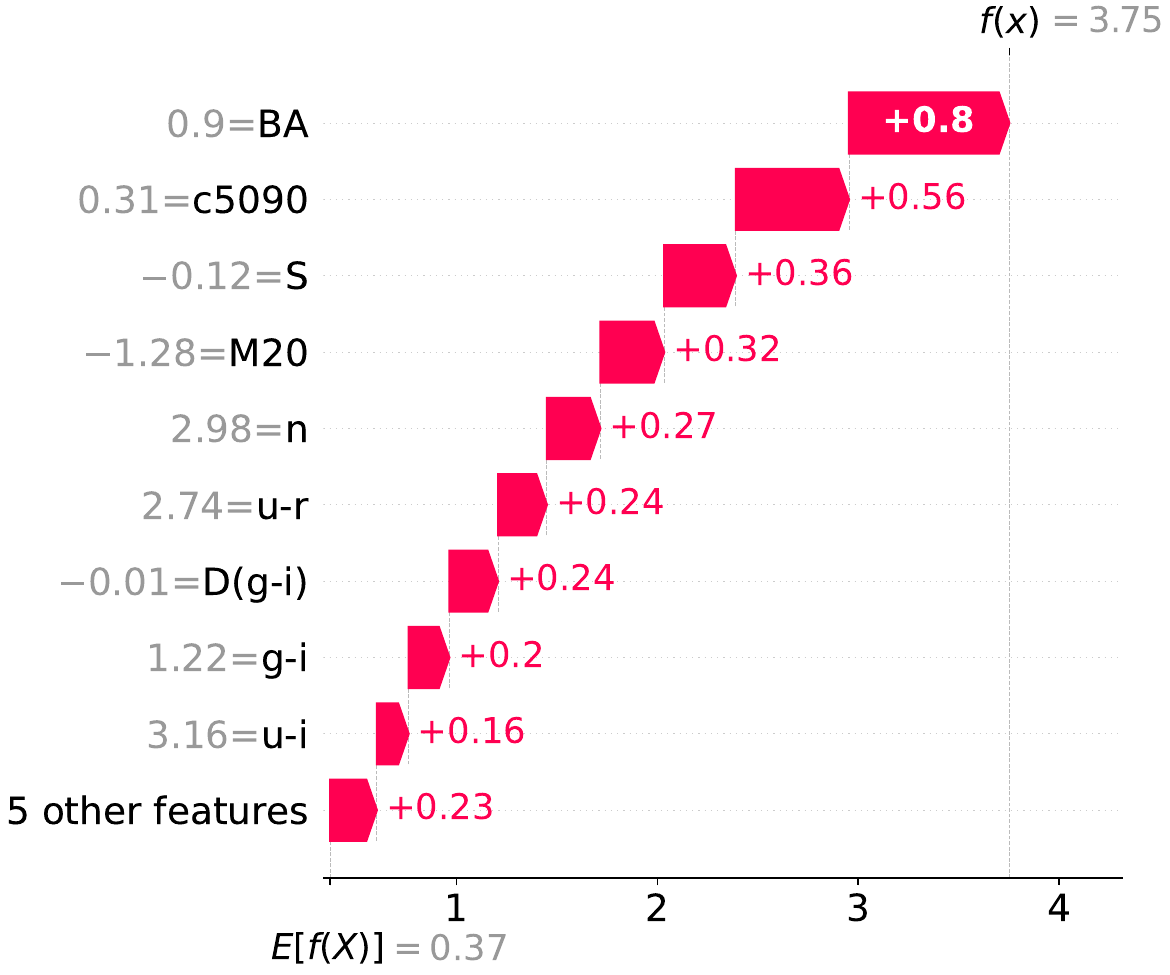}}
     \includegraphics[width=0.31\textwidth, height=0.215\textwidth]{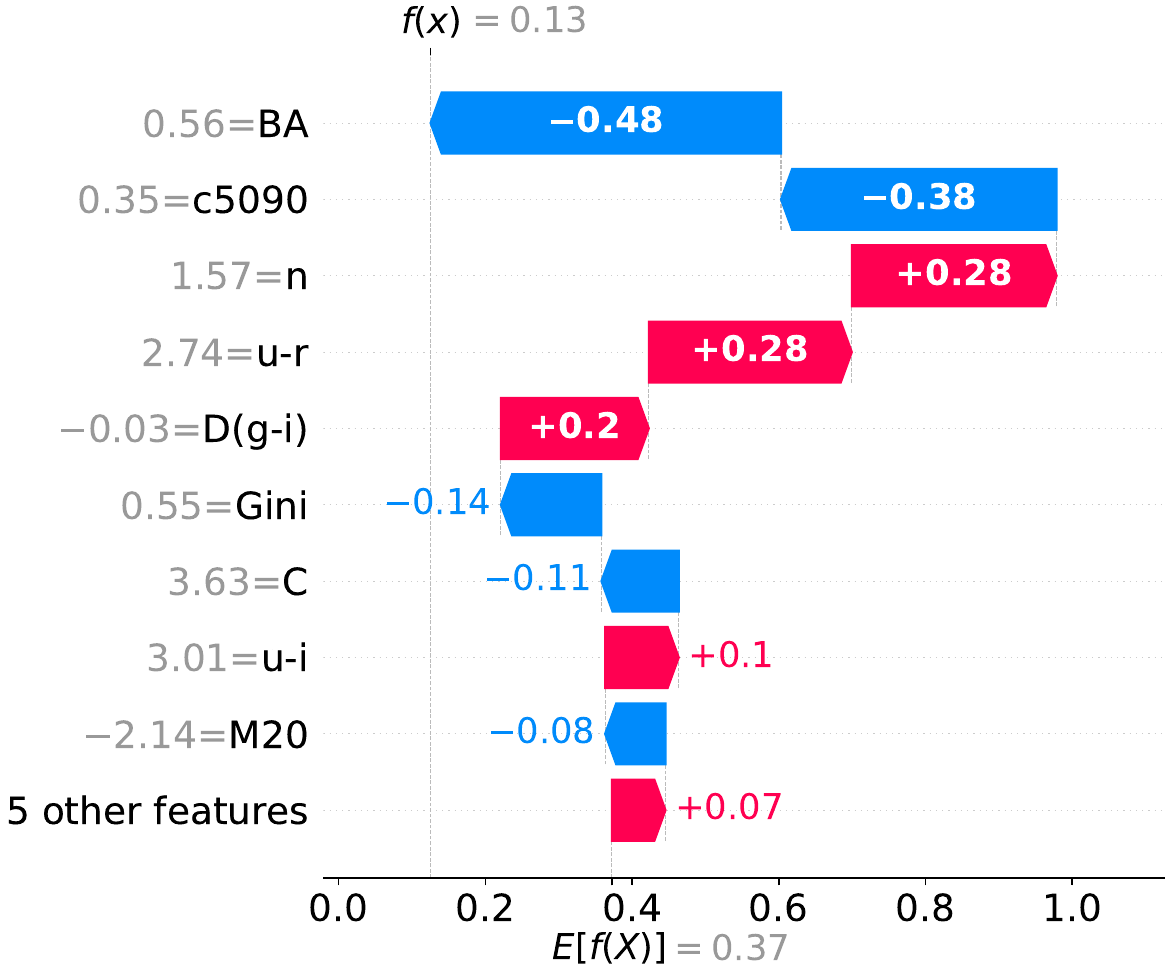}
     \includegraphics[width=0.31\textwidth, height=0.215\textwidth]{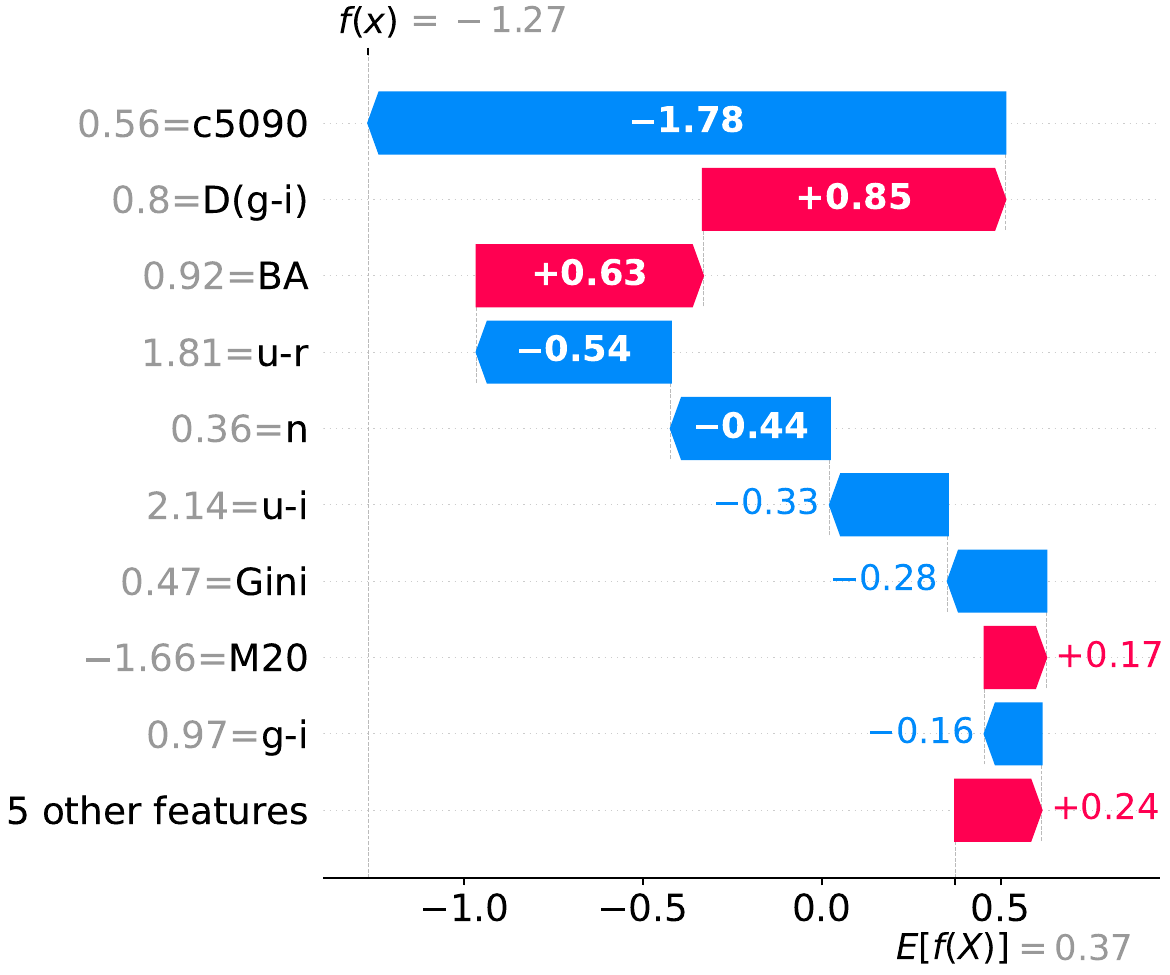} \\
     \rotatebox[origin=l]{90}{$\qquad\quad$\textbf{Class 1 (S0$^-$--S0a)}}$\,$
     \includegraphics[width=0.31\textwidth, height=0.215\textwidth]{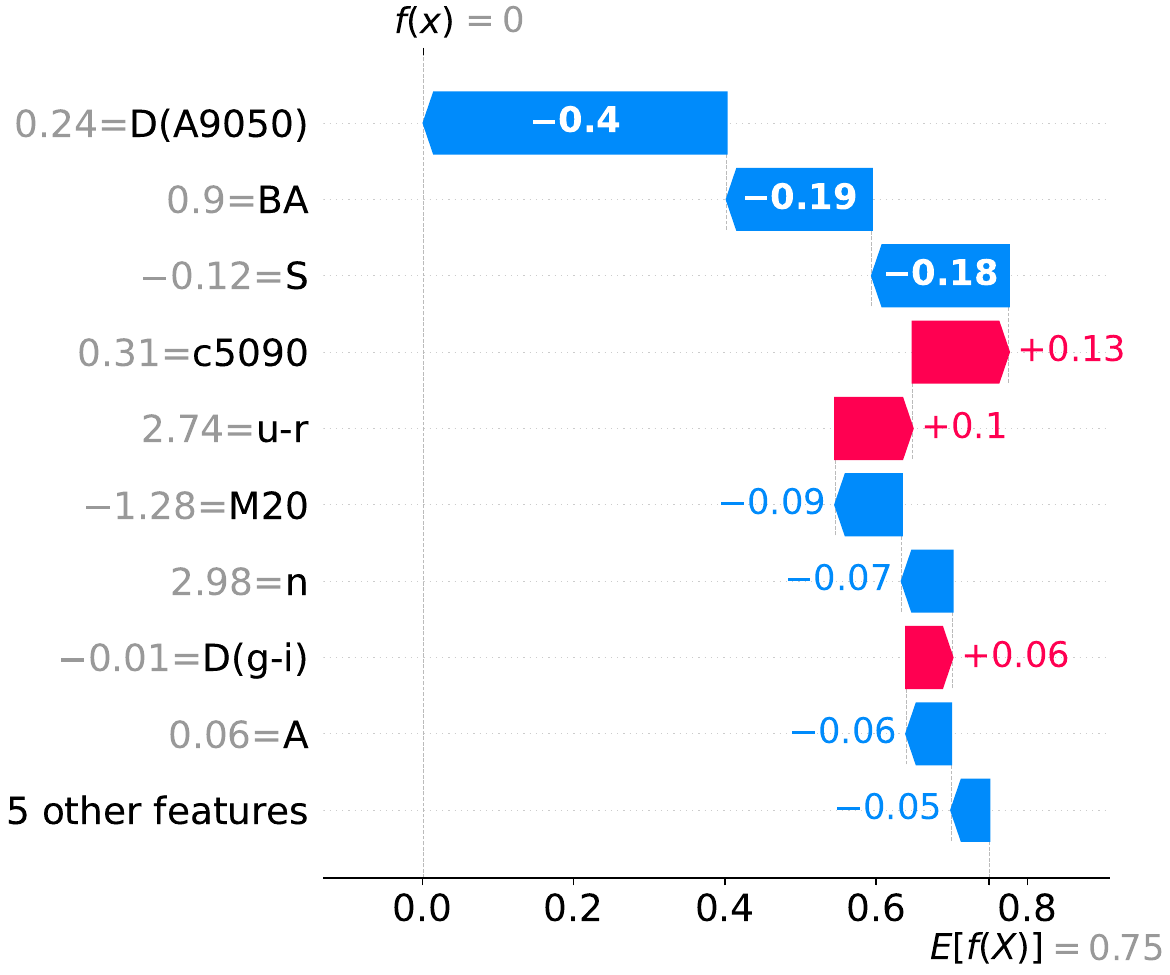}
     \adjustbox{cframe=red}{\includegraphics[width=0.31\textwidth, height=0.215\textwidth]{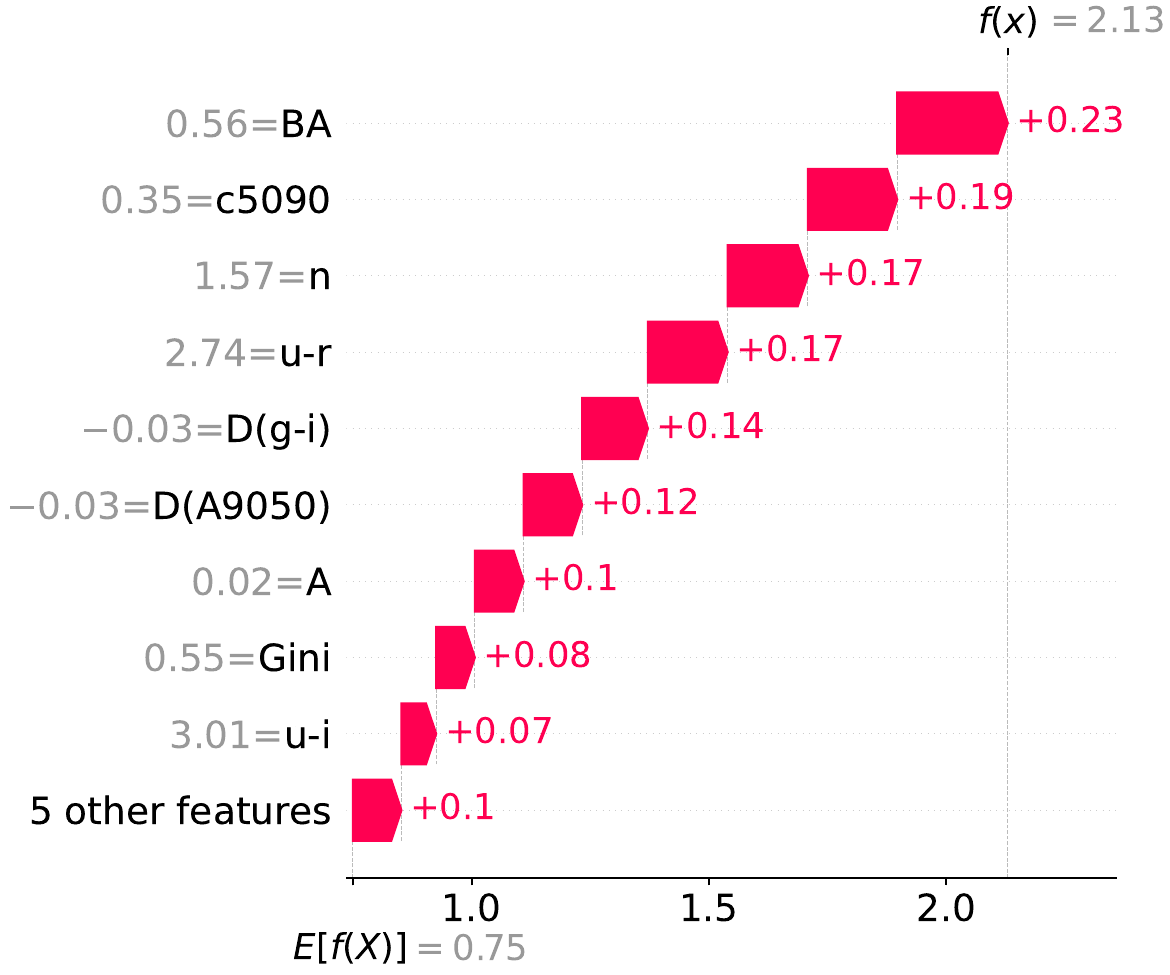}}
     \includegraphics[width=0.31\textwidth, height=0.215\textwidth]{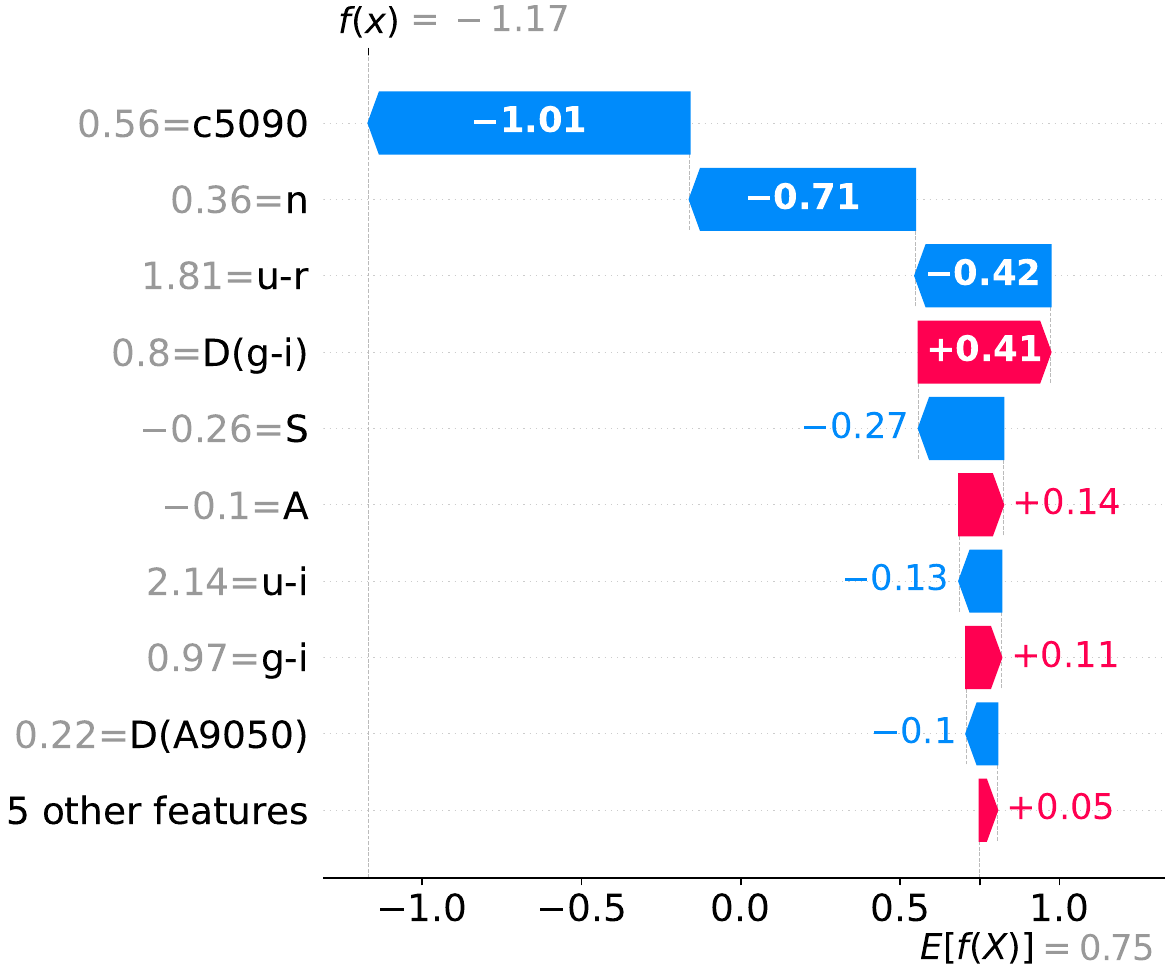} \\
     \rotatebox[origin=l]{90}{$\qquad\qquad$\textbf{Class 2 (Sa--Sb)}}$\,$
     \includegraphics[width=0.31\textwidth, height=0.215\textwidth]{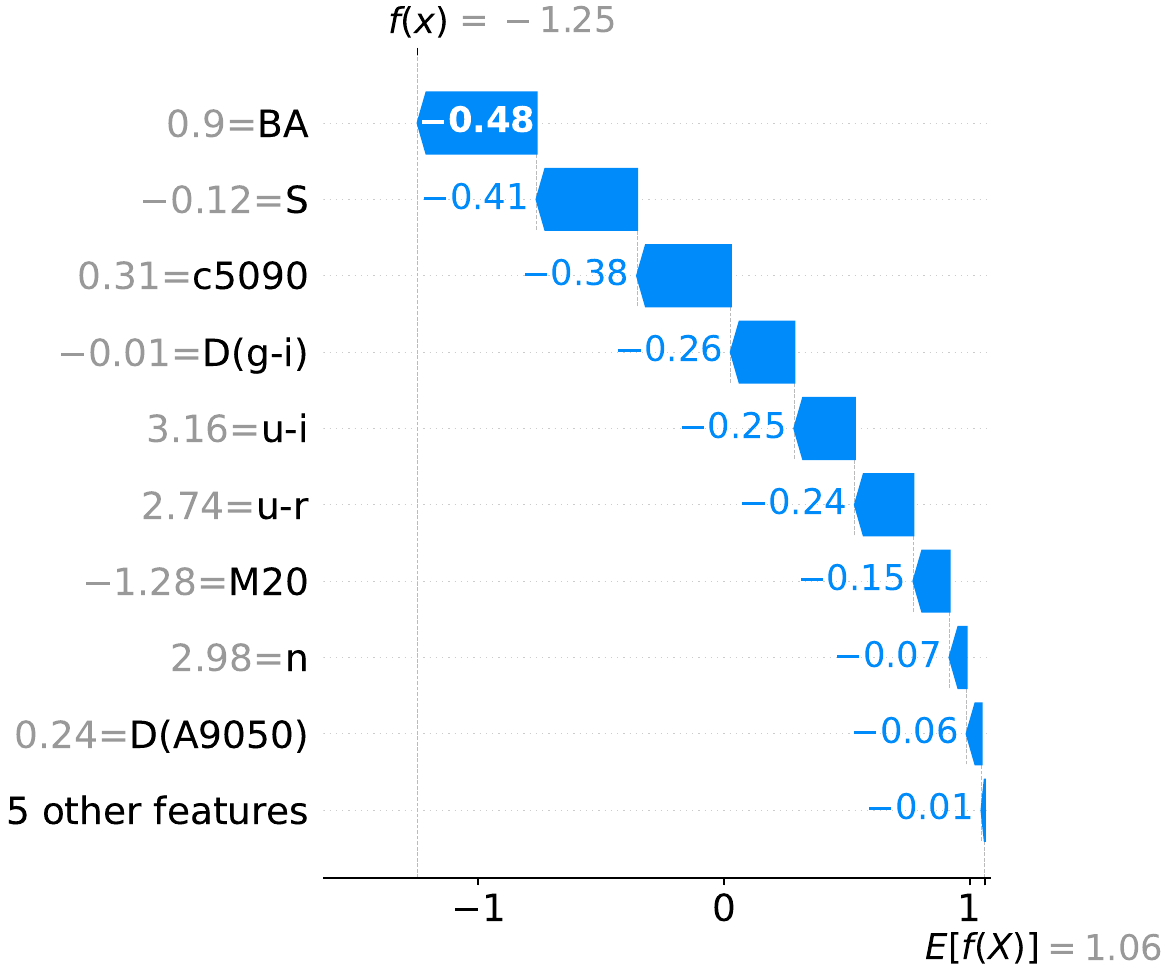}
     \includegraphics[width=0.31\textwidth, height=0.215\textwidth]{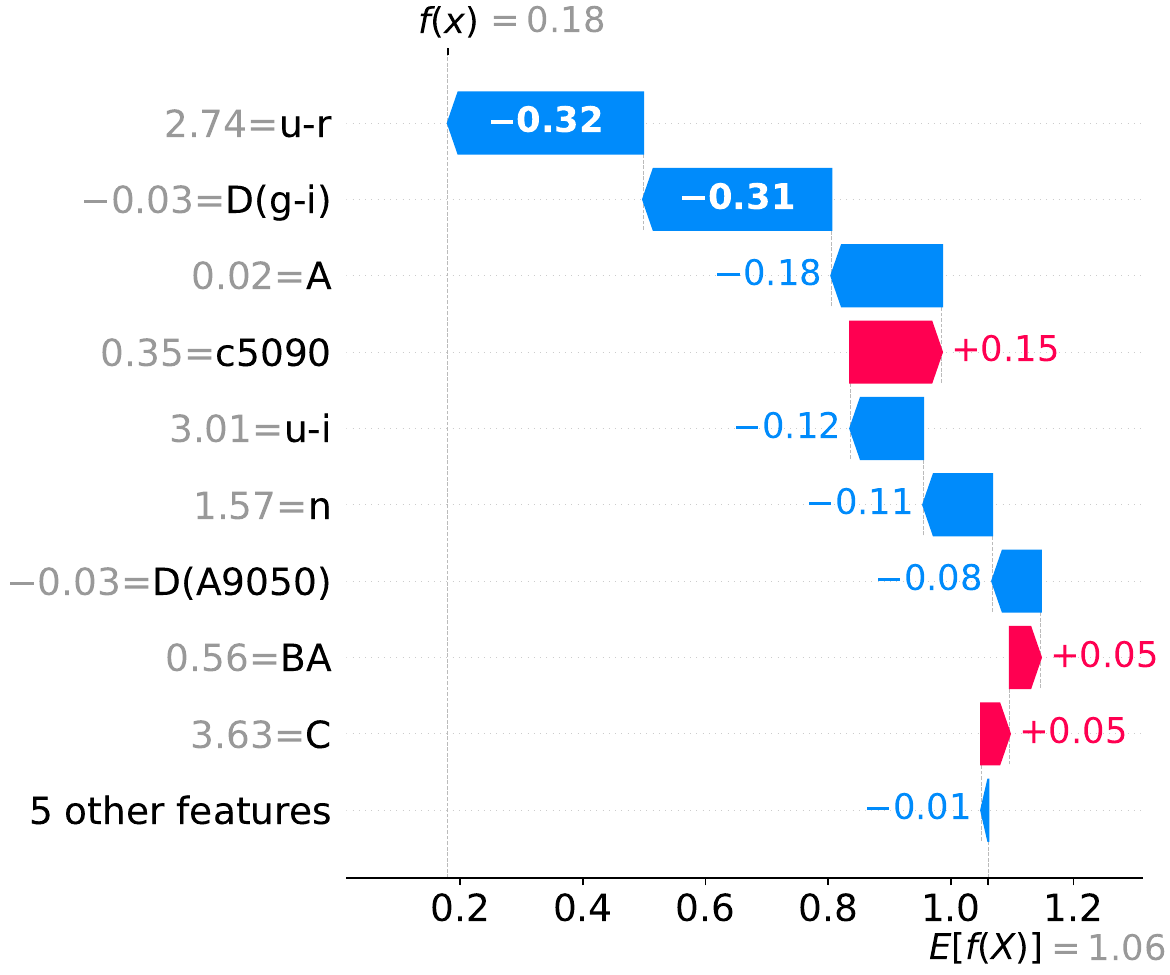}
     \includegraphics[width=0.31\textwidth, height=0.215\textwidth]{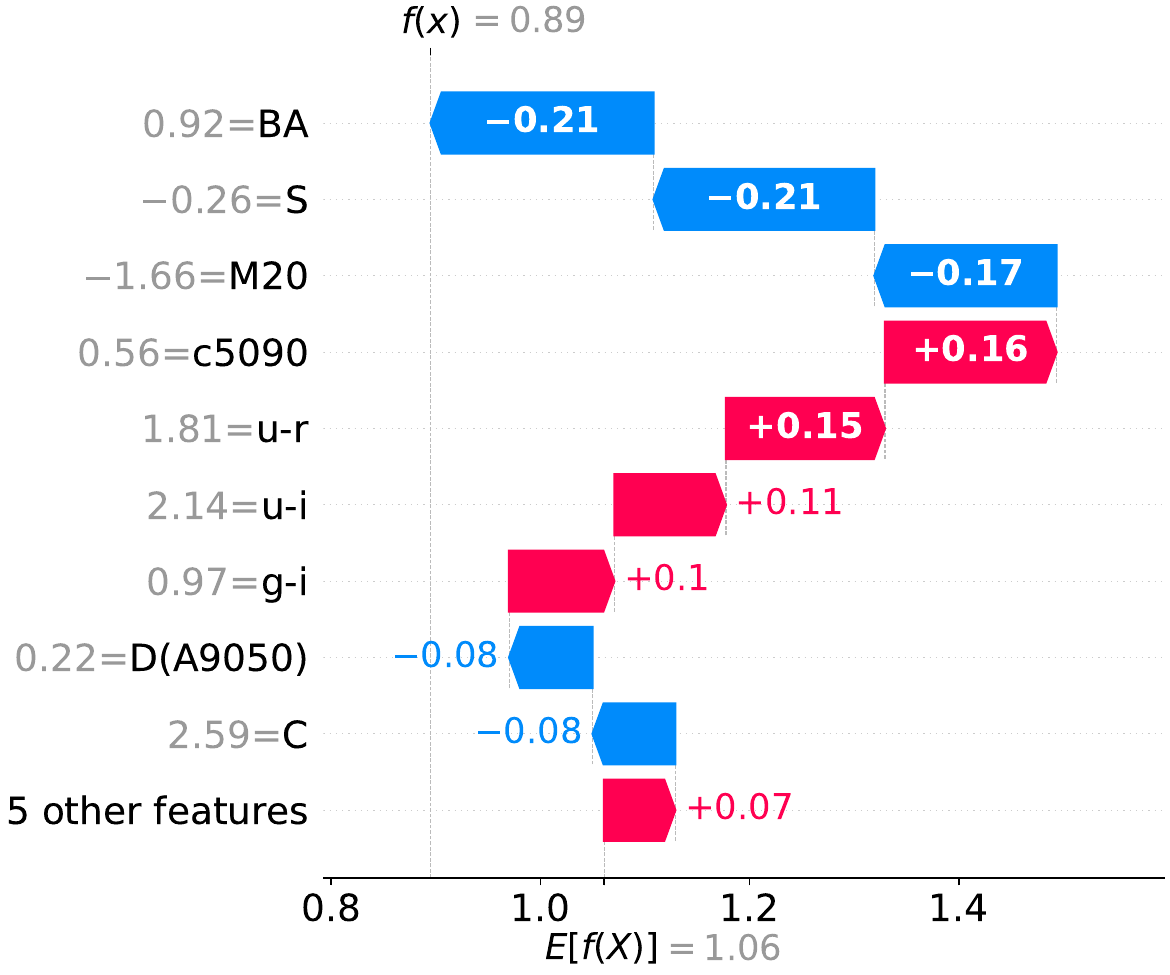} \\
     \rotatebox[origin=l]{90}{$\qquad\quad\,\,$\textbf{Class 3 (Sbc--Scd)}}$\,$
     \includegraphics[width=0.31\textwidth, height=0.215\textwidth]{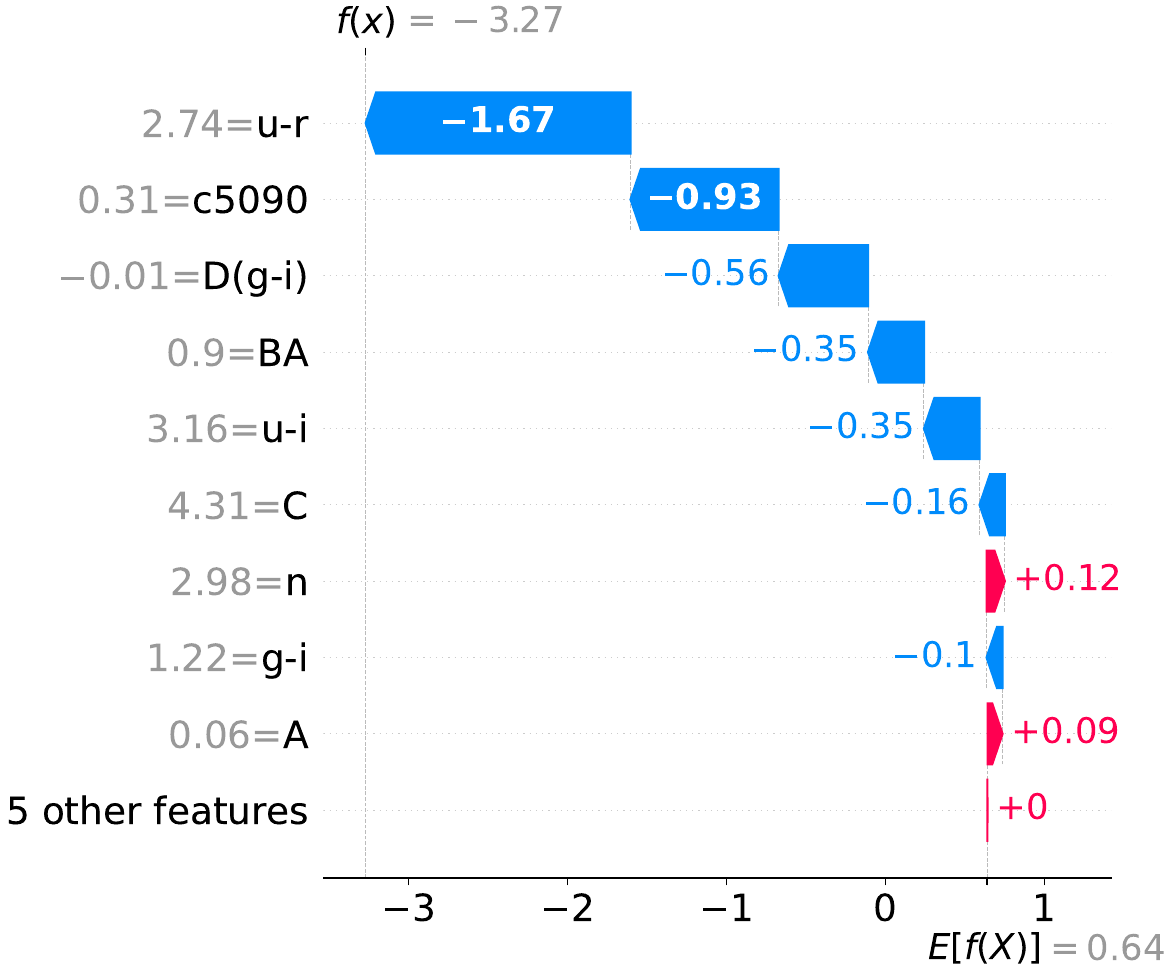}
     \includegraphics[width=0.31\textwidth, height=0.215\textwidth]{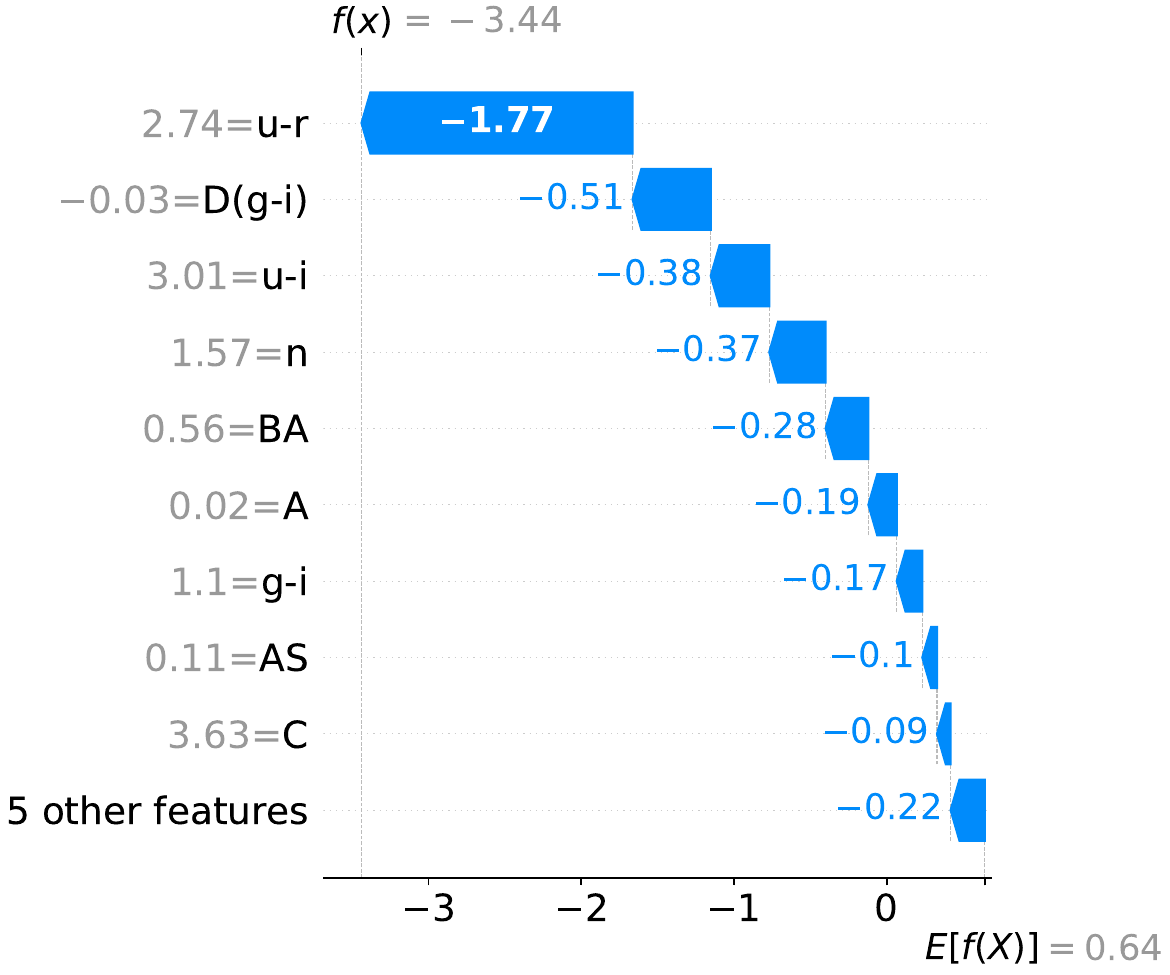}
     \adjustbox{cframe=red}{\includegraphics[width=0.31\textwidth, height=0.215\textwidth]{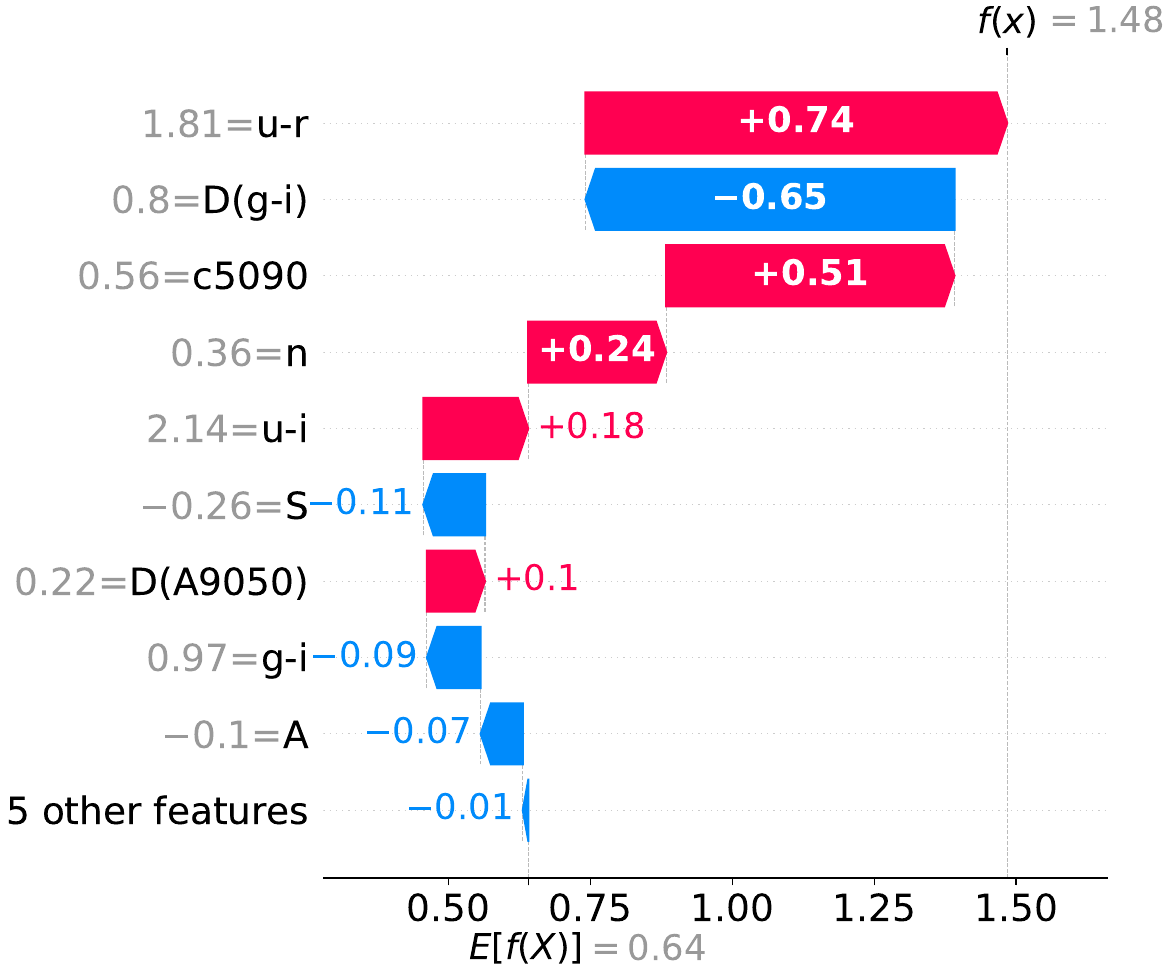}} \\
     \rotatebox[origin=l]{90}{$\qquad\quad\,\,\,\,$\textbf{Class 4 (Sd--Irr)}}$\,$
     \includegraphics[width=0.31\textwidth, height=0.215\textwidth]{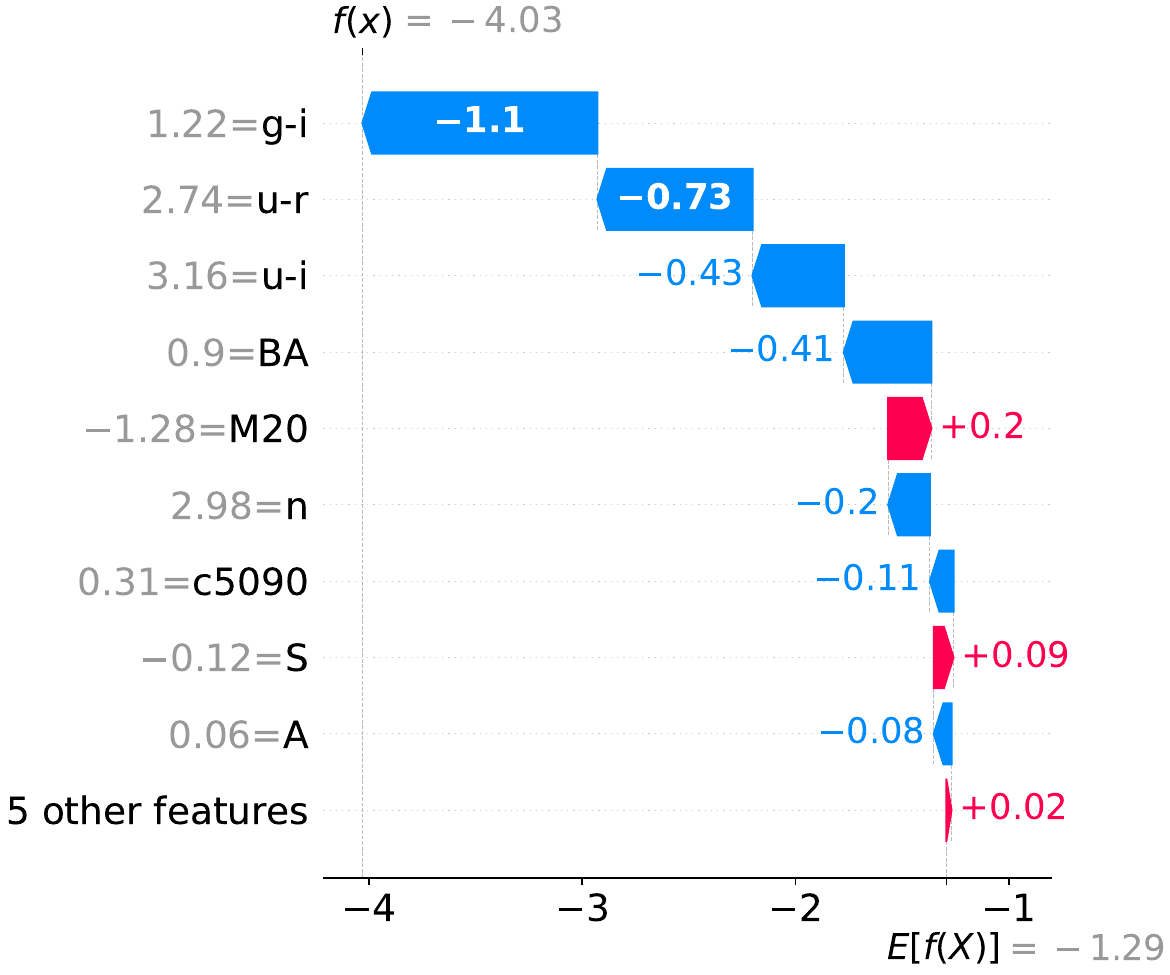}
     \includegraphics[width=0.31\textwidth, height=0.215\textwidth]{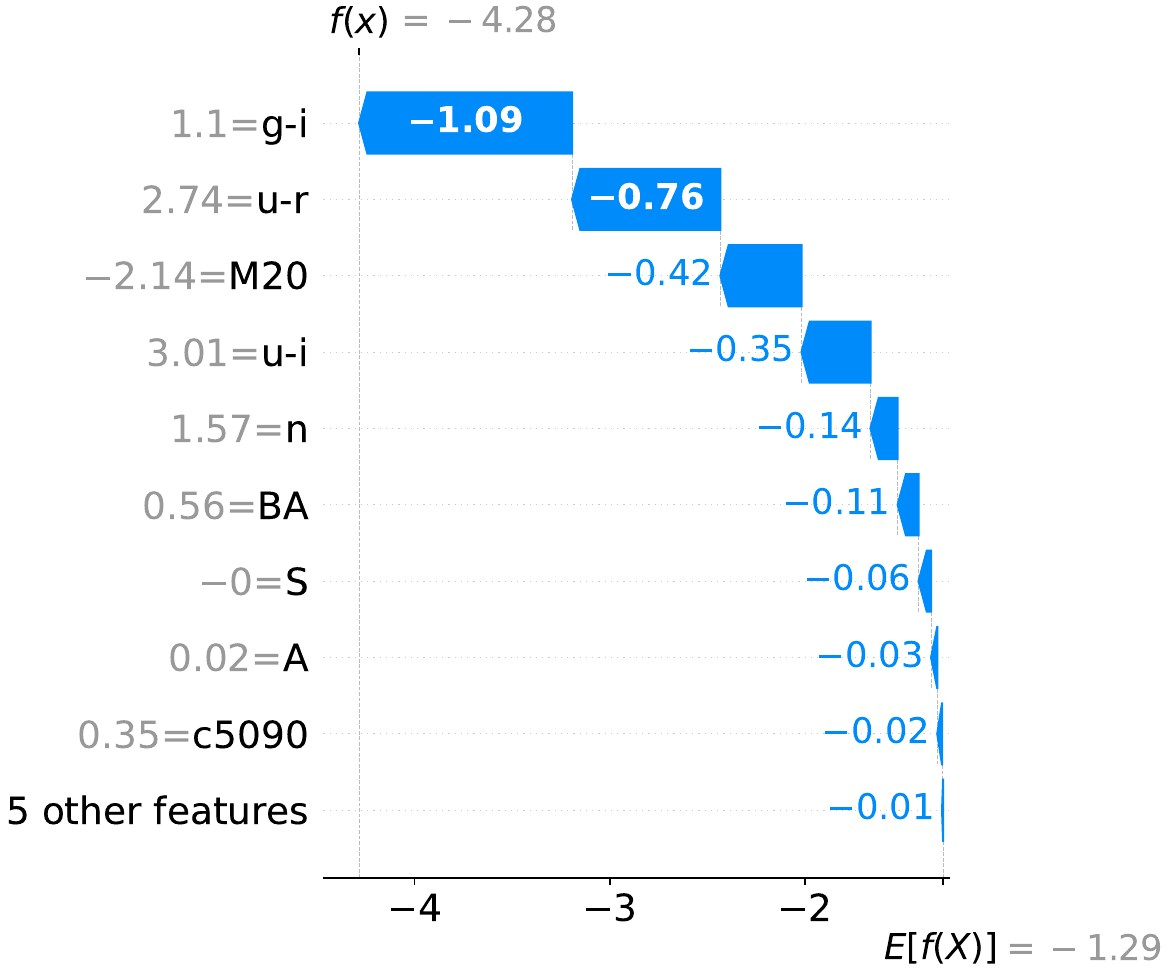}
     \includegraphics[width=0.31\textwidth, height=0.215\textwidth]{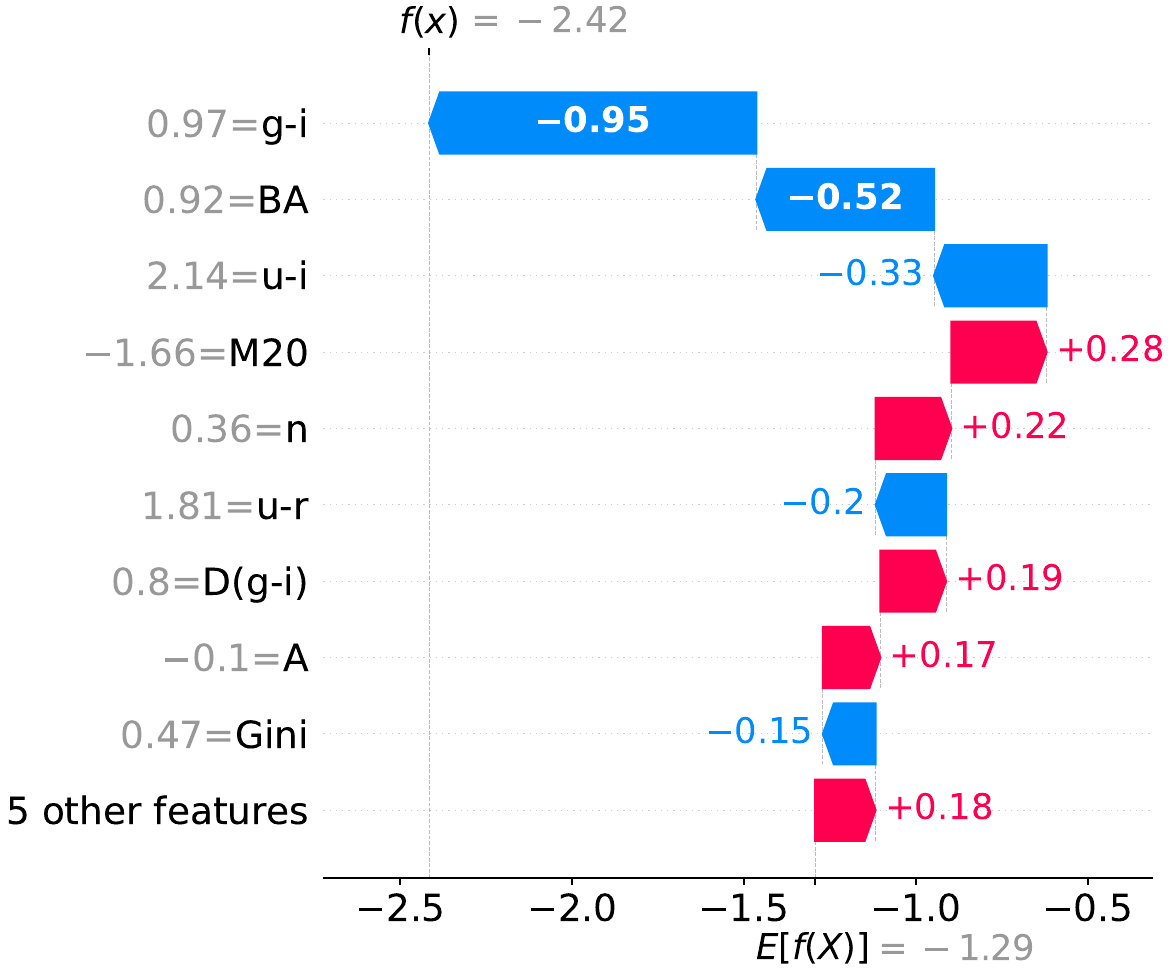}
\caption{SHAP waterfall plots depicting three scenarios: a correct prediction with high confidence (left), an incorrect prediction with high confidence (middle), and an error prediction three classes away (right). In all cases, the catalogued Class is 0. The rows show, from top to bottom, the galaxy image and the plots for Classes 0 to 4. The plots highlighted in red represent the model's predicted classes. See the main text for discussion.}
\label{fig:waterfall}
\end{figure*}

Figure~\ref{fig:waterfall} presents the waterfall plots of three different cases: i) a case where the model correctly predicts the galaxy class with high confidence (left column), ii) a case where the model makes an incorrect prediction (with high probability), misclassifying the galaxy by one class (middle column), and iii) an extreme (and rare) case where the model misclassifies the galaxy by three classes away from the expected class (right column). These cases correspond to the 5cats direct classification task using the S2+C parameter configuration. The first row shows the galaxy images indicating their ``real'' class (as given in the catalogue), the XGBoost predicted class, and the prediction probability. Each column displays the waterfall plots for each of the five classes, from Class 0 (top) to Class 4 (bottom). Notice that the ``real'' (catalogued) class in the three examples corresponds to Class 0 (elliptical), and that the red frames (boxes in red colour) indicate the waterfall plot of the model's predicted class.

In these plots, the $x$-axis represents the contribution of each input parameter to the difference between the model's output, $f(x)$ (vertical line), and the baseline prediction, $E[f(x)]$ (mean predicted probabilities of the corresponding class; e.g., \citealt{Lundberg2020}). The $y$-axis lists the input parameters and their respective values (in grey) for the specific instance, ordered by the magnitude of their SHAP values in descending order. The bars in the plots indicate the SHAP values of each feature, where the length corresponds to the magnitude of the feature's contribution to the model's output, and their direction and colour indicates whether its contribution increases (red) or decreases (blue) the model's output compared to the baseline prediction.

The left column in Fig.~\ref{fig:waterfall} presents an example where the XGBoost model correctly predicts the catalogued class (Class 0). The red box (second row) shows that all input parameter values for this galaxy increase the probability of it being classified as Class 0. This indicates that all feature values for this galaxy are typical of Class 0 (elliptical galaxies), with $BA=0.90$ and $c_{5090}=0.31$ contributing the most. For this class, $E[f(x)]=0.37$ and $f(x)=3.75$, resulting in a net contribution of $+3.38$ from all parameters. The subsequent plots in the first column (rows 3 to 6) display the Classes 1 to 4. It is noticeable how the net contribution of the galaxy parameters decreases the probability of this galaxy being classified by the model as any of those classes.

The middle column of Figure~\ref{fig:waterfall} illustrates an example where the catalogued class of the galaxy is Class 0, but the XGBoost model predicts it as Class 1 (S0$^-$--S0a). In this case, the waterfall plot for Class 0 (middle column, second row) indicates that the net contribution of the galaxy parameters results in a low probability of this galaxy being Class 0, with $BA=0.56$ and $c_{5090}=0.35$ having a more negative impact on the model's prediction. In contrast, the waterfall plot for Class 1 (red box, third row) shows that all input parameters augment the probability of this galaxy being Class 1, this time with $BA=0.56$ and $c_{5090}=0.35$ having the most positive impact, followed by the S\'ersic index and a colour parameter. In this case, $E[f(x)]=0.75$ and $f(x)=2.13$, suggesting that the galaxy's parameter values align more with Class 1. After closer inspection, we agree with the model's prediction that this galaxy belongs to Class 1, thus allowing us to identify a case of visual misclassification. However, even though the model classified this galaxy correctly as Class 1, it was considered a prediction error because, for the evaluation of the model, we assume that the provided catalogued classification is the ground truth.

We also present an example where the catalogued class is Class 0, but the XGBoost model predicts it as Class 3 (right column of Fig.~\ref{fig:waterfall}). The waterfall plot for Class 0 (right column, second row) reveals that the net contribution of all galaxy parameters diminishes the probability of this galaxy being Class 0, with $c_{5090}=0.56$ having the most negative impact. In contrast, the waterfall plot for Class 3 (right column, fifth row) shows that $\Delta\left( g-i \right)=0.80$, $S=-0.26$, $g-i=0.97$, and $A=-0.10$ decrease the probability of being a Class 3 (Sbc--Scd) galaxy. Even so, $u-r=1.81$, $c_{5090}=0.56$, $n=0.36$, $u-i=2.14$, and $\Delta A_{9050}=0.22$ greatly increase the probability in favour of Class 3. The contribution of the remaining parameters is lower. Consequently, the model misclassified this galaxy as Class 3.

Although this galaxy has a visual appearance of an elliptical (Class 0), a more careful inspection to the parameter values show that the light distribution, as indicated by $c_{5090}=0.56$ and $n=0.36$, is not consistent with an early-type galaxy. Instead, it shows a flatter light distribution, resembling fainter galaxies where the structural properties and the colour properties depart from the average values of more massive galaxies. For instance, the colour parameter $u-r=1.81$ suggests a bluer colour, which is atypical for early-type galaxies (Classes 0 and 1) and thus consistent with a discordant-morphology galaxy. This is a rare case. As shown in the confusion matrix (bottom panel of Fig.~\ref{fig:CM_2p_14p}), there are very few such instances, with around four for the catalogued Class 0, and one each for the catalogued Classes 1 and 3 in the test subset.

\subsection{Identification of Possible Error Sources and their Impact on the Results}
\label{sec:errorSource}
An evaluation of our trained models achieved overall performance metrics (accuracy, precision, recall, and F1-score) of up to 60-65\% for a five-group classification, with the best results obtained when using a combination of both structural and star formation parameters (see Table~\ref{tab:metrics}). To better understand these performance levels, we carried out a model interpretation study to elucidate the role of different parameters on the model's classification and examined the model's prediction for specific galaxies (see Section~\ref{sec:model_interp}). From this analysis, we have identified the following possible error sources that could be affecting the model's performance:

\begin{itemize}
    \item inaccuracies in the visual classification of galaxies,
    \item the presence of galaxies with discordant morphology, and
    \item geometric and projection effects influencing the colours of galaxies.
\end{itemize}

\begin{figure}
     \centering
     \begin{tabular}{cc}
        \includegraphics[width=0.7\columnwidth,angle=0]{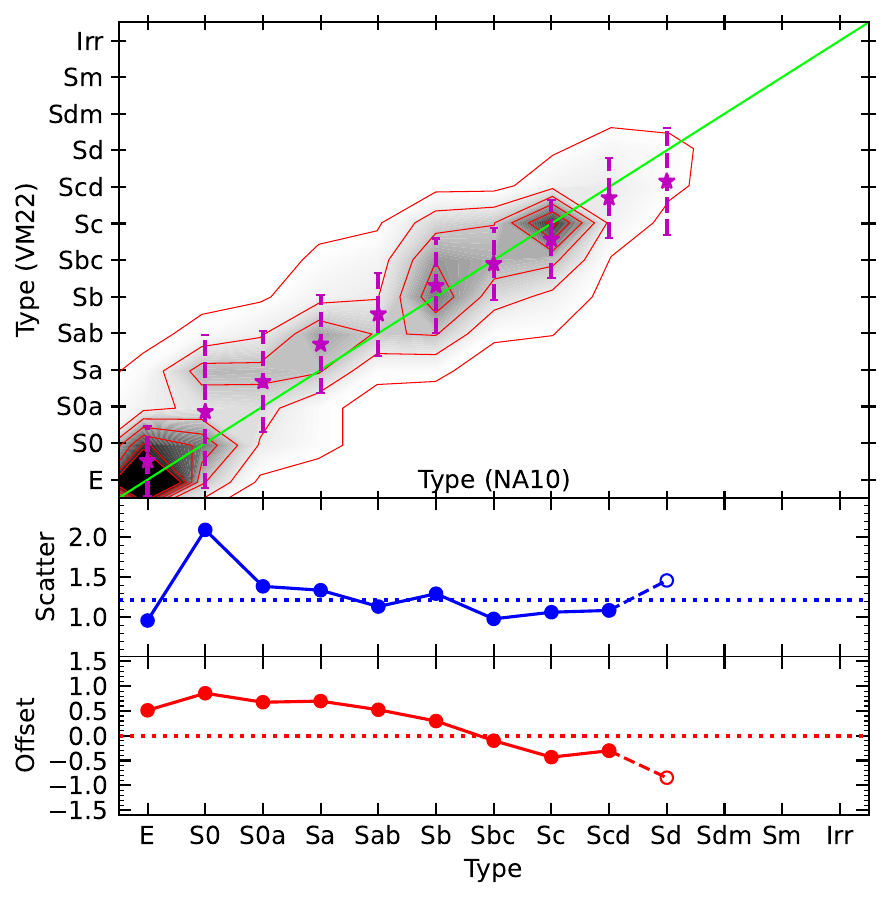}
     \end{tabular}
\caption{Comparison of morphological classifications from the \citetalias{VazquezMata2022} and \citetalias{Nair2010} samples for the galaxy subset in common ($\sim$3,000 galaxies). Both samples are consistent within $\pm$ 1.3 T-type, with S0 having the larger scatter.}
\label{fig:comparison}
\end{figure}

A first important error source impacting the performance of our ML models is the accuracy of the visual classification. The galaxy sample used in this work (see Section~\ref{sec:galaxy_sample}) is the result of combining two independent visual classifications of galaxies in the local universe: the MaNGA (\citetalias{VazquezMata2022}) sample and the \citetalias{Nair2010} sample. As a measure of the morphological variability in our sample introduced by different classifiers, Figure~\ref{fig:comparison} compares the morphological classifications reported for the subset of galaxies common to both samples. The green line in the upper panel shows the one-to-one correspondence between the classifications. The 2D density isocontours are displayed in linear scale, with the outermost contour enclosing at least 10 galaxies in the bin, while the error bars indicate the standard deviation around the average for each morphological type in the \citetalias{Nair2010} classification. The lower panels provide a measure of the scatter and offset between individual classifications for a given morphological type. The blue-dotted line indicates the median, and the red-dotted line is placed at zero as reference.

The observed median scatter indicates that the consistency of the classification between the \citetalias{Nair2010} and MaNGA samples lies within $\pm$ 1.3 T-Type, with a larger scatter in the S0 morphological type. Therefore, our merged sample inherits this $\pm$ 1.3 T-Type scatter in the visual morphological classification, hence impacting at that level the model predictions. Since this level of scatter is consistent with findings in other studies comparing different classifiers (e.g., \citealt{Naim1995}), we do not attempt to make any correction for this source of inaccuracy. It should be noted that this scatter is considered intrinsic to our sample, and all the prediction scores reported in our experiments always include that error source.

%------------------------Figure----------------------------------%

\begin{figure}
     \centering
     \begin{tabular}{cc}
        \includegraphics[width=0.97\columnwidth,angle=0]{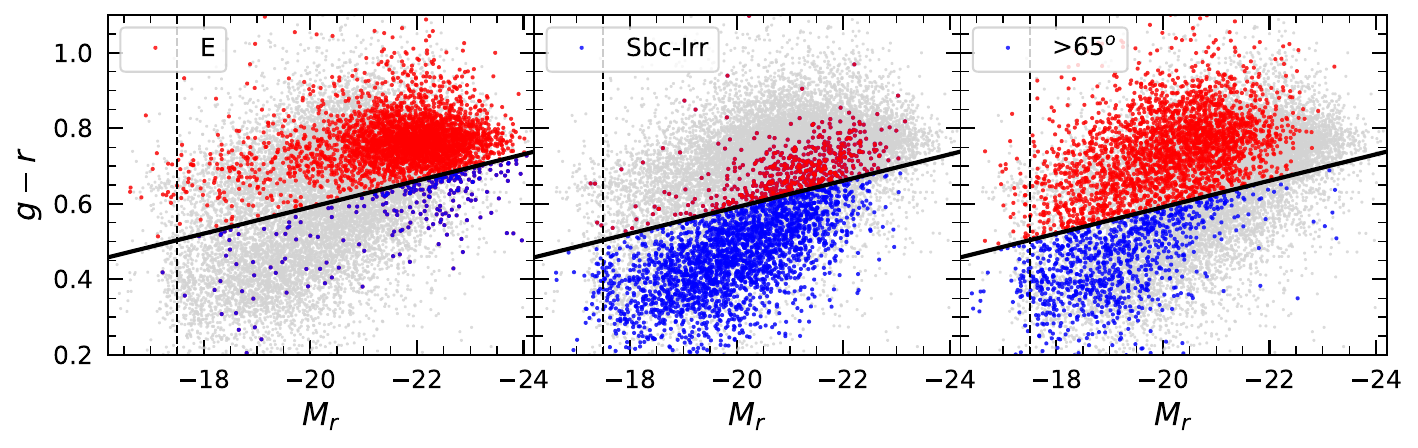}
     \end{tabular}
\caption{Colour--magnitude diagrams illustrating morphologically discordant E--S0a early-types (left; blue-dots), Sbc--Irr late-types (middle; red dots), and Sa--Irr galaxies with inclination values Inc $>$ 65 degrees (right; red-dots).}
\label{fig:color_magnitude}
\end{figure}

Another error source arises from the presence of galaxies within our sample that deviate from the typical relationship between colour and morphology, which defines the red and blue sequences in the colour--magnitude diagram (see Fig.~\ref{fig:color_magnitude}). A careful inspection of the structural and colour distributions within each morphological type revealed a fraction of late-type discs unexpectedly showing red colours \citep[][]{Yamauchi2004, Ishigaki2007}, as well as early-type galaxies exhibiting unusual blue colours \citep[][]{Ferreras2005, Lee2006}. These morphologically discordant galaxies could be contributing to the model's classification errors, particularly when colour is a relevant parameter for the model.

The Galaxy Zoo \citep[GZ;][]{Lintott2008,Lintott2011} project reported that about 6\% of the low-redshift early-type galaxy population is blue \citep[][]{Schawinski2009}, with the fraction of blue ellipticals increasing to about 12\% towards lower-mass galaxies situated in lower-density regions \citep[][]{Bamford2009,Park2007,Park2009}. The GZ project also revealed a significant population of red spirals \citep[][]{Bamford2009,Skibba2009}, suggesting that around 20\% of spiral galaxies lie on the red sequence, with an increasing fraction in intermediate-density environments.

Furthermore, late-type spirals may appear in the red sequence due to effects unrelated to stellar population, such as inclination-dependent dust reddening \citep[][]{Maller2009,Masters2010}, which may also be impacting the model predictions when the colour information is relevant for the classification.

Figure~\ref{fig:color_magnitude} shows the loci of discordant blue ellipticals in our sample, identified as those with $g-r$ colours below the division line (blue dots) between the red and blue regions as described in Appendix~\ref{app:col-mag}. Similarly, the middle panel illustrates the loci of discordant red late-type spirals (Sbc--Irr) with inclinations $<$ 65$^{\circ}$ (above the black line), while the right panel displays the loci of S0a--Irr galaxies reddened by inclination effects (inclinations $>$ 65$^{\circ}$, above the division line). As a reference, the vertical line at $M_r$=-17.5 indicates the transition into the sub-dwarf galaxy regime, which could be another source of error. However, as the number of galaxies in this regime is very low, we did not attempt to make any correction.

From this analysis we identified 238 blue elliptical galaxies, 532 red late-type galaxies (Sbc--Irr), and 2,605 reddened spiral galaxies due to strong inclination effects (inc $>$ 65$^{\circ}$), amounting to 3,375 galaxies or $\sim$19\% of the original data sample. To better understand the impact of these discordant and highly inclined galaxies on the model performance, we excluded them from our sample and re-trained the 5cats direct model with the S2+C parameter configuration. This exclusion lead to a 1-2\% improvement in overall performance metrics, including accuracy, precision, recall, AUC-ROC, and F1-score. 

\begin{figure}
     \centering
     \begin{tabular}{cc}
        \includegraphics[width=0.72\columnwidth,angle=0]{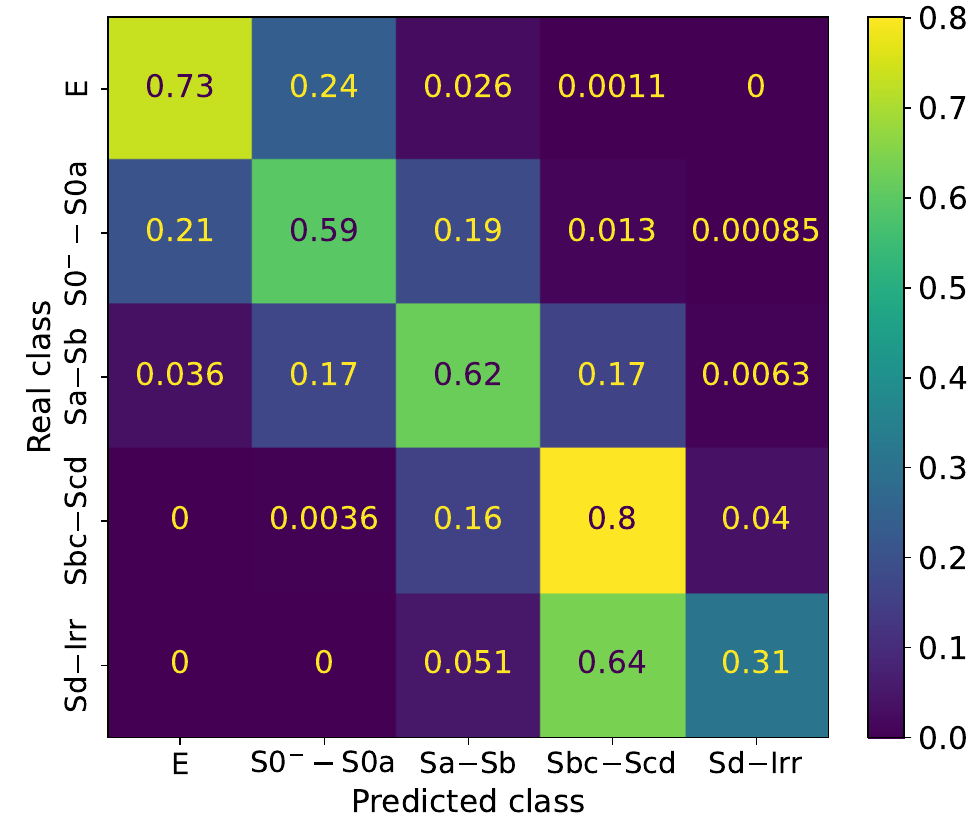}
     \end{tabular}
\caption{Confusion matrix for the experiment excluding discordant and highly inclined galaxies, calculated with the test subset. Colours are according to the accuracy of the classification. There is an improvement for Classes 1 and 3.}
\label{fig:CM_discordant}
\end{figure}

For per-class performance metrics (see Table~\ref{tab:discordant_class_report} in Appendix~\ref{app:metrics_without_discordant}), we found some trends of improvement. Specifically, Class 0 maintains its performance, Class 1 improves by 2--6\%, Class 2 has a 5\% gain in precision, Class 3 improves up to 7\%, and Class 4 has an enhanced precision of 4\%. The CM (see Fig.~\ref{fig:CM_discordant}) also shows improvements, with a 6\% and 7\% increase in the success prediction rate for Classes 1 and 3, respectively. Additionally, although misclassification rates remain generally similar across classes, there is a noticeable 5\% reduction in Class 1 being misclassified as Class 2, and a 7\% reduction in Class 3 being misclassified as Class 2.

Figures \ref{fig:CM_2p_14p}, \ref{fig:CM_hier}, and \ref{fig:CM_discordant} present the CMs summarizing the results of the different tests carried out in this work. For elliptical galaxies (Class 0), the trained models show a reasonable good success prediction rate of 73\%. However, the adjacent off-diagonal values consistently remain at 22-24\%, suggesting that a proportion of lenticular galaxies (S0$^-$, S0) are overlapping their structural properties. \citet{Cheng2011} proposed to combine the concentration ($C_{9050}$), the geometric axial ratio ($b/a$), and the bulge-to-total light fraction ($B/T$) as an optimum parameter space to segregate early-type (E, S0, and Sa) galaxies, achieving a completeness (recall) and reliability (precision) consistent with our results. More recently, \citetalias{VazquezMata2022} applied the classification scheme proposed by \citet{Cheng2011} to the MaNGA sample, further suggesting an alternative segregation region in the $C - B/T - b/a$ parameter space, consistent with the results of the SHAP analysis in this work and with other detailed studies using 2D image decomposition analysis of lenticular galaxies \citep[][]{Laurikainen2011}. 

For the combination of lenticular (S0$^-$, S0) and transition S0a galaxies (Class 1), the true prediction rate after various experiments accounts for 53\%, increasing to $\sim$60\% after excluding discordant and highly inclined galaxies from the training and test samples. This performance is explained by the structural similarities that a fraction of S0 galaxies share with Class 0 (left off-diagonal value), but also to similarities that another fraction of S0 share with early-type spirals (right off-diagonal rate) and intermediate- or late-type spirals. S0 galaxies exhibit not only a disk-like structure similar to spirals but also a wide variety of bulge-to-disk ($B/D$) ratios, comparable to those seen along the spiral sequence \citep[c.f.][]{Laurikainen2011}. It is worth mentioning that our ML-based conclusions on the diverse nature of S0 galaxies may fit, in a broader sense, with the findings of \citet{Graham2023,Graham2024}. They argue that S0 galaxies can be divided into two distinct types: low-mass, primordial S0s, and high-mass, dust-rich S0s formed through major wet mergers. This differentiation is based on the relationship between spheroids and their central massive black holes.

Figure~\ref{fig:comparison} shows that the scatter in the visual classification for these galaxies is $\pm$ 2 T-Types, the highest among all types, reflecting the difficulty of capturing the wide variety of structural and colour properties through a visual classification, which impacts the success rates of the models.

The galaxies of types Sa, Sab, and Sb (Class 2) have a success rate roughly constant at a level of 64\% across all our experiments. This can be attributed to the structural similarities shared with Class 1 (left off-diagonal element) as well as the overlap in structural and colour properties with galaxies of types Sbc, Sc, and Scd (Class 3; right off-diagonal element). For Class 3 galaxies (Sbc--Scd), the correct prediction rate increases from 73\% to 80\% after removing discordant and highly inclined galaxies from our sample. Notably, the misclassification rate in the left off-diagonal element also decreases, indicating that the colour properties of Classes 2 and 3 can be better separated when discordant galaxies and inclination effects are properly corrected. 

For the under-represented Class 4 galaxies (Sd--Irr), the true prediction rate ($\sim$32\%) is the lowest among all classes and across the different experiments. This class also presents the highest misclassification rate ($\sim$60\%) in the left off-diagonal element, suggesting the trained models lack enough representative examples to accurately characterize the morphological features of these galaxies. It is thus important to further work and improve the under-representation of galaxies in this class. However, the methods to address such imbalance in the galaxy sample are out of the scope of the present paper.

Finally, given that the model that excludes discordant and highly inclined galaxies yields improved performance across the evaluated metrics compared to the original models (with all galaxies; see Section~\ref{sec:results}), we recommend applying this model only when discordant galaxies can be omitted from the dataset.

\subsection{Comparison to other works} \label{sec:comparison}
To provide context for the performance of our trained models, we compare our results with other studies. Although they are not directly comparable due to variations in dataset characteristics and/or classification tasks, examining the trends and insights from various works can offer valuable perspectives on the efficacy of different approaches.

Table~\ref{tab:comparison_works} presents this comparison, in different columns (from left to right) including the authors of different works from the literature, datasets used, input data (galaxy features used for training), the number of adopted classes, the ML methods employed, and their corresponding classification accuracies. Note that we focus our comparison with works that employed galaxy parameters as input data, since they are more closely related to our methodology. 

%------------------------Table----------------------------------%
\begin{table*}
    \caption{Accuracy comparison with other works. Note the difference in datasets employed and the classification task across the studies.}
    \begin{threeparttable}[b]
        \centering
        \label{tab:comparison_works}
        \begin{tabular}{ >{\centering\arraybackslash}m{2.4cm} >{\centering\arraybackslash}m{3cm} >{\centering\arraybackslash}m{3.1cm} >{\centering\arraybackslash}m{3cm} >{\centering\arraybackslash}m{2cm} >{\centering\arraybackslash}m{1.9cm} }
            \hline
            Author & Dataset & Input data & Classification & ML method & Top Accuracy (\%) \\
            \hline
            \multirow{4}{2.4cm}{Present work\centering} & \multirow{4}{3cm}{NA10+VM22 (17,966 galaxies)\centering} & \multirow{4}{3cm}{S2+C parameters\centering} & 2cats & \multirow{4}{2cm}{XGBoost\centering} & 88.0 \\
             &  &  & 5cats &  & 65.0 \\
             &  &  & Hier1 &  & 65.0 \\
             &  &  & Hier2 &  & 64.0 \\
            \hline
            \multirow{2}{2.4cm}{\citet{Vavilova2021}\centering} & SDSS-DR9 (6,163 galaxies) & \multirow{2}{3cm}{$[u, g, r, i, z, u-r, g-i, r-z, c_{5090}]$\centering} & \multirow{2}{3cm}{ET, LT\centering} & \multirow{2}{2cm}{SVM\centering} & 96.4 \\
            & GZ2 ($\sim$8,500 galaxies) &  &  &  & 76.0 \\
            \hline
            \multirow{9}{2.4cm}{\citet{Barchi2020}\centering} & GZ1 (58,030 galaxies) & \multirow{9}{3cm}{$[C_{7535}, A_3, S_3, H, \mathrm{GPA}]$\centering} & \multirow{2}{3cm}{E, Sp\centering} & \multirow{9}{2cm}{MLP\centering} & 98.6 \\
             & NA10 (14,034 galaxies) &  &  &  & 87.0 \\
             & \multirow{7}{3cm}{GZ2 (67,637 galaxies)\centering} &  & E, Sp, SB &  & 78.8 \\
             &  &  & E, Sa, Sb, Sc, SBa, SBb, SBc &  & 66.0 \\
             &  &  & E, Sa, Sb, Sc, Sd, SBa, SBb, SBc, SBd &  & 66.2 \\
             &  &  & Er, Ei, Ec, Sa, Sb, Sc, Sd, SBa, SBb, SBc, SBd &  & 57.7 \\
            \hline
            \multirow{3}{2.4cm}{\citet{deDiego2020}\centering} & OTELO catalogue (1,834 galaxies) & \multirow{3}{3cm}{$[u-r,g-r,r-i,r-z,r-J,r-H_b,r-K_s,r,n]$\centering} & \multirow{3}{3cm}{ET, LT\centering} & \multirow{3}{2cm}{MLP\centering} & 98.5 \\
             & COSMOS (34,688 galaxies) &  &  &  & 96.7 \\
            \hline
            \multirow{6}{2.4cm}{\citet{Ferrari2015}\centering} & EFIGI (4,458 galaxies) & \multirow{6}{3cm}{$[C, A_3, S_3, H, \sigma_\psi]$\centering} & \multirow{6}{3cm}{E, Sp\centering} & \multirow{6}{2cm}{LDA\centering} & 93.8 \\
             & NA10 (14,034 galaxies) &  &  &  & 90.2 \\
             & LEGACY (804,974 galaxies) &  &  &  & 87.7 \\
             & LEGACY-$zr$ (337,097 galaxies) &  &  &  & 93.8 \\
            \hline
        \end{tabular}
    \end{threeparttable}
\end{table*}

From Table~\ref{tab:comparison_works}, we see that most of the works studied binary classification tasks, such as distinguishing between elliptical (E) vs. spiral (Sp) galaxies \citep[][]{Barchi2020,Ferrari2015} or early-type (ET) vs. late-type (LT) galaxies \citep[][]{Vavilova2021,deDiego2020}. Our 2cats classification task closely aligns with the ET vs. LT classification, where we attained an accuracy of 88.0\% using the S2+C parameter configuration. In comparison, \citet{deDiego2020} reached accuracies of 96.7\% and 98.5\% for the COSMOS \citep[][]{Scoville2007} and OTELO \citep[][]{Bongiovanni2019} datasets, respectively, utilizing photometric parameters (optical and near-infrared photometry) and Sérsic index ($n$) as input data with a Multi-Layer Perceptron \citep[MLP;][]{Rosenblatt1958,Fukushima1975,Fukushima1983}. Similarly, \citet{Vavilova2021} used photometric data and a Support Vector Machine \citep[SVM;][]{Cortes1995,Vapnik1995,Hearst1998} for ET vs. LT classification, achieving an accuracy of 96.4\% in a sample of 6,163 galaxies from SDSS-DR9 visually classified by them. However, the accuracy drops to 76.0\% when their model is applied to a sample of $\sim$8,500 galaxies from the Galaxy Zoo 2 \citep[GZ2;][]{Willett2013} catalogue.

The differences in performance between these works and the present study can be attributed in part to the dataset characteristics, such as data quality and distribution of the classes. For example, \citet{Vavilova2021} attributed the performance difference between the SDSS-DR9 and GZ2 samples to inconsistencies in human labeling, as well as the GZ2 sample containing more discordant galaxies than their own visually classified sample. Additionally, the method for galaxy classification before training differs across the works. In \citet{deDiego2020}, galaxies were classified based on the best fit between observed fluxes and spectral index distribution (SED) templates. By contrast, both \citet{Vavilova2021} and our dataset (\citetalias{Nair2010}+\citetalias{VazquezMata2022}) relied on human visual classification, which is influenced by human biases but can capture subtle morphological features that SED-based classification might miss.

Furthermore, the dataset employed by \citet{deDiego2020} was highly imbalanced, with only 5.4\% ET and 94.6\% LT galaxies, whereas our dataset has a more balanced distribution of 42\% ET and 58\% LT galaxies (see Table~\ref{tab:recat}). Class imbalance can affect model performance toward the dominant class, potentially increasing the accuracy. Nevertheless, from Table 5 of \citet{deDiego2020}, they achieved an F1-score (macro average) of 93\%, reflecting good performance across both classes. In comparison, we achieved an F1-score of 87\%, which still remains consistent.

Beyond dataset characteristics, differences in input parameters also influence model performance. While \citet{deDiego2020} employed both optical and near-infrared photometry, \citet{Vavilova2021} used optical photometry, and we combined structural parameters with optical photometric data. The inclusion of structural features in our work provides additional morphological information that purely photometric models may miss. On the other hand, the near-infrared photometry can capture a wider range of galaxy properties than optical photometry alone. However, given the differences on dataset characteristics across these works, a direct comparison of the performance of each input parameter configuration is not feasible.

The studies conducted by \citet{Barchi2020} and \citet{Ferrari2015} reported accuracies ranging from 87.7\% to 98.6\% for an E vs. Sp classification task. Specifically, \citet{Barchi2020} used the Galaxy Zoo 1 \citep[GZ1;][]{Lintott2008,Lintott2011} catalogue and a five-parameter configuration, incorporating features such as concentration ($C_{7535}$=$\log(R_{75}/R_{35})$), asymmetry ($A_3=1-s(I, I_{180})$, with $s()$ being the Spearman’s rank correlation coefficient; \citealt{Press2005}), clumpiness ($S_3=1-s(I, I_{S})$), entropy \citep[$H$;][]{Bishop2006}, and Gradient Pattern Analysis \citep[GPA;][]{Rosa2018}, alongside an MLP. They also evaluated their model using the \citetalias{Nair2010} dataset. \citet{Ferrari2015} employed parameters like $C$, $A_3$, $S_3$, $H$, and spirality \citep[$\sigma_\Psi$;][]{Shamir2011}, using a Linear Discriminant Analysis \citep[LDA;][]{Murtagh1987,Duda2000,Bishop2006} to assess their model across four catalogues: EFIGI \citep[][]{Baillard2011}, \citetalias{Nair2010}, SDSS-DR7 \citep[][]{Abazajian2009} complete LEGACY database, and a volume-limited subsample (LEGACY-zr). 

A key difference between these works and our study is the classification task. While both \citet{Barchi2020} and \citet{Ferrari2015} focused on distinguishing between elliptical and spiral galaxies, our task differentiates between ET (elliptical+lenticular) and LT (spiral) galaxies. Additionally, the GZ1 sample used by \citet{Barchi2020} is class-imbalanced, with 13\% E and 87\% Sp, which, as discussed earlier, can affect model performance. Although dataset characteristics also play a role, the \citetalias{Nair2010} dataset is more closely related to our work. When comparing our 2cats results (88\% accuracy) with \citet{Barchi2020} and \citet{Ferrari2015} using the \citetalias{Nair2010} catalogue (87.0\% and 90.2\% accuracy, respectively), they are on a par, though it's worth noting that their studies excluded lenticular galaxies. 

Table~\ref{tab:comparison_works} further illustrates that only two studies, \citet{Barchi2020} and the present work, have undertaken classification tasks involving more than two classes. As the number of classes increases, the accuracy tends to decline, which is unsurprising given the increased complexity of classification tasks, even for visual (human) classification. For instance, \citet{Barchi2020}, employing an MLP and galaxy parameters, reported an accuracy of 98.6\% for E vs. Sp classification, which dropped to 78.8\% for their three-class classification involving the classes E, Sp, and SB (barred galaxies). Furthermore, they obtained accuracies of 66.0\%, 66.2\%, and 57.7\% for seven-, nine-, and eleven-class tasks, respectively. In comparison, we achieved 65.0\% accuracy for our 5cats direct classification using the S2+C parameter configuration and XGBoost. Although our accuracy is lower than the seven- and nine-class tasks of \citet{Barchi2020}, it is important to highlight that the classification tasks are different. Our classification takes into account lenticular and irregular galaxies, whereas \citet{Barchi2020} incorporate barred galaxies. Additionally, the datasets differ, with \citet{Barchi2020} using GZ2 and our work utilizing \citetalias{Nair2010}.

Finally, it is important to highlight that variations in accuracy across the listed works may arise from differences in datasets, input features, classification tasks, and the ML methods used. For example, \citet{deDiego2020} observed a 1.4\% difference in accuracy when changing datasets, while \citet{Vavilova2021} found a $\sim$20\% variation, and \citet{Ferrari2015} obtained a 6.1\% variance across datasets. In our study, we found accuracy differences ranging from 0.6\% to 3.5\% for the 2cats classification and from 3.0\% to 9.9\% for the 5cats direct classification, depending on the parameter configuration used for training (see Table~\ref{tab:metrics}). In addition, \citet{Vavilova2021} evaluated five different ML methods for binary galaxy classification (ET vs. LT), with accuracy variations ranging from 0.9\% to 7.0\% across models. \citet{Barchi2020} also explored three traditional ML methods and found $\sim$0.2\% accuracy differences across methods for the E vs. Sp classification, with this difference reaching up to 3.0\% for finer classification tasks. Therefore, our findings are overall consistent with other studies. 

Beyond this consistency, our work offers two key contributions. First, we provide a detailed evaluation of various combinations of structural and star formation parameters, offering a more nuanced understanding of their roles in automated classification. Second, through the application of interpretation tools, we gained valuable insights into how these parameters influence model predictions, revealing patterns that correspond to physically expected behaviours of galaxies. These contributions not only reinforce existing research but also offer new perspectives on the interplay between ML models and astrophysical parameters in galaxy classification.

\section{Conclusions} \label{sec:conclusion}
In the present paper, we have used a sample of local galaxies assembled by merging two independent samples, the \citetalias{Nair2010} and \citetalias{VazquezMata2022} catalogues, both of which include detailed visual morphological classifications. We also incorporated a comprehensive set of structural and star formation parameters, uniformly estimated from SDSS images, to train XGBoost models for automatic morphological classification. Specifically, the structural parameters used included $C,\,A,\,S,\,A_S,\,\mathrm{Gini},\,M_{20},\,n,\,c_{5090},\,BA$, and $\Delta A_{9050}$, while the star formation parameters included $g-i,\,u-r,$ $u-i$, and $\Delta (g - i)$. 

We conducted a series of experiments with the goal of analyzing classification performance. These experiments explored different combinations of the structural and star formation parameters (as detailed in Table~\ref{tab:configs}) and diverse morphological type groupings (see Table~\ref{tab:recat}). Additionally, we investigated two classification approaches: 1) a direct approach, where a single model was trained to directly classify galaxies into the specified classes, and 2) a hierarchical approach, which decomposed the five-class task into three steps: first, distinguishing early- from late-type galaxies; second, further classifying early-type galaxies; and third, sub-classifying only late-type galaxies. We evaluated the performance of both direct and hierarchical models in galaxy classification using a range of standard metrics, including the confusion matrix, accuracy, precision, recall, F1-score, and AUC-ROC.

One of the main contributions of this paper is the comprehensive evaluation of different combinations of structural and star formation parameters for automated galaxy classification. This analysis provides a deeper understanding of how these parameters affect model performance, allowing us to identify the configurations that yield the best classification results across multiple metrics. Another important contribution is the use of advanced interpretation tools, such as SHAP, to analyze the impact of galaxy parameters on model predictions. By linking the model’s decision-making process to the physical properties of galaxies, we gained valuable insights into how these features influence the classification results. This analysis not only highlighted the relative importance of different parameters but also allowed us to connect the model’s outputs with the physically expected behaviour of galaxies. Through this approach, we identified the key features driving the model’s decisions across different galaxy classes, revealing how they align with the distinct physical characteristics of each morphological type.

In the following, we summarize our findings:

\begin{itemize}
    \item Among all the parameter configurations tested, we found that the S2+C configuration (using all fourteen parameters) led to better performance. Specifically, for the 2cats classification, S2+C improved performance by up to 3.5\%, 2.3\%, 2.2\%, and 4.1\% for accuracy, precision, recall, and AUC-ROC, respectively. Similarly, in the 5cats direct classification, the S2+C configuration yielded improvements ranging from 3.0-9.9\%, 3.6-9.2\%, 6.0-10.7\%, and 3.0-6.1\% for accuracy, precision, recall, and AUC-ROC metrics, respectively.

    \item The 2cats classification task achieved an accuracy of 88\%, with precision, recall, and F1-score all at 88\%, and an AUC-ROC at 95\%. This indicates the model's ability to distinguish between early-type and late-type galaxies. In contrast, the more challenging 5cats direct classification resulted in an accuracy of 65\%, with precision, recall, and F1-score at 64-65\%, and an AUC-ROC of 90\%. The drop in performance is expected due to the increased granularity and overlap between certain morphological types, which presents a more challenging task for the classifier.
    
    \item The experiments with hierarchical classification for the five-class task yielded a performance comparable to the 5cats direct classification, using the S2+C parameter configuration. Performance differences across metrics were minimal, with variations of less than 3.0\%. Given these similar outcomes, both approaches can be equally effective for the five-class galaxy classification. However, the hierarchical approach is more complex to implement and evaluate, and it also leads to a higher computational cost.

    \item Our model interpretation analysis, using SHAP for the 5cats direct classification with the S2+C parameter configuration, revealed that, overall, the $u-r$ colour parameter emerged as the most significant contributor to the classification task, followed by $BA$, $c_{5090}$, $g-i$, and $\Delta \left( g-i \right)$. Furthermore, on a per-class basis, we found that the structural parameters $BA$, $c_{5090}$, and $n$ had more impact on the model's prediction for Classes 0 (E), 1 (S0$^-$--S0a), and 2 (Sa--Sb). In contrast, the photometric parameters $u-r$, $g-i$, and $\Delta \left( g-i \right)$ were more relevant for Classes 3 (Sbc--Scd) and 4 (Sd--Irr).

    \item After the model interpretation analysis of individual galaxies, we identified possible error sources that could have an impact on the automated classification performance: 1) inaccuracies in the visual classification on the dataset, 2) discordant-morphology galaxies (i.e. blue early-type or red late-type galaxies), and 3) dust reddening due to inclination effects. Although we found consistency in visual classifications between the \citetalias{Nair2010} and \citetalias{VazquezMata2022} catalogues within $\pm$1.3 T-Type, we did not attempt to make any correction for this source of inaccuracy and considered it intrinsic to our sample. After excluding $\sim$19\% of the sample to remove discordant and highly inclined galaxies, we observed a 1-2\% improvement in overall performance metrics and significant gains in true prediction rates for Classes 1 and 3, with increases of 6\% and 7\%, respectively. These findings suggest that addressing intrinsic dataset issues can enhance model performance.

    \item Finally, upon comparing our findings with those of other studies, we observed that our results exhibit consistency with existing research. However, there are still areas of improvement that can be addressed in future work. For example, tackling the class imbalance in the 5cats task by applying techniques like class-weight adjustments, oversampling, or undersampling, among others, could lead to better performance, especially for the highly under-represented Class 4 (Sd--Irr). Moreover, the error sources we identified (see Section~\ref{sec:errorSource}) indicate the importance of refining the input data and addressing these issues, as the quality of the dataset and accuracy of the labels are crucial for model training. By mitigating these errors, we can improve classification performance. Another possible area for improvement is the incorporation of additional parameters that capture not only global galaxy properties but also more localized features (e.g., bulge-to-total light fraction) or spatially-resolved properties within galaxies, like those inferred from integral-field spectroscopic surveys (c.f. MaNGA, \citealt{Bundy2015}; \citealt{CanoDiaz2016,CanoDiaz2019}; SAMI, \citealt{Croom2012}; \citealt{Brough2017}; \citealt{vandeSande2021}). This would allow an exploration of the connection between internal physical properties and their relation to the global morphology of galaxies.
\end{itemize}

Looking ahead to the application of our model in the context of the next generation surveys like Euclid and LSST, it is important to emphasize that although higher quality and deeper imaging will be available, that is not enough to either guarantee for a correct estimate of any structural parameter nor for any model performance. Previous works in this direction (\citealt{Nevin2019} for SDSS, \citet{Tarsitano2018} for the Dark Energy Survey, and \citealt{Martin2022} for the upcoming LSST data) clearly suggest that a more detailed identification of biases and uncertainties present in imaging data is necessary to fully exploit the capabilities of the surveys \citep[see also][]{Walmsley2024}. Future work could involve training models on simulated data with varying $\left<S/N\right>$ levels, hence helping models to perform better with noisier observations, or to mimic the $\left<S/N\right>$ level of a particular survey.

\section*{Acknowledgements}

GAA and JAVM thank support from the CONAHCYT Postdoctoral program \textit{Estancias Postdoctorales por Mexico}. HMHT acknowledges CONAHCYT project CF-G-543 entitled ``Arqueología y filogenética de dinosaurios galácticos: formación y evolución de galaxias masivas apagadas''. GAA, GFP, HMHT, and JAVM acknowledge support from CONAHCYT project CF-2023-G-1052 entitle ``Sinergia y Retos del Censo LSST del Observatorio Vera Rubin para la Astrofísica, la Ciencia de datos, la Química y otras disciplinas''. RD gratefully acknowledges support by the ANID BASAL project FB210003. YJT acknowledges financial support from the Spanish MINECO grant PID2022-136598NB-C32 and from the State Agency for Research of the Spanish MCIU through the Center of Excellence Severo Ochoa award to the Instituto de Astrofísica de Andalucía (SEV-2017-0709) and grant CEX2021-001131-S funded by MCIN/AEI/ 10.13039/501100011033. GM acknowledges support from the UK STFC under grant ST/X000982/1. W.J.P. has been supported by the Polish National Science Center project UMO-2023/51/D/ST9/00147. CS acknowledges support from the Agencia Nacional de Investigaci\'on y Desarrollo (ANID) through Basal project FB210003. We thank Dr. Alister Graham for his valuable comments and suggestions on the present work. The authors thank the computing credits provided through ``Proyectos de investigación en la Nube UNAM-AWS''. The authors acknowledge the LSST-MX Consortium for the management to facilitate their participation in the Rubin Observatory (\url{www.fisica.ugto.mx/~lsstmx/}).

%%%%%%%%%%%%%%%%%%%%%%%%%%%%%%%%%%%%%%%%%%%%%%%%%%
\section*{Data Availability}

The code and data underlying this article can be shared on reasonable request to the corresponding author.

%%%%%%%%%%%%%%%%%%%% REFERENCES %%%%%%%%%%%%%%%%%%

\bibliographystyle{mnras}
\bibliography{bibfile}

%%%%%%%%%%%%%%%%%%%%%%%%%%%%%%%%%%%%%%%%%%%%%%%%%%

%%%%%%%%%%%%%%%%% APPENDICES %%%%%%%%%%%%%%%%%%%%%

\appendix

\section{Performance metrics description} \label{app:performance_metrics}

In this appendix, we provide definitions and explanations of the performance metrics used to evaluate the classification tasks presented in this study. These metrics are commonly used in ML and are essential for assessing various aspects of classification performance:

\begin{itemize}
    \item Confusion Matrix (CM): compares the true or real classes to the predicted classes. By convention in ML, the rows and columns correspond to the real and predicted (by the ML method) classes, respectively. The entry \textit{i,j} in a CM is the number of observations actually in class \textit{i}, but predicted to be in class \textit{j}. Hence, the diagonal of the CM indicates the successful predictions of the ML model, while off-diagonal values indicate the failing predictions. The best performance is a diagonal matrix.

    \item Accuracy: fraction of correctly predicted objects to the total number of objects. The best performance is 1.

    \item Precision: fraction of true positive predictions to the total positive predictions, denoting how many of the positive predictions are actually correct. Hence, this is a contamination indicator. The best value is 1.

    \item Recall: ratio of true positive predictions to the total real positive instances. This is a completeness proxy, since it indicates how well the model recognizes the real positive instances. The best value is 1.

    \item F1-score: can be interpreted as a harmonic mean of the precision and recall. The best value is 1 and the worst value is 0.

    \item AUC-ROC: refers to the Area Under the ROC Curve \citep[][]{Bradley1997,Fawcett2005}. A Receiver Operating Characteristic (ROC) curve represents the true positive rate (TPR) as function of the false positive rate (FPR) for various probability thresholds. AUC-ROC provides a single value that quantifies the overall performance of the classification model across all possible thresholds. This is a separability indicator, since it shows how well a ML model is capable of distinguishing between classes. The higher the AUC-ROC value (best value is 1), the better the classification performance and the discrimination ability of the model. ROC curve and AUC-ROC are most commonly associated with binary classification, where TPR and FPR are unambigously defined. For a multiclass task, these metrics can still be obtained by two supported approaches: the \textit{one-vs-one} (OvO) algorithm, which compares every unique pairwise combination of classes, and the \textit{one-vs-rest} (OvR) algorithms, which compares each class against all the others (assumed as one). Then, there will be one ROC curve and one AUC value for each class, which can be averaged (e. g., macro average) to summarize overall model performance. In this work, both OvO and OvR approaches yield the same result.
\end{itemize}

\section{Per-class performance metrics} \label{app:perclass_metrics}

\subsection{Direct classification tasks}
Table~\ref{tab:class_report} presents the per-class precision, recall, and F1-score metrics for the 2cats (upper) and 5cats (bottom) direct classifications using the S2+C configuration on the test subset. The table also includes the macro average and the weighted average for each metric, where the weighted average takes into account the presence of each class in the true data sample. Additionally, it shows the accuracy and AUC-ROC (macro average). For the 2cats classification, there is a balanced performance across both classes in the three metrics (precision, recall, and F1-score), similar to the results observed in Sec.~\ref{sec:direct_class}. This consistency indicates that the model is performing well and is reliable in predicting both galaxy types, without favoring one class over the other. 

In the 5cats direct classification (bottom panel of Table~\ref{tab:class_report}), performance varies significantly among the classes, indicating that the model performs better for some galaxy classes than others. Consistent with our findings in the CM (Fig.~\ref{fig:CM_2p_14p}), Class 0 (elliptical) and Class 3 (Sbc--Scd) have the best performance, with F1-scores of 72\% and 70\%, respectively. Next in performance are the Classes 2 (Sa--Sb), 1 (S0$^-$--S0a), and 4 (Sd--Irr), which have lower performance than Class 0 by 8-9\%, 12-20\%, and 22-41\%, respectively. The poor performance of Class 4 (e.g. 39\% in F1-score) could be attributed to its significant under-representation in the dataset, therefore the model has limited data to learn from.

As observed in Sec.~\ref{sec:direct_class}, precision and recall are similar for each class, especially for Class 0 and Class 2, where they differ by only 1\%, indicating a balanced performance. For Class 1 and Class 3, the difference is slightly larger at 6\%, suggesting a low variability in the model's performance for these two classes. However, Class 4 shows the largest difference with a 17\% gap between precision (49\%) and recall (32\%), highlighting the model's struggle with this under-represented class. This significant disparity indicates that while the model is conservative and identifies many of the galaxies it labels as Class 4 (better precision), it misses a large portion of true instances (lower recall), leading to inconsistent performance for this class.

On average, the precision and recall differ by 3\% for the macro average and by 1\% for the weighted average, suggesting a relatively balanced model overall, though individual class performance may vary. The macro averages for the three metrics range from 59\% to 62\%, and the weighted average goes between 64\% and 65\%, indicating moderate overall performance across all classes.

%------------------------Table----------------------------------%
\begin{table}
    \centering
    \caption{Precision, recall, F1-score, accuracy, and AUC-ROC evaluated over the test subset for the 2cats (upper) and 5cats (bottom) classifications using the S2+C configuration.}
    \label{tab:class_report}
    \begin{tabular}{ l *{3}{c} }
        \hline
        \multicolumn{4}{c}{\textbf{2cats with S2+C}} \\
        \hline
        \multicolumn{1}{c}{Class} & Precision & Recall & F1-score \\
        \hline
        0: E--S0a & 0.85 & 0.86 & 0.85 \\
        1: Sa--Irr & 0.90 & 0.89 & 0.89 \\
        \hline
        Macro Avg & 0.87 & 0.87 & 0.87 \\
        Weighted Avg & 0.88 & 0.88 & 0.88 \\
        \multicolumn{2}{l}{Accuracy: 0.88} & \multicolumn{2}{c}{AUC-ROC: 0.95} \\
        \hline
        \hline
        \multicolumn{4}{c}{\textbf{5cats with S2+C}} \\
        \hline
        0: E & 0.71 & 0.73 & 0.72 \\
        1: S0$^-$--S0a & 0.59 & 0.53 & 0.56 \\
        2: Sa--Sb & 0.63 & 0.64 & 0.64 \\
        3: Sbc--Scd & 0.67 & 0.73 & 0.70 \\
        4: Sd--Irr & 0.49 & 0.32 & 0.39 \\
        \hline
        Macro Avg & 0.62 & 0.59 & 0.60 \\
        Weighted Avg & 0.64 & 0.65 & 0.64 \\
        \multicolumn{2}{l}{Accuracy: 0.65} & \multicolumn{2}{c}{AUC-ROC: 0.90} \\
        \hline
    \end{tabular}
\end{table}

\subsection{Hierarchical classification tasks}
Table~\ref{tab:hier_class_report} shows the precision, recall and F1-score per class metrics calculated from the test subset for the Hier1 (upper panel) and Hier2 (bottom panel) classifications. This table also shows the accuracy and AUC-ROC (macro average). The performance metrics of both hierarchical classifications are quite similar, with differences of up to 2\% only for Classes 0 and 1. Comparing these results with the 5cats direct classification (bottom panel of Table~\ref{tab:class_report}) we observe that the three classifications yield very similar results. The differences among the three classifications are up to 3\% in the metrics. Therefore, as discuss in Sec.~\ref{sec:hier_class}, Hier1, Hier2, and 5cats are comparable, regardless of their different classification approach.

%------------------------Table----------------------------------%
\begin{table}
    \centering
    \caption{Precision, recall, F1-score, accuracy, and AUC-ROC evaluated over the test subset for the Hier1 (upper) and Hier2 (bottom) hierarchical classifications (Fig.~\ref{fig:CM_hier}).}
    \label{tab:hier_class_report}
    \begin{tabular}{ l *{3}{c} }
        \hline
        \multicolumn{4}{c}{\textbf{Hier1}} \\
        \multicolumn{4}{c}{(2cats: S2+C, Early: S2+C, Late: S2+C)} \\
        \hline
        \multicolumn{1}{c}{Class} & Precision & Recall & F1-score \\
        \hline
        0: E & 0.71 & 0.73 & 0.72 \\
        1: S0$^-$--S0a & 0.58 & 0.56 & 0.57 \\
        2: Sa--Sb & 0.64 & 0.63 & 0.63 \\
        3: Sbc--Scd & 0.67 & 0.73 & 0.70 \\
        4: Sd--Irr & 0.50 & 0.33 & 0.40 \\
        \hline
        Macro avg & 0.62 & 0.60 & 0.61 \\
        Weighted avg & 0.64 & 0.65 & 0.64 \\
        \multicolumn{2}{l}{Accuracy: 0.65} & \multicolumn{2}{c}{AUC-ROC: 0.87} \\
        \hline\hline
        \multicolumn{4}{c}{\textbf{Hier2}} \\
        \multicolumn{4}{c}{(2cats: S2+C, Early: Structural2, Late: S2+C)} \\
        \hline
        0: E & 0.69 & 0.74 & 0.71 \\
        1: S0$^-$--S0a & 0.58 & 0.54 & 0.56 \\
        2: Sa--Sb & 0.64 & 0.63 & 0.63 \\
        3: Sbc--Scd & 0.67 & 0.73 & 0.70 \\
        4: Sd--Irr & 0.50 & 0.33 & 0.40 \\
        \hline
        Macro avg & 0.62 & 0.59 & 0.60 \\
        Weighted avg & 0.64 & 0.64 & 0.64 \\
        \multicolumn{2}{l}{Accuracy: 0.64} & \multicolumn{2}{c}{AUC-ROC: 0.87} \\
        \hline
    \end{tabular}
\end{table}

\section{Colour--magnitude diagram} 
\label{app:col-mag}

\begin{figure}
     \centering
     \begin{tabular}{cc}
        \includegraphics[width=0.8\columnwidth,angle=0]{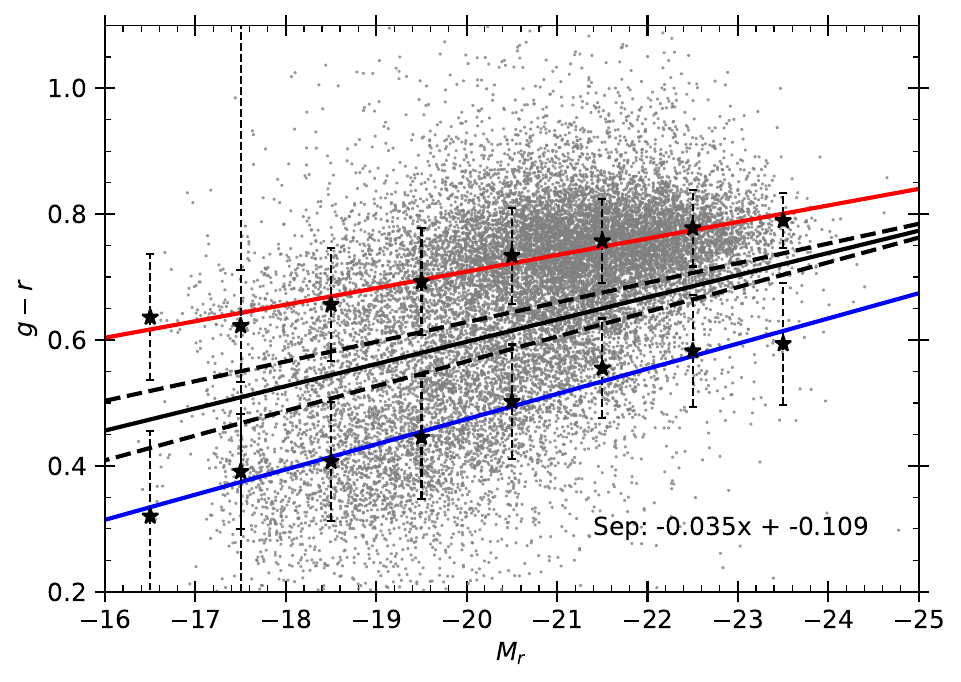}
     \end{tabular}
\caption{The colour--magnitude diagram for the data sample galaxies in the present paper (grey symbols). The blue and red lines show the blue, and red regions, respectively. The region between the dashed black lines corresponds to the green valley and the solid black line corresponds to the division line between the red and blue regions. This line is used to find the discordant galaxies presented in Figure~\ref{fig:color_magnitude}.}
\label{fig:color_magnitude2}
\end{figure}

Figure~\ref{fig:color_magnitude2} shows the colour--magnitude ($\left(g-r\right)$ vs. $M_r$) diagram for the entire data sample in this work (grey symbols), where $M_r$ is the absolute magnitude in the $r-$band, emphasizing the red region (overdense region at $(g-r)$ > 0.6; red line fit), the blue region (overdense region at $(g-r)$ < 0.6; blue line fit), and the division line (solid black line). Colours and magnitudes were corrected as described in Sec.~\ref{sec:m_params}. The fitting to the red and blue regions were estimated by setting a random division between them at $(g-r)$ = 0.6, and then estimating the $mean$ in colour given a $M_r$ bin in the red and blue regions, and also estimating the corresponding standard deviation ($SD$). The \textit{mean - SD} values for the red sequence, as well as the \textit{mean + SD} values for the blue sequence were fit. Given the fact that these lines are not perpendicular, we considered the bisector of the angle formed at their intersection. This bisector is then used as the new division line between the red and blue region. We repeat this process until the resulting division line does not varies significantly, and we adopt it to separate the red region from the blue region. As reference, we plot final $SD$ error bars and their corresponding fit in dashed black lines. The division line is adopted in the lower panels in solid black line.

\section{Performance metrics excluding discordant and highly inclined galaxies} \label{app:metrics_without_discordant}

The experiment excluding the discordant galaxies in the dataset ($\sim$3,375 galaxies) for the 5cats direct classification using the S2+C parameter configuration achieves an accuracy of 66\% and AUC-ROC of 91\%, after retraining the model and evaluating it on the test subset. Table~\ref{tab:discordant_class_report} shows the per-class performance metrics for this experiment, in addition to the macro and weighted averages. Comparing these results with Table~\ref{tab:class_report}, we observe some improvements, particularly on Classes 1 and 3.

%------------------------Table----------------------------------%
\begin{table}
    \centering
    \caption{Precision, recall, F1-score, accuracy, and AUC-ROC evaluated over the test subset for the experiment excluding discordant and highly-inclined galaxies.}
    \label{tab:discordant_class_report}
    \begin{tabular}{ l *{3}{c} }
        \hline
        \multicolumn{1}{c}{Class} & Precision & Recall & F1-score \\
        \hline
        0: E & 0.70 & 0.73 & 0.72 \\
        1: S0$^-$--S0a & 0.61 & 0.59 & 0.60 \\
        2: Sa--Sb  & 0.68 & 0.62 & 0.65 \\
        3: Sbc--Scd & 0.67 & 0.80 & 0.73 \\
        4: Sd--Irr & 0.53 & 0.31 & 0.39 \\
        \hline
        Macro avg & 0.64 & 0.61 & 0.62 \\
        Weighted avg & 0.66 & 0.66 & 0.66 \\
        \multicolumn{2}{l}{Accuracy: 0.66} & \multicolumn{2}{c}{AUC-ROC: 0.91} \\
        \hline
    \end{tabular}
\end{table}

%%%%%%%%%%%%%%%%%%%%%%%%%%%%%%%%%%%%%%%%%%%%%%%%%%

% Don't change these lines
\bsp	% typesetting comment
\label{lastpage}
\end{document}